\pdfoutput=1
\documentclass[aps,prd,amsmath,floats,floatfix, twocolumn,
superscriptaddress,nofootinbib,showpacs]{revtex4-1}

\usepackage[T1]{fontenc}
\usepackage[utf8]{inputenc}
\usepackage{lmodern}
\usepackage{enumitem}  

\usepackage{verbatim}
\usepackage{bbold}
\usepackage[dvipsnames, usenames]{xcolor}
\definecolor{linkcolor}{rgb}{0.0,0.3,0.5}
\usepackage[hypertexnames=false, unicode, colorlinks=true, linkcolor=linkcolor,
citecolor=linkcolor, filecolor=linkcolor,urlcolor=linkcolor,
pdfusetitle]{hyperref}
\usepackage[all]{hypcap}
\usepackage{graphicx}
\usepackage{xspace}
\usepackage{tensor} 
\usepackage{amssymb}
\usepackage[normalem]{ulem} 
\usepackage{bm} 
\usepackage{mathrsfs}  
\usepackage[caption=false]{subfig}
\usepackage{tikz}
\usetikzlibrary{shapes.misc, positioning}
\usepackage{microtype}

\usepackage[english]{babel}
\usepackage{blindtext}

\graphicspath{%
  {figs/}%
}

\DeclareMathAlphabet{\mathpzc}{OT1}{pzc}{m}{it}

\newcommand{\sma}[1]{\textcolor{WildStrawberry}{[Sizheng: #1]}}

\newcommand{\Note}[1]{\textcolor{blue}{\textbf{[#1]}}}

\newcommand{\etal}{\textit{et al.\ }}
\newcommand{\mycomment}[1]{}

\numberwithin{equation}{section}
\renewcommand{\theequation}{\arabic{section}.\arabic{equation}}

\usepackage{orcidlink}

\begin{document}

\title{Fully relativistic three-dimensional Cauchy-characteristic matching for physical degrees of freedom}

\newcommand{\Cornell}{\affiliation{Cornell Center for Astrophysics
    and Planetary Science, Cornell University, Ithaca, New York 14853, USA}}
\newcommand\CornellPhys{\affiliation{Department of Physics, Cornell
    University, Ithaca, New York 14853, USA}}
\newcommand\Caltech{\affiliation{TAPIR 350-17, California Institute of
    Technology, 1200 E California Boulevard, Pasadena, CA 91125, USA}}
\newcommand{\AEI}{\affiliation{Max Planck Institute for Gravitational Physics
    (Albert Einstein Institute), Am M\"uhlenberg 1, Potsdam 14476, Germany}} %
\newcommand{\UMassD}{\affiliation{Department of Mathematics,
    Center for Scientific Computing and Visualization Research,
    University of Massachusetts, Dartmouth, MA 02747, USA}}
\newcommand\Olemiss{\affiliation{Department of Physics and Astronomy,
    The University of Mississippi, University, MS 38677, USA}}
\newcommand{\Bham}{\affiliation{School of Physics and Astronomy and Institute
    for Gravitational Wave Astronomy, University of Birmingham, Birmingham, B15
    2TT, UK}}
\newcommand{\Fullerton}{\affiliation{Nicholas and Lee Begovich Center for
    Gravitational-Wave Physics and Astronomy, California State University
    Fullerton, Fullerton, CA 92834, USA}}
\newcommand{\Icts}{\affiliation{International Centre for Theoretical Sciences,
    Tata Institute of Fundamental Research, Bangalore 560089, India}}


\author{Sizheng Ma\orcidlink{0000-0002-4645-453X}}
\email{sma@caltech.edu}
\Caltech

\author{Jordan Moxon\orcidlink{0000-0001-9891-8677}}
\Caltech

\author{Mark A. Scheel\orcidlink{0000-0001-6656-9134}}
\Caltech

\author{Kyle C. Nelli\orcidlink{0000-0003-2426-8768}}
\Caltech

\author{Nils Deppe\orcidlink{0000-0003-4557-4115}}
\CornellPhys
\Cornell
\Caltech

\author{Marceline S.~Bonilla\orcidlink{0000-0003-4502-528X}}
\Fullerton

\author{Lawrence E.~Kidder\orcidlink{0000-0001-5392-7342}}
\Cornell

\author{Prayush Kumar\orcidlink{0000-0001-5523-4603}}
\Icts

\author{Geoffrey Lovelace\orcidlink{0000-0002-7084-1070}}
\Fullerton

\author{William Throwe\orcidlink{0000-0001-5059-4378}}
\Cornell

\author{Nils L.~Vu\orcidlink{0000-0002-5767-3949}}
\Caltech

\hypersetup{pdfauthor={Ma et al.}}

\date{\today}

\begin{abstract}
A fully relativistic three-dimensional Cauchy-characteristic matching (CCM) algorithm is implemented for physical degrees of freedom in a numerical relativity code SpECTRE. The method is free of approximations and can be applied to any physical system. We test the algorithm with various scenarios involving smooth data, including the propagation of Teukolsky waves within a flat background, the perturbation of a Kerr black hole with a Teukolsky wave, and the injection of a gravitational-wave pulse from the characteristic grid. Our investigations reveal no numerical instabilities in the simulations. In addition, the tests indicate that the CCM algorithm effectively directs characteristic information into the inner Cauchy system, yielding higher precision in waveforms and smaller violations of Bondi-gauge constraints, especially when the outer boundary of the Cauchy evolution is at a smaller radius.

\end{abstract}

\maketitle


\section{Introduction}
\label{sec:introduction}

Since the detection of GW150914 \cite{Abbott:2016blz}, gravitational wave (GW) astronomy has become a flourishing field. Accurate modeling of GW signals is a key ingredient in extracting signals from detector noise and understanding the properties of sources. To date, numerical relativity
(NR) remains the only \textit{ab initio} method to simulate the major sources of the GW signals: the
coalescence of binary black hole (BBH) systems.

Generally speaking, the formulations of NR can be classified into two groups: Cauchy \cite{Baumgarte_Shaprio:NumRel,SXSCatalog} and characteristic \cite{winicour2012characteristic,1992anr..conf...20B,Bishop:1996gt,Bishop:1997ik,Bishop:1998uk,Barkett:2019uae,Moxon:2020gha,Moxon:2021gbv} formalism, depending on how spacetime is foliated\footnote{The third group adopts hyperboloidal slicing \cite{frauendiener2004conformal,Buchman:2009ew,Zenginoglu:2008pw,Vano-Vinuales:2014koa,Vano-Vinuales:2014ada}. Its discussion is beyond the scope of this paper.}. For the Cauchy approach, a spacelike foliation is adopted, and Einstein's equations are split into evolution and constraint sets. This formalism has successfully led to high-accuracy simulations of BBH systems \cite{SXSCatalog}. 

On the other hand, in the characteristic approach, spacetime is sliced into a sequence of null hypersurfaces that extend to future null infinity. Einstein's equations are formulated in terms of the unambiguous geometric treatment of gravitational radiation in curved spacetimes due to  Bondi \etal \cite{1962RSPSA.269...21B} Sachs \cite{Sachs:1962wk} and Penrose \cite{PhysRevLett.10.66}. Meanwhile, future null infinity is rigorously encompassed on the characteristic grid via a compactified coordinate system and treated as a perfect absorbing outer boundary. In this way, one is able to extract faithful GWs with the characteristic formalism at future null infinity without any ambiguity \cite{Babiuc:2005pg,Boyle:2009vi,Chu:2015kft,Taylor:2013zia,Iozzo:2020jcu}. However, the characteristic method cannot evolve the near-field region of BBHs when caustics of null rays are present \cite{corkill1983numerical,friedrich1983characteristic,Stewart1986,Frittelli:1997rw,Bhagwat:2017tkm,Baumgarte:2023tdh}, because for this method coordinates are chosen to follow null rays so caustics result in coordinate singularities. Therefore, in practice, one can use the Cauchy evolution to simulate the near-zone of the systems and construct metric data on a timelike worldtube \cite{Winicour:2011jn,Barkett:2019uae,Moxon:2020gha,Moxon:2021gbv}. Then the characteristic system propagates the worldtube data nonlinearly to future null infinity, which in turn yields GW information there. This procedure of extracting GWs is known as \textit{Cauchy-characteristic evolution} (CCE) \cite{winicour2012characteristic,1992anr..conf...20B,Bishop:1996gt,Bishop:1997ik,Bishop:1998uk,Szilagyi:2000xu,Barkett:2019uae,Moxon:2020gha,Moxon:2021gbv}. Studies of characteristic evolution and CCE date back to the 1980s. Isaacson \etal \cite{doi:10.1063/1.525904} and Winicour \cite{1983JMP....24.1193W,doi:10.1063/1.526472} considered a prototype of CCE by shrinking the worldtube to a timelike geodesic and investigated the GWs emitted by an axially symmetric ideal fluid. More complete and complicated CCE systems were developed later \cite{gomez1994null,Bishop:1996gt,Bishop:1997ik,Gomez:2001pb,Bishop:2003bz,Reisswig:2006nt,Gomez:2007cj,Gomez:2007cj,Babiuc:2008qy,Reisswig:2012ka}. Early applications of the characteristic evolution were focused on simulating generic three-dimensional (3D) single-black-hole spacetimes \cite{Gomez:1998uj}, Einstein-perfect fluid systems \cite{Papadopoulos:1999kt,Bishop:1999yg,Siebel:2001ii,Siebel:2001dp,Bishop:2004mt}, Einstein-Klein-Gordon systems \cite{1992JCoPh..98...11G,1992JMP....33.1445G,Barreto:1996rc,Siebel:2001dp,Barreto:2004fn}, (nonlinear) perturbations of BHs \cite{Papadopoulos:2001zf,Campanelli:2000in,Husa:2001pk,Zlochower:2003yh}, event horizons \cite{Winicour:1999ba,Husa:1999nm,Gomez:2000kn}, fissioning white holes \cite{Gomez:2002ev}, extreme mass ratio inspirals \cite{Bishop:2003bs}, stellar core collapse \cite{Siebel:2003sp}, as well as linearized systems \cite{Bishop_2005}. By using finite-difference methods, PITT null \cite{Bishop:1996gt,Bishop:1997ik,Bishop:1998uk} was the first code to implement CCE and characteristic evolution, which led to the first CCE simulation of BBH systems \cite{Reisswig:2009us,Reisswig:2009rx,Babiuc:2010ze,Babiuc:2011qi}. The code was also used to extract GWs emitted by rotating stellar core collapse \cite{Reisswig:2010cd}. Later, a spectral algorithm for CCE was built as a module in SpEC \cite{Handmer:2014qha,Handmer:2015dsa,Handmer:2016mls,Barkett:2019uae} and SpECTRE \cite{Moxon:2020gha,Moxon:2021gbv}, developed by the SXS collaboration \cite{SXSWebsite,SpECwebsite,SXSCatalog,Kidder:2016hev,spectrecode}. Bhagwat \etal \cite{Bhagwat:2017tkm} used SpEC CCE to investigate the start time of BBH ringdown. And SpECTRE CCE has been applied to computing memory effects \cite{Mitman:2020pbt,Mitman:2020bjf}, fixing the Bondi-Metzner-Sachs frame of GWs \cite{Mitman:2021xkq,MaganaZertuche:2021syq}, extracting GWs emitted by black hole-neutron star binaries \cite{Foucart:2020xkt}, computing GW echoes \cite{Ma:2022xmp}, and constructing a NR surrogate model \cite{Yoo:2023spi}.

Although CCE has led to high-accuracy and unambiguous GWs at future null infinity, CCE's data flow is one-way, meaning that the Cauchy evolution does not depend at all on the characteristic evolution.  This is inaccurate because for a nonlinear set of equations like general relativity, outgoing radiation at arbitrarily far distances can backscatter off the spacetime curvature and eventually affect the source; the Cauchy evolution (with or without CCE) fails to capture this backscatter.  To explain this in more detail, note that to perform a Cauchy simulation, the spatial Cauchy domain is typically truncated at a finite distance from the source, with suitable boundary conditions provided at the artificial outer boundary\footnote{In this paper we restrict our discussions to the outer boundary.}. Ideally speaking, perfect boundary conditions would make the artificial boundary as transparent as possible so that the numerical solution is identical to one that would be evolved on an infinite domain, and these boundary conditions would ideally include nonlinear backscatter. On the contrary, if poor boundary conditions are prescribed, not only will the backscatter be incorrect, but also spurious reflection can be introduced at the boundary and contaminate the whole simulation. In SpEC \cite{SpECwebsite} and SpECTRE \cite{Kidder:2016hev,spectrecode}, the generalized harmonic (GH) evolution system \cite{Lindblom:2005qh} is adopted for the Cauchy simulation, whose boundary conditions can be divided into three subsets: constraint-preserving, physical, and gauge boundary conditions \cite{Kidder:2004rw}. Effort has been made to improve the accuracy of these boundary conditions, such as Refs.~\cite{Buchman:2006xf,Buchman:2007pj,Rinne:2007ui,Rinne:2008vn}. In particular, the boundary conditions on the physical degrees of freedom are expected to encode the information of the backscattered (incoming) GWs that enter the Cauchy domain. Accurate modeling of the backscattered radiation is not a trivial task. Although there were some attempts \cite{Buchman:2006xf,Buchman:2007pj} to improve the physical boundary conditions, in most SpEC production simulations \cite{SXSCatalog} the incoming GWs at the boundary are treated by freezing the Weyl scalar $\psi_0$ to zero \cite{Kidder:2004rw}, which effectively eliminates all backscatter from beyond the outer boundary.

It was pointed out that the characteristic evolution is a natural system to compute the value of the backscattered radiation in an exact and efficient way, e.g., see Ref.~\cite{winicour2012characteristic} and references therein. A matching of the internal Cauchy system and the exterior characteristic system is expected to provide accurate physical boundary conditions for the Cauchy module. In this way, the interface between the two grids is transparent and GWs can pass cleanly off of and onto the Cauchy grid. This algorithm is known as \textit{Cauchy-characteristic matching} (CCM). Historically, the idea of CCM was outlined in Refs.~\cite{d'inverno_1992} and \cite{Bishop_1993}. Then the algorithm was applied to the evolution of a scalar field on a flat background \cite{1994CQGra..11.1463C,PhysRevLett.76.4303}, and around a Schwarzschild BH \cite{Papadopoulos:1996pr} (with metric being fixed). The CCM simulation of gravitational systems was also visited by a series of papers \cite{PhysRevD.52.6863,PhysRevD.52.6868,PhysRevD.54.4919,PhysRevD.56.772,dInverno:2000uvh} that assumed cylindrical symmetry \cite{PhysRevD.52.6863,PhysRevD.52.6868,dInverno:2000uvh} and axial symmetry \cite{PhysRevD.54.4919,PhysRevD.56.772}. Meanwhile, CCM was used to study an Einstein-perfect fluid system \cite{PhysRevD.58.044019} and an Einstein-Klein-Gordon system \cite{Gomez:1996dy} with spherical symmetry. Going to the 3D regime, Bishop \etal investigated a scalar wave  \cite{1997JCoPh.136..140B}. Szilagyi \etal \cite{Szilagyi:2002kv} performed the matching in linearized harmonic coordinates. An alternative to CCM is Cauchy-perturbative matching \cite{BinaryBlackHoleGrandChallengeAlliance:1997aaw,Rupright:1998uw,Rezzolla:1998xx}, where the exterior region is not evolved fully nonlinearly with a characteristic code but instead is treated as a linearized Schwarzschild BH. This algorithm led to a simulation of a 3D Teukolsky wave \cite{PhysRevD.26.745} propagating on a flat background \cite{BinaryBlackHoleGrandChallengeAlliance:1997aaw}. Later, this topic was revisited \cite{Zink:2005pe} in 2005 after years of progress in numerical relativity. However, until now, all the existing matching algorithms for the gravitational sector are based on either assumptions (symmetries) or approximations (perturbative matching, linearized equations); a full matching in three spatial dimensions is still missing. Further, although the existence and uniqueness of CCE solutions have been established in \cite{Frittelli:1999yr,Gomez:2003ew}, and a linearized characteristic system was found to be symmetric hyperbolic\footnote{See also \cite{3fc386c8-ab39-37dd-8ec6-272c32120063} for the well-posedness of the characteristic formulation for the Maxwell equations. } \cite{Frittelli:2004pk}, the full CCE system is only weakly hyperbolic \cite{Giannakopoulos:2020dih,Giannakopoulos:2021pnh,Giannakopoulos:2023zzm}, rendering CCM not well-posed.

As a step toward addressing those questions, in this paper, we perform fully relativistic 3D CCM for physical degrees of freedom of gravitational fields without any approximation. The code is implemented in SpECTRE \cite{Kidder:2016hev,spectrecode}. Unlike CCE, the data in CCM flows in both directions, meaning that the Cauchy and characteristic systems need to be evolved simultaneously. The communication from the Cauchy to the characteristic system has been discussed extensively \cite{Barkett:2019uae,Moxon:2020gha,Moxon:2021gbv}. In this paper, we will be explaining how to feed the information of the characteristic module back to the Cauchy system.

This paper is organized as follows. In Sec.~\ref{sec:GH}, we review the Cauchy evolution in SpECTRE, with particular attention given to its physical boundary conditions. Next in Sec.~\ref{sec:CCE} we discuss some basic information about the characteristic module in SpECTRE. Then a thorough algorithm to complete the matching procedure is introduced in Sec.~\ref{sec:CCM}. Our code is tested with several types of physical systems in Sec.~\ref{sec:tests}. Finally, we summarize the results in Sec.~\ref{sec:conclusion}.

Throughout this paper we use Latin indices $i, j, k, \ldots$ to denote 3D spatial components; and Greek indices $\mu,\nu,\ldots$ for 4D spacetime components. We generally avoid using abstract indices, denoted by Latin letters  from the first part of the alphabet $a, b, \ldots$, to keep the text concise, unless  stated otherwise. 

\section{Summary of the Cauchy evolution and its boundary conditions}
\label{sec:GH}
The detailed communication (matching) algorithm depends on the formulation of the Cauchy evolution. For instance, the perturbative matching in Ref.~\cite{BinaryBlackHoleGrandChallengeAlliance:1997aaw} was performed through Dirichlet and Sommerfeld boundary conditions. In SpECTRE, the Cauchy data are evolved with the GH formalism \cite{Lindblom:2005qh}. Outer boundary conditions are imposed via the Bjørhus method \cite{doi:10.1137/0916035,Kidder:2004rw}: the time derivatives of the incoming characteristic fields are replaced on the boundary. In this section, we provide a brief overview of the Cauchy evolution and refer the reader to Ref.~\cite{Lindblom:2005qh} for more details. We specifically give more attention to the physical subset of the boundary conditions \cite{Kidder:2004rw}.

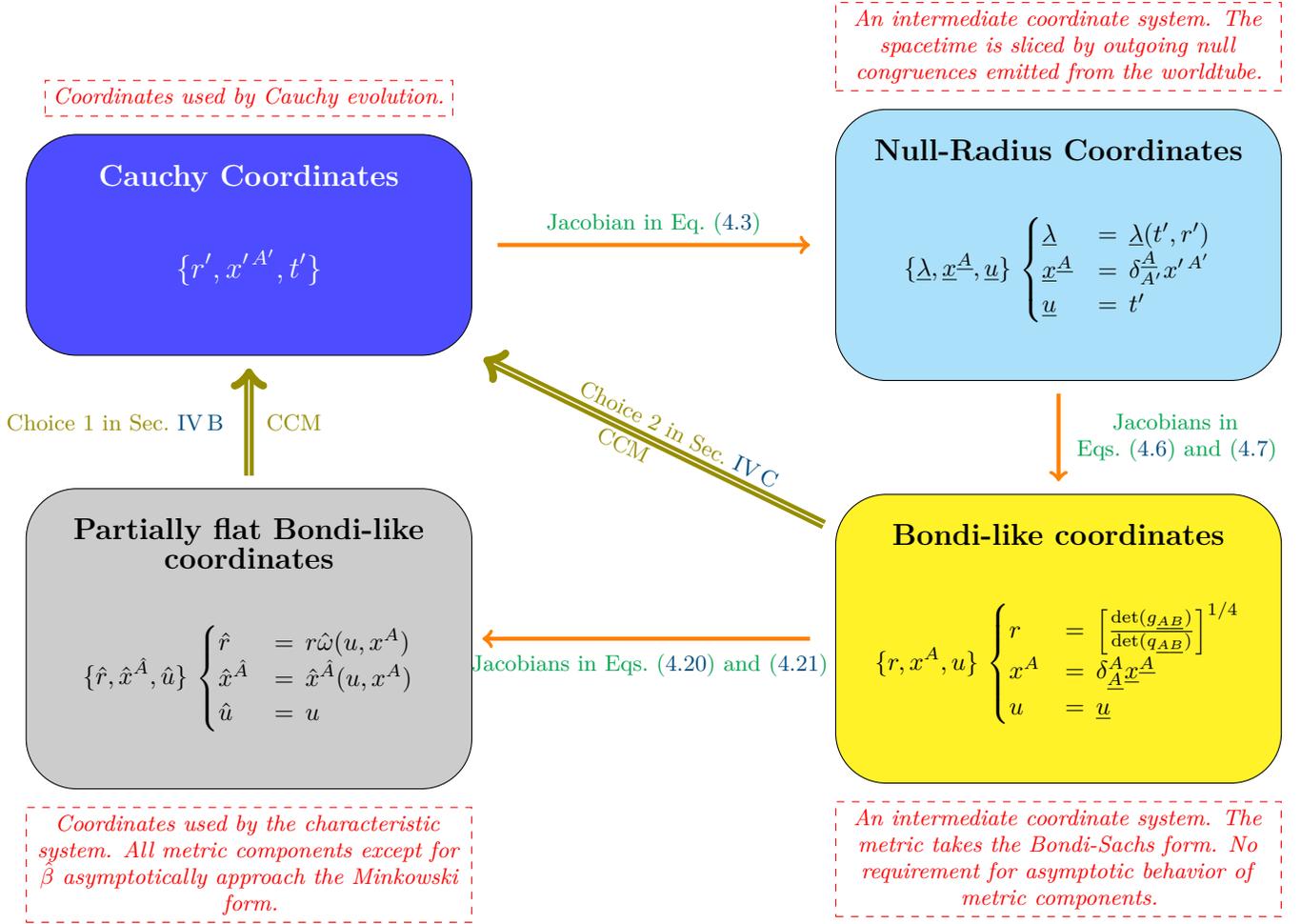
\begin{figure*}[tb]
\begin{tikzpicture}
\node (cauchy) [draw, rounded corners=20pt,
                 text width=0.3\linewidth,    
                 align=flush center, 
                 inner sep=12 pt,
                 fill=blue!70,text=white,
                 label={[draw, dashed, align=center,label distance=7pt, color=red] 90:{\textit{\small Coordinates used by Cauchy evolution.}}},
                 minimum height=80pt]%
                 at (-1,0)
{\textbf{\large Cauchy Coordinates}\par
 \large 
$$\{r^\prime, x^{\prime\, A^\prime},t^\prime\}$$};
\node (NullRadius) [draw, rounded corners=20pt,
                     text width=0.3\linewidth,    
                     align=flush center, 
                     inner sep=12 pt,
                     fill=cyan!30,text=black,
                     label={[draw, dashed, align=center,label distance=7pt, color=red]{\begin{minipage}{6cm}\centering\small \textit{An intermediate coordinate system. The spacetime is sliced by outgoing null congruences emitted from the worldtube.} \end{minipage}}},
                     minimum height=80pt]%
                    at (10.3,0)
{\textbf{\large Null-Radius Coordinates}\par
 \normalsize $$ \{\underline{\lambda}, \underline{x}^{\underline{A}},\underline{u}\}\, 
 \begin{cases}
     \underline{\lambda}&=\,\underline{\lambda}(t^\prime,r^\prime) \\
     \underline{x}^{\underline{A}}&=\,\delta_{A^\prime}^{\underline{A}} x^{\prime\, A^\prime} \\
     \underline{u}&=\,t^\prime
 \end{cases}$$};
\node (partially-bondi) [draw, rounded corners=20pt,
                 text width=0.3\linewidth,    
                 align=flush center, 
                 inner sep=12 pt,
                 fill=lightgray!80,text=black,
                 label={[draw, dashed, align=center,label distance=7pt, color=red]-90:{\begin{minipage}{6cm}\centering \small \textit{Coordinates used by the characteristic system. All metric components except for $\hat{\beta}$ asymptotically approach the Minkowski form.}\end{minipage}}},
                 minimum height=80pt]%
                 at (-1,-5.5)
{\textbf{\large Partially flat Bondi-like coordinates}\par
 \normalsize 
$$ \{\hat{r}, \hat{x}^{\hat{A}},\hat{u}\}\, 
 \begin{cases}
     \hat{r}&=\,r\hat{\omega}(u,x^A) \\
     \hat{x}^{\hat{A}}&=\,\hat{x}^{\hat{A}}(u,x^A) \\
     \hat{u}&=\,u
 \end{cases}$$};
\node (Bondilike) [draw, rounded corners=20pt,
                     text width=0.3\linewidth,    
                     align=flush center, 
                     inner sep=12 pt,
                     fill=yellow!90,text=black,
                     label={[draw, dashed, align=center,label distance=7pt, color=red]-90:{\begin{minipage}{6cm}\centering\small \textit{An intermediate coordinate system. The metric takes the Bondi-Sachs form. No requirement for asymptotic behavior of metric components.} \end{minipage}}},
                     minimum height=80pt]%
                    at (10.3,-5.5)
{\textbf{\large Bondi-like coordinates}\par
 \normalsize $$ \{r, x^{A},u\}\, 
 \begin{cases}
     r&=\,\left[\frac{{\rm det}(g_{\underline{A}\underline{B}})}{{\rm det}(q_{\underline{A}\underline{B}})}\right]^{1/4} \\
     x^A&=\,\delta^{A}_{\underline{A}} \underline{x}^{\underline{A}} \\
     u&=\,\underline{u}
 \end{cases}$$};
\begin{scope}[transform canvas={yshift=0pt}]
\draw[ultra thick, -> ,color=orange,
     shorten >=10pt,shorten <=10pt
     ] (cauchy) -- (NullRadius)
    node[midway,above,yshift=-0pt, color=Green] 
     {$$ \mbox{\small Jacobian in Eq.~\eqref{eq:Jacobian_cauchy_null_radius}} $$};
\end{scope}
\begin{scope}[transform canvas={yshift=0pt}]
\draw[ultra thick, -> ,color=orange,
     shorten >=4pt,shorten <=10pt
     ] (Bondilike) -- (partially-bondi) node[midway,below,yshift=-2pt, color=Green] 
     {$$ \mbox{\small Jacobians in Eqs.~\eqref{eq:Jacobian_bondi_like_inertial} and \eqref{eq:jacobian_bondi_inertial} } $$};
\end{scope}
\begin{scope}[transform canvas={yshift=0pt}]
\draw[ultra thick, double, -> ,color=olive,
     shorten >=5pt,shorten <=5pt
     ] (partially-bondi) -- (cauchy) node[midway,right,yshift=0pt, color=olive] 
     {$$ \mbox{\small CCM } $$}
     node[midway,left,yshift=0pt, color=olive] 
     {$$ \mbox{\small Choice 1 in Sec.~\ref{sec:tetrad_transformation_Scenario_I} } $$};
\end{scope}
\begin{scope}[transform canvas={yshift=0pt}]
\draw[ultra thick, double, -> ,color=olive,
     shorten >=5pt,shorten <=5pt
     ] (Bondilike) -- (cauchy) node[midway,right,yshift=25pt, color=olive,xshift=-35pt, rotate=-26] 
     {$$ Choice 2 in Sec.~\ref{sec:tetrad_transformation_Scenario_II} $$}
     node[midway,left,yshift=-9pt, color=olive,xshift=4pt, rotate=-26] 
     {$$ CCM $$};
\end{scope}
\begin{scope}[transform canvas={yshift=0pt}]
\draw[ultra thick, -> ,color=orange,
     shorten >=5pt,shorten <=5pt
     ] (NullRadius) -- (Bondilike) node[midway,right,yshift=0pt, color=Green] 
     {\begin{minipage}{3cm}\centering\small Jacobians in \\ Eqs.~\eqref{eq:Jacobian_null_radius_bondi_like} and \eqref{eq:Jacobian_bondi_like_null_radius}\end{minipage}};
\end{scope}
\end{tikzpicture}
\caption{ Coordinate systems used in the SpECTRE CCE and CCM modules. The interior Cauchy evolution uses the Cauchy coordinates, whereas the exterior characteristic system adopts the partially flat Bondi-like coordinates. To achieve their communication, two intermediate coordinate systems are introduced.
}
\label{fig:coordinates}
\end{figure*}

The Cauchy evolution relies on the $3+1$ decomposition of a metric tensor $g_{\mu^\prime \nu^\prime}$
\begin{align}
    &ds^2=g_{\mu^\prime \nu^\prime}dx^{\prime\, \mu^\prime}dx^{\prime\, \nu^\prime}=(-\alpha^2+\beta^{i^\prime}\beta^{i^\prime}\gamma_{i^\prime j^\prime})dt^{\prime 2}\notag \\
    &+2\beta^{i^\prime}\gamma_{i^\prime j^\prime}dx^{\prime \,j^\prime}dt^\prime+\gamma_{i^\prime j^\prime}dx^{\prime \,i^\prime}dx^{\prime \,j^\prime}, \label{eq:3+1_metric}
\end{align}
with $\alpha$ the lapse function, $\beta^{i^\prime}$ the shift function, and $\gamma_{i^\prime j^\prime}$ the spatial metric\footnote{In Ref.~\cite{Lindblom:2005qh}, the authors used $\psi_{a^\prime b^\prime}$ and $g_{i^\prime j^\prime}$ to refer to the spacetime metric and the spatial metric, respectively.}. We use primes on the coordinates
to distinguish them from different coordinate systems that will be introduced later; see Fig.~\ref{fig:coordinates}. Then the vacuum Einstein equations, $R_{\mu^\prime \nu^\prime}=0$, can be cast into a first-order symmetric hyperbolic (FOSH) evolution system
\begin{align}
    \partial_{t^\prime} u^{\alpha^\prime}+\tensor[]{A}{^{k^\prime \alpha^\prime}_{\beta^\prime}}\partial_{k^\prime} u^{\beta^\prime}=F^{\alpha^\prime}, \label{FOSH}
\end{align}
where $u^{\alpha^\prime}=\{g_{\mu^\prime \nu^\prime},\Pi_{\mu^\prime \nu^\prime},\Phi_{i^\prime \mu^\prime \nu^\prime}\}$ is a collection of dynamical variables, $\Pi_{\mu^\prime \nu^\prime}=\alpha^{-1}(\beta^{i^\prime}\partial_{i^\prime}g_{\mu^\prime \nu^\prime}-\partial_{t^\prime}g_{\mu^\prime \nu^\prime})$ and $\Phi_{i^\prime \mu^\prime \nu^\prime}=\partial_{i^\prime}g_{\mu^\prime \nu^\prime}$ are related to the time and spatial derivatives of the metric.

The FOSH system in Eq.~\eqref{FOSH} is symmetric hyperbolic, and its characteristic fields $u^{\hat{\alpha}^\prime}=\tensor[]{e}{^{\hat{\alpha}^\prime}_{\beta^\prime}}u^{\beta^\prime}$ play an important role in imposing boundary conditions. Here the left eigenvectors $\tensor[]{e}{^{\hat{\alpha^\prime}}_{\beta^\prime}}$ are defined by
\begin{align}
    \tensor[]{e}{^{\hat{\alpha}^\prime}_{\mu^\prime}}s_{k^\prime}\tensor[]{A}{^{k^\prime\mu^\prime}_{\beta^\prime}}=v_{(\hat{\alpha}^\prime)}\tensor[]{e}{^{\hat{\alpha}^\prime}_{\beta^\prime}},
\end{align}
where $s^{k^\prime}$ is the outward-directed unit normal to the boundary of the computational domain:
\begin{align}
    &s^{t^\prime}=0,&s^{k^\prime}=\frac{\gamma^{i^\prime k^\prime}\partial_{i^\prime}r^\prime}{\sqrt{\gamma^{i^\prime j^\prime}\partial_{i^\prime}r^\prime\partial_{j^\prime}r^\prime}}, \label{eq:def_s}
\end{align}
and $v_{(\hat{\alpha}^\prime)}$ are the eigenvalues. As pointed out by Kidder \etal \cite{Kidder:2004rw}, a convenient way to impose the Bjørhus boundary conditions \cite{doi:10.1137/0916035} can be achieved via
\begin{align}
    d_{t^\prime}u^{\hat{\alpha}^\prime}=D_{t^\prime}u^{\hat{\alpha}^\prime}+v_{(\hat{\alpha}^\prime)}\left(d_\perp u^{\hat{\alpha}^\prime}-\left. d_\perp u^{\hat{\alpha}^\prime}\right|_{\rm BC}\right), \label{Bjorhus_bc}
\end{align}
with 
\begin{align}
    &d_{t^\prime}u^{\hat{\alpha}^\prime}\equiv\tensor[]{e}{^{\hat{\alpha}^\prime}_{\beta^\prime}}\partial_{t^\prime}u^{\beta^\prime}, && d_\perp u^{\hat{\alpha}^\prime}\equiv\tensor[]{e}{^{\hat{\alpha}^\prime}_{\beta^\prime}}s^{k^\prime}\partial_{k^\prime}u^{\beta^\prime},
\end{align}
and
\begin{align}
    D_{t^\prime}u^{\hat{\alpha}^\prime}\equiv\tensor[]{e}{^{\hat{\alpha}^\prime}_{\beta^\prime}}(-\tensor[]{A}{^{{k^\prime}\beta^\prime}_{\alpha^\prime}}\partial_{k^\prime}u^{\alpha^\prime}+F^{\beta^\prime}).
\end{align}
Here Eq.~\eqref{Bjorhus_bc} replaces the normal derivative $d_\perp u^{\hat{\alpha}^\prime}$ by its desired value $\left. d_\perp u^{\hat{\alpha}^\prime}\right|_{\rm BC}$ on the boundary while leaving the tangential derivative unchanged. 

The boundary conditions in Eq.~\eqref{Bjorhus_bc} must be imposed on each incoming characteristic field $v_{(\hat{\alpha}^\prime)}<0$ \cite{10.2307/1999902,die/1367969881,1996ArRMA.134..155S}. As discussed in Refs.~\cite{Kidder:2004rw,Lindblom:2005qh}, for the fully first-order generalized harmonic formulation there are fifty evolved variables in the Cauchy domain ($g_{\mu^\prime \nu^\prime},\Pi_{\mu^\prime \nu^\prime},\Phi_{i^\prime \mu^\prime \nu^\prime}$), and there are (for typical values of the shift vector at the outer boundary) forty incoming characteristic fields on the outer boundary. Thirty-four of these incoming fields, namely $u^{\hat{0}}_{\mu^\prime \nu^\prime}$, $u^{\hat{2}}_{i^\prime\mu^\prime \nu^\prime}$, and four components of $u^{\hat{1}-}_{\mu^\prime \nu^\prime}$ (see Eqs.~(32-34) in Ref.~\cite{Lindblom:2005qh} for their expressions), are directly related to the influx of constraint violations. So the appropriate boundary conditions for those forty fields are those that preserve the constraints; we use Eqs.~(63) through (65) of \cite{Lindblom:2005qh}, which prevent influx of constraint violations without any approximation.  Thus CCM matching is unnecessary for these fields, and furthermore CCM matching for these fields is not well-motivated because constraints are local and must be preserved independent of the solution in the characteristic domain.

This leaves six incoming characteristic fields, which are the remaining six components of $u^{\hat{1}-}_{\mu^\prime \nu^\prime}$ that are not related to constraints.  These fields require incorporating additional information into the Cauchy system. Four of these correspond to incoming gauge modes, and two represent incoming physical degrees of freedom, as described in \cite{Lindblom:2005qh}. In this paper we use CCM to set the physical boundary conditions, and we leave the gauge boundary conditions for future work. We note that, after matching the physical and gauge boundary conditions, our CCM system converges to the exact infinite domain problem in the continuum limit.

The physical boundary condition for $u^{\hat{1}-}_{\mu^\prime \nu^\prime}$ reads
\begin{align}
  d_{t^\prime} u^{\hat{1}-}_{\mu^\prime \nu^\prime}=P^{{\rm P}\rho^\prime \tau^\prime}_{\mu^\prime \nu^\prime}&\left[D_{t^\prime} u^{\hat{1}-}_{\rho^\prime \tau^\prime}\right.-(\alpha+s_{j^\prime}\beta^{j^\prime}) \notag \\
   &\left.\times(w_{\rho^\prime \tau^\prime}^{-}-\left.w_{\rho^\prime \tau^\prime}^{-}\right|_{\rm BC}-\gamma_2s^{ i^\prime}c_{i^\prime \rho^\prime \tau^\prime}^{3})\right], \label{eq:bc_bjorhus}
\end{align}
where the constraint fields $c_{i^\prime \rho^\prime \tau^\prime}^{3}$ can be found in Eq.~(57) of Ref.~\cite{Lindblom:2005qh}; and the physical projection operators $P^{{\rm P}\rho^\prime \tau^\prime}_{\mu^\prime \nu^\prime}$ are given by
\begin{align}
    P^{{\rm P}\rho^\prime \tau^\prime}_{\mu^\prime \nu^\prime}\equiv\left(\tensor[]{P}{_{\mu^\prime}^{\rho^\prime}}\tensor[]{P}{_{\nu^\prime}^{ \tau^\prime}}-\frac{1}{2}P_{\mu^\prime \nu^\prime}P^{\rho^\prime \tau^\prime}\right),
\end{align}
with
\begin{align}
    P_{\mu^\prime \nu^\prime}=g_{\mu^\prime \nu^\prime}+n_{\mu^\prime}n_{\nu^\prime}-s_{\mu^\prime}s_{\nu^\prime}, \label{eq:projection_operator}
\end{align}
as well as the normal vector of the time slice $n_{\mu^\prime}$. 

Crucially, $w_{\rho^\prime \tau^\prime}^{-}$ in Eq.~(\ref{eq:bc_bjorhus}) are
the inward propagating components of the Weyl tensor $C_{\mu^\prime \eta^\prime \nu^\prime \alpha^\prime}$,
\begin{align}
    w_{\rho^\prime \tau^\prime}^{-}=P^{{\rm P}\mu^\prime \nu^\prime}_{\rho^\prime \tau^\prime}(n^{\eta^\prime}+s^{\eta^\prime})(n^{ \alpha^\prime}+s^{\alpha^\prime})C_{\mu^\prime \eta^\prime \nu^\prime \alpha^\prime}, \label{eq:wab_projection}
\end{align}
where $n^{\alpha'}$ is the spacetime unit normal vector to the spatial hypersurface and $\left.w_{\rho^\prime \tau^\prime}^{-}\right|_{\rm BC}$ are the desired
values of $w_{\rho^\prime \tau^\prime}^{-}$ at the outer boundary.  The effect of the
boundary condition, Eq.~(\ref{eq:bc_bjorhus}), is to drive $w_{\rho^\prime \tau^\prime}^{-}$
toward $\left.w_{\rho^\prime \tau^\prime}^{-}\right|_{\rm BC}$. We find it is convenient to write $w_{\rho^\prime \tau^\prime}^{-}$ in terms of the Weyl scalar $\psi_0^\prime$:
\begin{align}
    w_{\rho^\prime \tau^\prime}^{-}=2(\psi_0^\prime\bar{m}_{\rho^\prime}\bar{m}_{\tau^\prime}+\bar{\psi}_0^\prime m_{\rho^\prime}m_{\tau^\prime}), \label{eq:wab}
\end{align}
where we have used an identity [Eq.~\eqref{eq:projection_operator}]
\begin{align}
P_{\rho^\prime \tau^\prime}=m_{\rho^\prime}\bar{m}_{\tau^\prime}+m_{\tau^\prime}\bar{m}_{\rho^\prime},
\end{align}
and the definition of $\psi_0^\prime$:
\begin{align}
    \psi_0^\prime=C_{\mu^\prime \nu^\prime \rho^\prime \tau^\prime}l^{\mu^\prime}m^{\nu^\prime}l^{\rho^\prime}m^{\tau^\prime}. \label{eq:GH_psi0_def}
\end{align}
Here $\{l^{\mu^\prime},k^{\mu^\prime},m^{\mu^\prime}\}$ refer to the null tetrad within the Newman-Penrose formalism. The choice of the null vectors $l^{\mu^\prime}$ (outgoing) and $k^{\mu^\prime}$ (ingoing) are determined \emph{uniquely} by Eqs.~\eqref{eq:projection_operator} and \eqref{eq:wab_projection} (namely the Cauchy system):
\begin{subequations}
    \label{eq:gh_tetrad}
\begin{align}
    &l^{\mu^\prime}=\frac{1}{\sqrt{2}}(n^{\mu^\prime}+s^{\mu^\prime}), \label{eq:gh_tetrad_l} \\
    &k^{\mu^\prime}=\frac{1}{\sqrt{2}}(n^{\mu^\prime}-s^{\mu^\prime}). \label{eq:gh_tetrad_k}
\end{align}
\end{subequations}
However, the choice of $m^{\mu^\prime}$ is not unique. The requirements on $m^{\mu^\prime}$ read:
\begin{align}
    m^{\mu^\prime}l_{\mu^\prime}=0,\quad m^{\mu^\prime}k_{\mu^\prime}=0,\quad m^{\mu^\prime}\bar{m}_{\mu^\prime}=1. \label{eq:GH_m}
\end{align}
As we shall show later, the only allowed gauge freedom on $m^{\mu^\prime}$ is a rotation: $m^{\mu^\prime}\to m^{\mu^\prime}e^{i\Theta}$, but the values of $w_{\rho^\prime \tau^\prime}$ in Eq.~\eqref{eq:wab_projection} do not depend on the gauge variable $\Theta$. Therefore, in our following calculations, we will take advantage of this fact and choose $m^{\mu^\prime}$ as close as possible to that of the characteristic system, in order to simplify calculations.

As mentioned earlier, production SpEC simulations set $\left.w_{\rho^\prime \tau^\prime}^{-}\right|_{\rm BC}$ in Eq.~\eqref{eq:bc_bjorhus} to zero. But within the CCM framework, we shall use the characteristic system to determine
$\left.w_{\rho^\prime \tau^\prime}^{-}\right|_{\rm BC}$ from Eq.~(\ref{eq:wab}), where
the $\psi'_0$ in Eq.~(\ref{eq:wab})
will be computed from the characteristic evolution and interpolated back
to the Cauchy grid. We will explain more details in Sec.~\ref{sec:CCM} below.


\section{Summary of the characteristic evolution}
\label{sec:CCE}
In this section, we briefly summarize the SpECTRE characteristic system as described in Refs.~\cite{Moxon:2020gha,Moxon:2021gbv}.  The procedures for extracting the Cauchy quantities on the worldtube, evolving the characteristic variables in the exterior region, and computing waveform quantities at future null infinity are identical for CCE versus CCM. 
For more details of the characteristic algorithm, see Refs. \cite{Moxon:2020gha,Moxon:2021gbv}.

The SpECTRE characteristic system is based on Bondi-Sachs metric in partially flat Bondi-like coordinates $\{\hat{r},\hat{x}^{\hat{A}},\hat{u}\}$ \cite{Moxon:2020gha,Moxon:2021gbv} 
\begin{align}
    ds^2=&-\left(e^{2\hat{\beta}}\frac{\hat{V}}{\hat{r}}-\hat{r}^2\hat{h}_{\hat{A}\hat{B}}\hat{U}^{\hat{A}}\hat{U}^{\hat{B}}\right)d\hat{u}^2-2e^{2\hat{\beta}}d\hat{u}d\hat{r}\notag \\
    &-2\hat{r}^2\hat{h}_{\hat{A}\hat{B}}\hat{U}^{\hat{B}}d\hat{u}d\hat{x}^{\hat{A}}+\hat{r}^2\hat{h}_{\hat{A}\hat{B}}d\hat{x}^{\hat{A}}d\hat{x}^{\hat{B}},\label{eq:BS_PFB}
\end{align}
where $\hat{x}^{\hat{A}}$ stands for the pair of angular coordinates $\{\hat{\theta},\hat{\phi}\}$. With this coordinate system, a few gauge conditions have been imposed: $g_{\hat{r}\hat{r}}=0$, $g_{\hat{r}\hat{A}}=0$, and the determinant of the angular components $\hat{h}_{\hat{A}\hat{B}}$ is set to that of the unit sphere metric $q_{\hat{A}\hat{B}}$
\begin{align}
    {\rm det}(\hat{h}_{\hat{A}\hat{B}})={\rm det}(q_{\hat{A}\hat{B}})=\sin^2\hat{\theta}. \label{eq:BS_angular_gauge_condition}
\end{align}
Consequently, the system is characterized by 6 degrees of freedom (4 quantities): $\hat{W}=(\hat{V}-\hat{r})/\hat{r}^2$, $\hat{h}_{\hat{A}\hat{B}}$, $\hat{U}^{\hat{B}}$, and $\hat{\beta}$. Near future null infinity, the metric components need to follow falloff rates \cite{Moxon:2020gha,Moxon:2021gbv}:
\begin{subequations}
\label{eq:falloff_partial_inertial}
\begin{align}
    &\lim_{\hat{r}\to\infty}\hat{W}=\mathcal{O}(\hat{r}^{-2}), \\
    &\lim_{\hat{r}\to\infty}\hat{U}^{\hat{A}}=\mathcal{O}(\hat{r}^{-2}),\\
    &\lim_{\hat{r}\to\infty}\hat{h}_{\hat{A}\hat{B}}=q_{\hat{A}\hat{B}}+\mathcal{O}(\hat{r}^{-1}).
\end{align}
\end{subequations}
Note that the conditions in Eqs.~\eqref{eq:falloff_partial_inertial} are not sufficient for the metric to asymptotically approach the Minkowski metric, as true Bondi-Sachs coordinates do. To bring the partially flat Bondi-like coordinates to a true Bondi-Sachs system (up to BMS transformations), one needs to further impose 
\begin{align}
    &\lim_{\hat{r}\to\infty}\hat{\beta}=\mathcal{O}(\hat{r}^{-1}). \label{eq:true_BS}
\end{align}
In practice, it was found that most computations are more straightforward
in partially flat Bondi-like coordinates $\{\hat{r},\hat{x}^{\hat{A}},\hat{u}\}$
where Eq.~\eqref{eq:true_BS} is not satisfied. We transform into
true Bondi-Sachs coordinates only when necessary
for computing waveform quantities at future null infinity
\cite{Moxon:2020gha,Moxon:2021gbv}. See Fig.~\ref{fig:coordinates}
(and also Table I of \cite{Moxon:2020gha}) for the various coordinate systems
used in CCE and CCM.

Following the algorithm outlined in Refs.~\cite{Moxon:2020gha,Moxon:2021gbv},
the characteristic system needs to take boundary data on a time-like worldtube from the inner Cauchy system. Therefore, one has to perform gauge transformations to convert the Cauchy $3+1$ metric in Eq.~\eqref{eq:3+1_metric} to the Bondi-Sachs metric in Eq.~\eqref{eq:BS_PFB}. The procedure involves three steps, and we summarize them in Fig.~\ref{fig:coordinates}. First, the space-like foliation of Eq.~\eqref{eq:3+1_metric} is converted to a null foliation. To achieve this goal, one needs to construct a class of null vectors $\partial_{\underline{\lambda}}$ at the worldtube surface.
\begin{align}
    \left(\partial_{\underline{\lambda}}\right)^{\underline{a}}=\delta_{a^\prime}^{\underline{a}}\frac{n^{a^\prime}+s^{a^\prime}}{\alpha-\gamma_{i^\prime j^\prime}\beta^{i^\prime}s^{j^\prime}}, \label{eq:null_wt}
\end{align}
where $\underline{\lambda}$ is an affine parameter, $a^\prime$ and $\underline{a}$ are abstract indices, the unit vector $s^{a^\prime}$ is defined in Eq.~\eqref{eq:def_s}, and $n^{a^\prime}$ still stands for the normal vector of the time slice. A new null coordinate system $\{\underline{u},\underline{\lambda},\underline{x}^{\underline{A}}\}$ is introduced, and quantities are transformed into this coordinate system.
This coordinate system is discussed in more detail in Sec.~\ref{subsubsec:cauchy_null_radius}.

The second step is to transform the null-radius coordinates to so-called Bondi-like coordinates $\{u,r,x^A\}$ by imposing the gauge condition in Eq.~\eqref{eq:BS_angular_gauge_condition}. At this stage, the metric is brought into Bondi-Sachs form
\begin{align}
    ds^2=&-\left(e^{2\beta}\frac{V}{r}-r^2h_{AB}U^AU^B\right)du^2-2e^{2\beta}dudr\notag \\
    &-2r^2h_{AB}U^Bdudx^A+r^2h_{AB}dx^Adx^B. \label{eq:BS_bondi_like}
\end{align}
The coordinates still differ from the partially flat Bondi-like coordinates because the falloff rates in Eqs.~\eqref{eq:falloff_partial_inertial} are now relaxed to
\begin{subequations}
\begin{align}
    &\lim_{r\to\infty}W=\mathcal{O}(r^{0}), \\
    &\lim_{r\to\infty}U^A=\mathcal{O}(r^{0}),\\
    &\lim_{r\to\infty}h_{AB}=\mathcal{O}(r^{0}).
\end{align}
\end{subequations}
The transformation to this coordinate system is discussed in detail in Sec.~\ref{subsubsec:null_radius_bondi_like}.

Finally, the Bondi-like coordinates are transformed to the partially-flat Bondi-like coordinates $\{\hat{r},\hat{x}^{\hat{A}},\hat{u}\}$ by removing the asymptotic value of $U^{A}$ at null infinity, $U^{(0)A}$. Here we define $U^{(0)A}$
by
\begin{align}
    U^A=U^{(0)A}+\mathcal{O}(r^{-1}). \label{eq:UA_asymptotic}
\end{align}
Details can be found in Sec.~\ref{subsubsec:bond_like_inertial}.

Once the worldtube quantities have been computed in
partially flat Bondi-like coordinates, they serve
as inner boundary conditions to evolve the characteristic system.  This evolution step is described in detail
in Refs.~\cite{Moxon:2020gha,Moxon:2021gbv} and is identical for CCM versus
CCE. Here we emphasize again that the characteristic evolution with the partially-flat Bondi-like coordinates is only weakly hyperbolic, which in turn makes CCM not well-posed \cite{Giannakopoulos:2023zzm}.

After determining all the metric components with the characteristic algorithm,
we can now
compute Weyl scalars. For CCM,
we need the Weyl scalar $\psi_0$ in the exterior characteristic region, which
will be used in the outer-boundary condition for the interior Cauchy system. To assemble $\psi_0$ from the metric components, we adopt the tetrad provided by Ref.~\cite{Moxon:2020gha}
\begin{subequations}
\label{eq:CCE_tetrad}
\begin{align}
    &m^{\mu}=-\frac{1}{\sqrt{2}r}\left(\sqrt{\frac{K+1}{2}}q^{\mu}-\frac{J}{\sqrt{2(1+K)}}\bar{q}^{\mu}\right), \label{eq:CCE_tetrad_m} \\
    &k^{\mu}=\sqrt{2}e^{-2\beta}\left[\delta^{\mu}_u-\frac{V}{2r}\delta^{\mu}_r+\frac{1}{2}\bar{U}q^{\mu}+\frac{1}{2}U\bar{q}^{\mu}\right],\label{eq:CCE_tetrad_k}\\
    &l^{\mu}=\frac{1}{\sqrt{2}}\delta^{\mu}_r. \label{eq:CCE_tetrad_l}
\end{align}
\end{subequations}
where $J$, $K$, and $U$ are spin-weighted scalars, defined by\footnote{We note that there is a typo in Eq.~(10e) of Ref.~\cite{Moxon:2020gha}; the correct expression is given in
Eq.~(\ref{eq:CCE_all_variables:K}).}
\begin{subequations}
\label{eq:CCE_all_variables}
\begin{align}
    &U\equiv U^{A}q_{A}, \quad J\equiv\frac{1}{2}q^{A}q^{B}h_{AB}, \\
    &K\equiv\frac{1}{2}q^{A}\bar{q}^{B}h_{AB}=\sqrt{1+J\bar{J}},\label{eq:CCE_all_variables:K}
\end{align}
\end{subequations}
The complex dyads $q^A$ and $q_A$ read
\begin{subequations}
\label{eq:q_hat_two_expressions}
\begin{align}
    &q^{A}\partial_{A}=-\partial_{\theta}-\frac{i}{\sin\theta}\partial_{\phi} \label{q_superA_expression}, \\
    &q_{A}dx^{A}=-d\theta-i\sin\theta d\phi. \label{q_subA_expression}
\end{align}
\end{subequations}
They obey the identity
\begin{align}
    &q^{A}q_{A}=0,&q^{A}\bar{q}_{A}=2. \label{eq:dyad_property}
\end{align}
Note that the tetrad vectors in Eqs.~\eqref{eq:CCE_tetrad} are constructed with the Bondi-like coordinates, of which the partially flat Bondi-like coordinates are subclasses. Therefore, Eqs.~\eqref{eq:CCE_tetrad} can be applied directly to the partially flat Bondi-like coordinates as long as all variables are replaced by their partially flat Bondi-like counterparts. With the tetrad vectors at hand, we can now derive a full expression of the Bondi-like $\psi_0$ in relation to Bondi quantities
\cite{Moxon:2020gha}
\begin{widetext}
\begin{align}
    \psi_0=\left(\frac{r\partial_{r}\beta-1}{4Kr}\right)\left[(1+K)\partial_{r}J-\frac{J^2\partial_{r}\bar{J}}{1+K}\right]+\frac{J(1+K^2)\partial_{r}J\partial_{r}\bar{J}}{8K^3}+\frac{1}{8K}\left[\frac{J^2\partial^2_{r}\bar{J}}{1+K}-(1+K)\partial_{r}^2J\right]-\frac{J\bar{J}^2(\partial_{r}J)^2+J^3(\partial_{r}\bar{J})^2}{16K^3}. \label{eq:psi0_CCE}
\end{align}
\end{widetext}
Similarly, Eq.~\eqref{eq:psi0_CCE} is also applicable to the partially-flat-Bondi-like $\hat{\psi}_0$ when the Bondi quantities are evaluated with the partially flat Bondi-like coordinates.

\section{Matching characteristic and Cauchy systems}
\label{sec:CCM}
We are now in a position to accomplish Cauchy-characteristic matching for the physical degrees of freedom.
The goal is to use the Weyl scalar $\psi_0$ obtained with the characteristic system to compute the boundary value $\left.w_{\rho^\prime \tau^\prime}^{ -}\right|_{\rm BC}$ that goes into the physical boundary condition of the Cauchy system [Eq.~(\ref{eq:bc_bjorhus})].  This is done by evaluating $\left.w_{\rho^\prime \tau^\prime}^{ -}\right|_{\rm BC}$ by inserting the characteristic system's $\psi_0$
into Eq.~\eqref{eq:wab}. Notice that the tetrad adopted by the characteristic system in Eqs.~\eqref{eq:CCE_tetrad} differs from the one used by Cauchy evolution in Eqs.~\eqref{eq:gh_tetrad}, so
we need to perform Lorentz transformations to obtain (a) the Cauchy Weyl scalar $\psi^\prime_0$ [defined in Eq.~\eqref{eq:GH_psi0_def}] and (b) the null covariant vector $m_{\mu^\prime}$ in Eq.~\eqref{eq:wab}. Necessary ingredients for the Lorentz transformations involve a set of Jacobian matrices across different coordinate systems. So in Sec.~\ref{subsec:Jacobians_for_ccm} we first work out the explicit expressions for these Jacobians,
and then in Secs.~\ref{sec:tetrad_transformation_Scenario_I} and \ref{sec:tetrad_transformation_Scenario_II} we carry out the transformations. Notice that the evaluation of $\psi_0$ with the characteristic system [Eq.~\eqref{eq:psi0_CCE}] could be done in either the partially flat Bondi-like coordinates [Eq.~\eqref{eq:BS_PFB}] or the Bondi-like coordinates [Eq.~\eqref{eq:BS_bondi_like}], and different choices
lead to different Lorentz transformations. In order to keep our discussions as general as possible, we consider both choices
in Secs.~\ref{sec:tetrad_transformation_Scenario_I} and \ref{sec:tetrad_transformation_Scenario_II}, respectively. We also illustrate these two options in 
Fig.~\ref{fig:coordinates} (see two arrows labeled by ``CCM''), serving as a roadmap for the CCM algorithm. The final step toward finishing the matching is to interpolate the values of $\psi^\prime_0$ and $m_{\mu^\prime}$ from the characteristic grid to the Cauchy grid. This is done in Sec.~\ref{sec:ccm_Cauchy_inertal}. 

For ease of future reference, we summarize the two primary matching steps below:
\begin{enumerate}
\item Construct $\psi^\prime_0$ and $m_{\mu^\prime}$ from  either the partially flat Bondi-like coordinates (Sec.~\ref{sec:tetrad_transformation_Scenario_I}) or the Bondi-like coordinates (Sec.~\ref{sec:tetrad_transformation_Scenario_II}).
\item Interpolate $\psi^\prime_0$ and $m_{\mu^\prime}$ from the characteristic grid to the Cauchy grid. See Sec.~\ref{sec:ccm_Cauchy_inertal}.
\end{enumerate}

\subsection{Jacobians for CCM}
\label{subsec:Jacobians_for_ccm}

As outlined in Sec.~\ref{sec:CCE} and summarized in
Fig.~\ref{fig:coordinates}, two intermediate coordinate systems are
introduced to convert the worldtube data from the Cauchy coordinates
to the partially flat Bondi-like coordinates. Below, we provide the
definition of these transformations and their Jacobians.

\subsubsection{Cauchy and null-radius coordinates}
\label{subsubsec:cauchy_null_radius}
The null-radius coordinates consist of
$\{\underline{u},\underline{\lambda},\underline{x}^{\underline{A}}\}$,
where $\underline{\lambda}$ is the affine parameter of the null vector
in Eq.~\eqref{eq:null_wt}. Meanwhile, the time and angular coordinates
are the same as the Cauchy coordinates:
\begin{equation}
  \begin{cases}
    \underline{u}=t^\prime, &\\
    \underline{x}^{\underline{A}}=\delta^{\underline{A}}_{A^\prime}x^{\prime\,A^{\prime}}, & \\
    \underline{\lambda}=\underline{\lambda}(t^\prime,r^\prime).&
  \end{cases}
  \label{eq:cauchy_null_radius_dependence}
\end{equation}
Consequently, the metric components in the null-radius coordinates are \cite{Moxon:2020gha}
\begin{align}
    &g_{\underline{\lambda}\underline{u}}=-1,\quad g_{\underline{\lambda}\underline{\lambda}}=0, \quad g_{\underline{\lambda}\underline{A}}=0, \quad g_{\underline{u}\underline{u}}=g_{t^{\prime}t^\prime}, \notag \\
    &g_{\underline{u}\underline{A}}=\delta_{\underline{A}}^{A^\prime} g_{t^{\prime}A^{\prime}}, \quad g_{\underline{A}\underline{B}}=\delta_{\underline{A}}^{A^\prime}\delta_{\underline{B}}^{B^\prime} g_{A^{\prime}B^{\prime}}. \label{eq:transformation_cauchy_null_radius}
\end{align}
Eqs.~\eqref{eq:transformation_cauchy_null_radius} lead to the Jacobian between two coordinate systems:
\begin{align}
\frac{\partial(t^\prime,r^\prime,x^{\prime\, A^\prime})}{\partial(\underline{u},\underline{\lambda},\underline{x}^{\underline{A}})}=
\begin{pmatrix}
1& \partial_{\underline{\lambda}}t^\prime & 0 \\
0 & \partial_{\underline{\lambda}}r^\prime & 0 \\
0 & 0 & \delta_{\underline{A}}^{A^\prime}
\end{pmatrix}
\label{eq:Jacobian_cauchy_null_radius}
\end{align}

\subsubsection{Null-radius and Bondi-like coordinates}
\label{subsubsec:null_radius_bondi_like}
To bring the null-radius coordinates to
Bondi-like coordinates, one needs to impose the gauge condition in Eq.~\eqref{eq:BS_angular_gauge_condition} and define the Bondi-like radius:
\begin{align}
    r=\left[\frac{{\rm det}(g_{\underline{A}\underline{B}})}{{\rm det}(q_{\underline{A}\underline{B}})}\right]^{1/4}.
\end{align}
Then the Bondi-like coordinates $\{u,r,x^A\}$ can be written as
\begin{equation}
  \begin{cases}
    u=\underline{u}, &\\
    x^A=\delta^A_{\underline{A}}\underline{x}^{\underline{A}}, & \\
    r=r(\underline{u},\underline{\lambda},\underline{x}^{\underline{A}}).&
  \end{cases}
  \label{eq:transformation_null_radius_bondi_like}
\end{equation}
Eqs.~\eqref{eq:transformation_null_radius_bondi_like} result in the Jocobian
\begin{align}
\frac{\partial(u,r,x^{ A})}{\partial(\underline{u},\underline{\lambda},\underline{x}^{\underline{A}})}=
\begin{pmatrix}
1&0 & 0 \\
\partial_{\underline{u}}r & \partial_{\underline{\lambda}}r & \partial_{\underline{A}}r \\
0 & 0 & \delta_{\underline{A}}^{A}
\end{pmatrix},
\label{eq:Jacobian_null_radius_bondi_like}
\end{align}
and its inverse
\begin{align}
\frac{\partial(\underline{u},\underline{\lambda},\underline{x}^{\underline{A}})}{\partial(u,r,x^{ A})}=
\begin{pmatrix}
1&0 & 0 \\
-\partial_{\underline{u}}r/\partial_{\underline{\lambda}}r & (\partial_{\underline{\lambda}}r)^{-1} & -\delta^{\underline{A}}_{A}\partial_{\underline{A}}r/\partial_{\underline{\lambda}}r \\
0 & 0 & \delta^{\underline{A}}_{A}
\end{pmatrix}.
\label{eq:Jacobian_bondi_like_null_radius}
\end{align}

\subsubsection{Bondi-like and partially flat Bondi-like coordinates}
\label{subsubsec:bond_like_inertial}
One difference between these two coordinate systems is that the
quantity $U^{A}$ is finite at future null infinity, but the quantity
$\hat{U}^{\hat{A}}$ vanishes. To remove the asymptotically constant part
of $U^{A}$, the angular coordinates $\hat{x}^{\hat{A}}$
must satisfy
\begin{align}
    \partial_u \hat{x}^{\hat{A}}=-\partial_A \hat{x}^{\hat{A}}U^{(0)A}, \label{eq:du_xhat}
\end{align}
where $U^{(0)A}$ is defined by Eq.~\eqref{eq:UA_asymptotic}. In practice, the quantities $\hat{x}^{\hat{A}}$ are evolved numerically on the characteristic grid along with the evolution of the characteristic metric components.
The Bondi-like radius $r$ also needs to be adjusted accordingly to meet the gauge condition in Eq.~\eqref{eq:BS_angular_gauge_condition}. Finally, the time coordinate $\hat{u}=u$ remains unchanged. In summary, the transformation is given by 
\begin{equation}
  \begin{cases}
    \hat{u}=u, &\\
    \hat{x}^{\hat{A}}=\hat{x}^{\hat{A}}(u,x^A), & \\
    \hat{r}=r\hat{\omega}(u,x^A),&
  \end{cases}
  \label{eq:transformation_bondi_like_inertial}
\end{equation}
where $\hat{\omega}(u,x^A)$ is a conformal factor 
\begin{align}
    \hat{\omega}=\frac{1}{2}\sqrt{\hat{b}\bar{\hat{b}}-\hat{a}\bar{\hat{a}}}, \label{eq:omegahat}
\end{align}
and two spin-weighted Jacobian factors $\hat{a}$ and $\hat{b}$ are given by
\begin{align}
    &\hat{a}=\hat{q}^{\hat{A}}\partial_{\hat{A}}x^Aq_A, & (\text{spin-weight 2}) \label{eq:ahat} \\
    &\hat{b}=\bar{\hat{q}}^{\hat{A}}\partial_{\hat{A}}x^Aq_A. & (\text{spin-weight 0}) \label{eq:bhat}
\end{align}
Since $\{q_A,\bar{q}_A\}$ ($\{\hat{q}_{\hat{A}},\bar{\hat{q}}_{\hat{A}}\}$) form a complete basis for the angular subspace spanned by $\{x^A\}$ ($\{\hat{x}^{\hat{A}}\}$), we can expand $\partial_{\hat{A}}x^A$ into\footnote{To obtain Eq.~\eqref{eq:expand_angular_jacobian_nohat}, one can exhaust all the possible linear combinations formed by the two bases $\{\hat{q}_{\hat{A}},\bar{\hat{q}}_{\hat{A}}\}$ and $\{q^A,\bar{q}^A\}$, and then determine the coefficients uniquely via Eqs.~\eqref{eq:dyad_property}, \eqref{eq:ahat} and \eqref{eq:bhat}.}
\begin{align}
\label{eq:expand_angular_jacobian_nohat}
\partial_{\hat{A}}x^A=
\frac{1}{4}
    \begin{pmatrix}
    \hat{q}_{\hat{A}},\bar{\hat{q}}_{\hat{A}}
    \end{pmatrix}
    \begin{pmatrix}
    \bar{\hat{a}} & \bar{\hat{b}} \\
    \hat{b} & \hat{a}
    \end{pmatrix}
    \begin{pmatrix}
    q^{A} \\
    \bar{q}^{A}
    \end{pmatrix},
\end{align}
where the expression is written in terms of matrix products. Note that the determinant of the middle $2\times2$ matrix (together with the factor of $1/4$) is equal to $-\hat{\omega}^2$ [see Eq.~\eqref{eq:omegahat}]. In practice, we find it is also convenient to define  spin-weighted factors that are related to the inverse of the Jacobian:
\begin{align}
    &a=q^{A}\partial_{A}\hat{x}^{\hat{A}}\hat{q}_{\hat{A}}, & (\text{spin-weight 2}) \\
    &b=\bar{q}^{A}\partial_{A}\hat{x}^{\hat{A}}\hat{q}_{\hat{A}}, & (\text{spin-weight 0})
\end{align}
as well as the conformal factor $\omega(\hat{u},\hat{x}^{\hat{A}})$ associated with them
\begin{align}
    \omega=\frac{1}{2}\sqrt{b\bar{b}-a\bar{a}}. \label{eq:oemga_b_a}
\end{align}
Similarly, the counterpart of Eq.~\eqref{eq:expand_angular_jacobian_nohat} reads
\begin{align}
\label{eq:expand_angular_jacobian}
\partial_{A}\hat{x}^{\hat{A}}=
\frac{1}{4}
    \begin{pmatrix}
    q_{A},\bar{q}_{A}
    \end{pmatrix}
    \begin{pmatrix}
    \bar{a} & \bar{b} \\
    b & a
    \end{pmatrix}
    \begin{pmatrix}
    \hat{q}^{\hat{A}} \\
    \bar{\hat{q}}^{\hat{A}}
    \end{pmatrix}.
\end{align}
At the same spacetime point, the identity $\partial_{\hat{A}}x^A\partial_{A}\hat{x}^{\hat{B}}=\delta_{\hat{A}}^{\hat{B}}$ leads to
\begin{align}
    &a=-\frac{\hat{a}}{\hat{\omega}^2}, & b=\frac{\bar{\hat{b}}}{\hat{\omega}^2}. \label{eq:a_ahat_b_bhat}
\end{align}
Plugging Eq.~\eqref{eq:a_ahat_b_bhat} into Eq.~\eqref{eq:oemga_b_a} we obtain another identity
\begin{align}
    \omega\hat{\omega}=1.
\end{align}

We then use Eq.~\eqref{eq:transformation_bondi_like_inertial} to get the Jacobian between the Bondi-like and the partially flat Bondi-like coordinates
\begin{align}
\frac{\partial(\hat{r},\hat{x}^{\hat{A}},\hat{u})}{\partial(r,x^A,u)}=
\begin{pmatrix}
\hat{\omega} & r\partial_A\hat{\omega} & r\partial_u\hat{\omega} \\
0 & \partial_A \hat{x}^{\hat{A}} & \partial_u \hat{x}^{\hat{A}} \\
0 & 0 & 1
\end{pmatrix}.
\label{eq:Jacobian_bondi_like_inertial}
\end{align}
Its inverse reads
\begin{align}
\frac{\partial(r,x^A,u)}{\partial(\hat{r},\hat{x}^{\hat{A}},\hat{u})}=
\begin{pmatrix}
\omega & r\delta^A_{\hat{A}}\partial_{A}\ln\omega & r\partial_u\ln\omega+rU^{(0)A}\partial_A\ln\omega \\
0 & \partial_{\hat{A}} x^{A} & U^{(0)A} \\
0 & 0 & 1
\end{pmatrix},
\label{eq:jacobian_bondi_inertial}
\end{align}
where we have used Eq.~\eqref{eq:du_xhat} to simplify the result.

\subsection{Choice 1: Transforming \texorpdfstring{$m_{\hat{\mu}}$}{mhatmu} and \texorpdfstring{$\hat{\psi}_0$}{psi0hat} to the Cauchy tetrad}
\label{sec:tetrad_transformation_Scenario_I}
We first consider Choice 1, as summarized in Fig.~\ref{fig:coordinates}, where the tetrad vector $m_{\hat{\mu}}$ and the Weyl scalar $\hat{\psi}_0$ are evaluated in the partially flat Bondi-like coordinates, using Eqs.~\eqref{eq:CCE_tetrad_m} and \eqref{eq:psi0_CCE}. Before transforming them into the Cauchy tetrad, we first observe a useful and important fact: The characteristic outgoing null tetrad vector $l^{\hat{a}}$ at the worldtube surface, as defined in Eq.~\eqref{eq:CCE_tetrad_l}, is by construction \emph{proportional to} that of the Cauchy system $l^{a^\prime}$, defined in Eq.~\eqref{eq:gh_tetrad_l}. Again, here $\hat{a}$ and $a^\prime$ stand for abstract indices. To see this, we write 
\begin{align}
l^{\hat{a}}&=\frac{1}{\sqrt{2}}\left(\partial_{\hat{r}}\right)^{\hat{a}}=\frac{1}{\sqrt{2}}\left(\partial_{\underline{\lambda}}\hat{r}\right)^{-1}\delta_{\underline{a}}^{\hat{a}}\left(\partial_{\underline{\lambda}}\right)^{\underline{a}}\notag \\
&=\frac{1}{\sqrt{2}}e^{2\hat{\beta}}\delta_{\underline{a}}^{\hat{a}}\left(\partial_{\underline{\lambda}}\right)^{\underline{a}},\label{eq:null_vector_CCE_rhat}
\end{align}
where the first equality comes from Eq.~\eqref{eq:CCE_tetrad_l}, the second equality is due to the combination of the Jacobian matrices in Eq.~\eqref{eq:Jacobian_bondi_like_null_radius} and \eqref{eq:jacobian_bondi_inertial}, and the final equality is based on a relationship [see Eqs.~(19a) and (33a) of Ref.~\cite{Moxon:2020gha}]
\begin{align}
    \hat{\beta}=-\frac{1}{2}\ln(\partial_{\underline{\lambda}}\hat{r}).
\end{align}
On the other hand, the null vector $\left(\partial_{\underline{\lambda}}\right)^{\underline{a}}$ in Eq.~\eqref{eq:null_vector_CCE_rhat} is proportional to the Cauchy outgoing null vector $l^{a^\prime}$ needed by the boundary condition [see Eqs.~\eqref{eq:null_wt} and \eqref{eq:gh_tetrad_l}], but with a different normalization. After combining Eq.~\eqref{eq:null_vector_CCE_rhat} with \eqref{eq:null_wt} and \eqref{eq:gh_tetrad_l}, we obtain:
\begin{align}
    l^{a^\prime}=(\alpha-\gamma_{i^\prime j^\prime}\beta^{i^\prime}s^{j^\prime})e^{-2\hat{\beta}}l^{\hat{a}}\delta_{\hat{a}}^{a^\prime}. \label{eq:l_GH_and_l_CCE_transform}
\end{align}
Therefore, the statement $l^{a^\prime}\propto l^{\hat{a}}$ is proven.
Under this constraint,  the allowed Lorentz transformation between the characteristic and Cauchy tetrads can be split into two categories
\begin{itemize}
    \item Type I: ($\bm{l}$ unchanged)
    \begin{align}
        \bm{l}&\to \bm{l}, & \bm{k} &\to \bm{k} + \bar{\kappa}\bm{m} + \kappa\bar{\bm{m}} + \kappa\bar{\kappa}\bm{l}, \notag \\
        \bm{m}&\to \bm{m} + \kappa\bm{l}, &\bar{\bm{m}} &\to \bar{\bm{m}} + \bar{\kappa}\bm{l}. \label{eq:lorentz_transformation_I}
    \end{align}
    \item Type II: (both $\bm{l}$ and $\bm{k}$ changed)
    \begin{align}
       \bm{l}&\to A\bm{l}, &\bm{k}&\to A^{-1}\bm{k}, \notag \\
       \bm{m}&\to e^{i\Theta}\bm{m}, &\bar{\bm{m}}&\to e^{-i\Theta}\bar{\bm{m}},\label{eq:lorentz_transformation_II}
    \end{align}
\end{itemize}
where the complex scalar $\kappa$ has a spin weight of 1, $A$ and $\Theta$ are real scalars. The Weyl scalar $\hat{\psi}_0$ transforms correspondingly:
\begin{itemize}
\item Type I:
\begin{align}
    \hat{\psi}_0\to\hat{\psi}_0.
\end{align}
\item Type II:
\begin{align}
    \hat{\psi}_0\to A^{2}e^{2i\Theta}\hat{\psi}_0. \label{eq:lorentz_psi0_ii}
\end{align}
\end{itemize}
Notice that $\hat{\psi}_0$ is not mixed with other Weyl scalars. In particular, it remains unchanged within the Type I category. Below we will take advantage of this observation to simplify the calculation.

As summarized in Fig.~\ref{fig:coordinates}, for Choice 1, we need to transform both $m_{\hat{\mu}}$ and $\hat{\psi}_0$
to the Cauchy tetrad in order to evaluate the inward propagating components of the Weyl tensor $\left.w_{\rho^\prime \tau^\prime}^{-}\right|_{\rm BC}$ [Eq.~\eqref{eq:wab}]
in the correct tetrad.  We treat the transformation of
$m_{\hat{\mu}}$ and $\hat{\psi}_0$ separately in the two following sections.

\subsubsection{Type I transformation of \texorpdfstring{$m_{\hat{\mu}}$}{typeimhatmu} }
\label{subsubsec:type_i}
The characteristic system's $m_{\hat{\mu}}$ [Eq.~\eqref{eq:CCE_tetrad_m}] is not aligned with that of the Cauchy system [Eq.~\eqref{eq:GH_m}]. This is because  our choice of the ingoing null vector $k^\mu$
for the characteristic system [Eq.~\eqref{eq:CCE_tetrad_k}] is not the same as $k^{\mu^\prime}$ used in the Bjørhus boundary condition, which
is defined uniquely by Eq.~\eqref{eq:gh_tetrad_k}. To transform the characteristic vector $\bm{m}$ to the corresponding choice in
the Cauchy boundary condition, it suffices to add some multiple of the
outgoing null vector $\bm{l}$ to $\bm{m}$; thus we need to perform a
type I transformation.
We want to emphasize that the value of $\hat{\psi}_0$ is not impacted by a type I transformation, so when performing such a transformation it is not necessary to keep track of the explicit Lorentz parameter [namely $\kappa$ in Eq.~\eqref{eq:lorentz_transformation_I}] that was used in the transformation. Accordingly,
in the vector expressions below, we will simply drop terms that are
proportional to the outgoing null vector $\bm{l}$, since these terms can be eliminated through a type I transformation. Whenever this is done we will indicate that such terms have been dropped
by a type I transformation by using the symbol $\approx$ instead of $=$.

By combining Jacobians in Eqs.~\eqref{eq:Jacobian_cauchy_null_radius}, \eqref{eq:Jacobian_bondi_like_null_radius} and \eqref{eq:jacobian_bondi_inertial}, we obtain the relationship
\begin{align}
    \partial_{\hat{A}}=(\partial_{\hat{A}}x^A)\delta^{A^\prime}_A \partial_{A^\prime}+\partial_{\underline{\lambda}}\times\left[\frac{\partial_{\hat{A}}r}{\partial_{\underline{\lambda}}r}-\frac{\partial_{\underline{A}}r}{\partial_{\underline{\lambda}}r}(\partial_{\hat{A}}x^A) \delta_A^{\underline{A}}\right] \label{typeI_Ahat_to_Ap}
\end{align}
Since $\partial_{\underline{\lambda}}$ is the outgoing null vector given in Eq.~\eqref{eq:null_vector_CCE_rhat}, the second term in Eq.~\eqref{typeI_Ahat_to_Ap} can be removed via a type I Lorentz transformation. We then insert Eqs.~\eqref{typeI_Ahat_to_Ap} and \eqref{eq:expand_angular_jacobian_nohat} into Eq.~\eqref{eq:q_hat_two_expressions}, which yields
\begin{align}
    q^{\hat{\mu}}\approx \frac{1}{2}\hat{a}\delta_{\mu^\prime}^{\hat{\mu}}\bar{q}^{\mu^\prime}+\frac{1}{2}\bar{\hat{b}}\delta_{\mu^\prime}^{\hat{\mu}}q^{\mu^\prime}, \label{eq:q_transformation_type_i}
\end{align}
where $\approx$ implies that a type I Lorentz transformation has been
performed, as described above.
Plugging Eq.~\eqref{eq:q_transformation_type_i} into Eq.~\eqref{eq:CCE_tetrad_m}, we obtain 
\begin{align}
    &m^{\hat{\mu}}\approx-\frac{\delta^{\hat{\mu}}_{\mu^\prime}}{\sqrt{2}\hat{r}}\left[\left(\sqrt{\frac{\hat{K}+1}{2}}\frac{1}{2}\hat{a}-\frac{\hat{J}}{\sqrt{2(1+\hat{K})}}\frac{1}{2}\hat{b}\right)\bar{q}^{\mu^\prime}\right.\notag \\
    &\left.+\left(\sqrt{\frac{\hat{K}+1}{2}}\frac{1}{2}\bar{\hat{b}}-\frac{\hat{J}}{\sqrt{2(1+\hat{K})}}\frac{1}{2}\bar{\hat{a}}\right)q^{\mu^\prime}\right].
\end{align}
Or equivalently
\begin{align}
    m^{\hat{\mu}}&\approx \delta^{\hat{\mu}}_{\mu^\prime}m^{\mu^{\prime}}, \label{eq:new_m_cce_to_GH_abstract}
\end{align}
with $m^{\mu^{\prime}}$ being the components of a new contravariant vector $m^{a^{\prime}}$
\begin{align}
    m^{a^{\prime}}&= \hat{M}_{\theta^\prime}\left(\partial_{\theta^\prime}\right)^{a^\prime}+ \hat{M}_{\phi^\prime}\frac{i}{\sin\hat{\theta}(\theta^\prime)}\left(\partial_{\phi^\prime}\right)^{a^\prime},\label{eq:new_m_cce_to_GH}
\end{align}
and
\begin{align}
    &4\hat{r} \hat{M}_{\theta^\prime}=(\hat{a}+\bar{\hat{b}})\sqrt{\hat{K}+1}-(\bar{\hat{a}}+\hat{b})\frac{\hat{J}}{\sqrt{(1+\hat{K})}}, \\
    &4\hat{r}\hat{M}_{\phi^\prime}=(\bar{\hat{b}}-\hat{a})\sqrt{\hat{K}+1}-(\bar{\hat{a}}-\hat{b})\frac{\hat{J}}{\sqrt{(1+\hat{K})}}.
\end{align}
At this stage, we have constructed a Cauchy tetrad vector $m^{a^{\prime}}$ in Eq.~\eqref{eq:new_m_cce_to_GH} that differs from the original characteristic tetrad vector $m^{\hat{a}}$ by only a type I transformation. Meanwhile, we can see $m^{a^{\prime}}$ has components only within the Cauchy angular subspace $\{\theta^\prime,\phi^\prime\}$. Therefore it meets all the requirements in Eq.~\eqref{eq:GH_m}. Consequently we can convert it to its covariant form $m_{a^{\prime}}$ and insert it into Eq.~\eqref{eq:wab} to evaluate $\left.w_{\rho^\prime \tau^\prime}^{-}\right|_{\rm BC}$.

Since Cartesian coordinates are used to evolve the Cauchy system, we write down the Cartesian components of two angular bases $\left(\partial_{\theta^\prime}\right)^{a^\prime}$ and $\left(\partial_{\phi^\prime}\right)^{a^\prime}$ for completeness:
\begin{align}
    &\left(\partial_{\theta^\prime}\right)^{a^\prime}=R^\prime_{\rm wt}\left(\cos\phi^\prime\cos\theta^\prime,\sin\phi^\prime\cos\theta^\prime,-\sin\theta^\prime\right),\\
    &\left(\partial_{\phi^\prime}\right)^{a^\prime}=R^\prime_{\rm wt}\sin\theta^\prime\left(-\sin\phi^\prime,\cos\phi^\prime,0\right).
\end{align}

\subsubsection{Type II transformation of \texorpdfstring{$\hat{\psi}_0$}{typeiihatpsi0}}
\label{subsubsec:type_ii}
Eq.~\eqref{eq:l_GH_and_l_CCE_transform} indicates that two outgoing null vectors $l^{a^\prime}$ and $l^{\hat{a}}$ are related by a Type II transformation [Eq.~\eqref{eq:lorentz_transformation_II}], with the Lorentz parameter $\hat{A}$ given by
\begin{align}
    \hat{A}=(\alpha-\gamma_{i^\prime j^\prime}\beta^{i^\prime}s^{j^\prime})e^{-2\hat{\beta}},
\end{align}
which leads to
\begin{align}
    \psi_0^{\prime}=\hat{A}^2\hat{\psi}_0. \label{eq:psi0_prime_psi0_hat}
\end{align}
On the other hand, there is one more gauge freedom: the rotation of $\bm{m}$ with a phase factor $e^{i\Theta}$.
However, the combination $\psi_0^\prime\bar{m}_{\rho^\prime}\bar{m}_{\tau^\prime}$ that appears in $w_{\rho^\prime \tau^\prime}^{ -}$ [Eq.~\eqref{eq:wab}] is invariant under such a phase rotation because $\psi_0^\prime$ is also transformed accordingly due to Eq.~\eqref{eq:lorentz_psi0_ii}. Physically speaking, the incoming characteristics $w_{\rho^\prime \tau^\prime}^{ -}$ do not depend on the choice of the angular tetrad vector. Therefore, we can neglect this gauge freedom while performing the matching.



\subsection{Choice 2: Transforming \texorpdfstring{$m_{\mu}$}{choice2mmu} and \texorpdfstring{$\psi_0$}{choice2psi0} to the Cauchy tetrad}
\label{sec:tetrad_transformation_Scenario_II}
Then we consider Choice 2, where the characteristic quantities $m_{\mu}$ and $\psi_0$ are evaluated in the Bondi-like coordinates. Similar to Sec.~\ref{sec:tetrad_transformation_Scenario_I}, below we treat the transformation of $m_{\mu}$ and $\psi_0$ separately.

\subsubsection{Type I transformation of \texorpdfstring{$m_{\mu}$}{mmu}}
\label{subsub:II_m}
By combining Jacobians in Eqs.~\eqref{eq:Jacobian_cauchy_null_radius} and \eqref{eq:Jacobian_bondi_like_null_radius}, we obtain
\begin{align}
    \partial_A\approx\delta_A^{A^\prime}\partial_{A^\prime},
\end{align}
which leads to
\begin{align}
    q^{\mu}&\approx \delta^{\mu}_{\mu^\prime}q^{\mu^{\prime}}.\label{eq:new_m_cce_to_GH_B_abstract}
\end{align}
Here we have used the definition of $q^\mu$ in Eq.~\eqref{q_superA_expression}. 
Inserting Eq.~\eqref{eq:new_m_cce_to_GH_B_abstract} into Eq.~\eqref{eq:CCE_tetrad_m} yields
\begin{align}
    m^\mu\approx\delta^{\mu}_{\mu^\prime}m^{\mu^{\prime}},
\end{align}
with $m^{\mu^{\prime}}$ being the components of the vector $m^{a^{\prime}}$
\begin{align}
    m^{a^{\prime}}&= M_{\theta^\prime}\left(\partial_{\theta^\prime}\right)^{a^\prime}+M_{\phi^\prime}\frac{i}{\sin\theta^\prime}\left(\partial_{\phi^\prime}\right)^{a^\prime},\label{eq:new_m_cce_to_GH_B}
\end{align}
and 
\begin{align}
    &2r M_{\theta^\prime}=\sqrt{K+1}-\frac{J}{\sqrt{(1+K)}}, \\
    &2r M_{\phi^\prime}=\sqrt{K+1}+\frac{J}{\sqrt{(1+K)}}.
\end{align}
We remark that the null vector $m^{\mu^\prime}$, which differs from $m^\mu$ by only a type I transformation,  is now in the Cauchy angular subspace $\{\theta^\prime,\phi^{\prime}\}$, as required by the Cauchy boundary condition in Eq.~\eqref{eq:GH_m}. Therefore, its covariant form can be used to construct $\left.w_{\rho^\prime \tau^\prime}^{-}\right|_{\rm BC}$ in Eq.~\eqref{eq:wab}.

In practice, the characteristic system is evolved with the partially flat Bondi-like coordinates, as summarized in Fig.~\ref{fig:coordinates}. Therefore, we need to transform the partially flat Bondi quantities $\hat{J}$ and $\hat{K}$ [Eq.~\eqref{eq:CCE_all_variables}] to the Bondi-like coordinates via \cite{Moxon:2020gha}
\begin{align}
    &J=\frac{\bar{b}^2\hat{J}+a^2\bar{\hat{J}}+2a\bar{b}\hat{K}}{4\omega^2},\label{eq:J_BS_like}\\
    &K=\sqrt{1+J\bar{J}},\label{eq:K_BS_like}
\end{align}
and then insert the results into Eq.~\eqref{eq:new_m_cce_to_GH_B} to construct the tetrad vector $m_{\mu^\prime}$ for matching.

\subsubsection{Type II transformation of \texorpdfstring{$\psi_0$}{typeiipsi0}}
\label{subsub:II_psi0}
In the meantime, after obtaining $J$ and $K$ from Eqs.~\eqref{eq:J_BS_like} and \eqref{eq:K_BS_like}, we can evaluate $\psi_0$ with Eq.~\eqref{eq:psi0_CCE}. Similar to the discussion in Sec.~\ref{subsubsec:type_ii}, the two outgoing null vectors $l^{\mu^\prime}$ and $l^{\mu}$ are related by a Type II transformation, and the corresponding Lorentz parameter $A$ reads
\begin{align}
    A=(\alpha-\gamma_{i^\prime j^\prime}\beta^{i^\prime}s^{j^\prime})e^{-2\beta}.
\end{align}
Consequently, the desired $\psi_0^\prime$ is given by
\begin{align}
    \psi_0^\prime=A^2\psi_0.
\end{align}






\subsection{Interpolating to the Cauchy coordinates}
\label{sec:ccm_Cauchy_inertal}
Now we have obtained the desired tetrad vector $m_{\mu^\prime}$ and the Weyl scalar $\psi_0^\prime$. But they are still evaluated on the characteristic grid for both choices\footnote{For Choice 2, the Bondi-like $J$ and $K$ obtained from Eqs.~\eqref{eq:J_BS_like} and \eqref{eq:K_BS_like}, as well as the Weyl scalar $\psi_0$ built upon them, are functions of the partially flat Bondi-like coordinates, rather than the Bondi-like coordinates.}. The final step to complete the matching is to interpolate the results to the Cauchy grid. Specifically, since the matching is performed on a 2D spherical surface, we need to construct a map from the partially flat Bondi-like angular coordinates $\hat{x}^{\hat{A}}=\{\hat{\theta},\hat{\phi}\}$ to the Cauchy angular coordinates $x^{\prime\,A^{\prime}}=(\theta^\prime,\phi^\prime)$ for each time step of simulations. Recall from Fig.~\ref{fig:coordinates} that the Bondi-like angular coordinates $x^A$ are constructed to be the same as $x^{\prime\,A^\prime}$; therefore the task is equivalent to constructing the dependence of $x^A$ on $\hat{x}^{\hat{A}}$. 

\begin{figure*}[!htb]
        \includegraphics[width=\textwidth,clip=true]{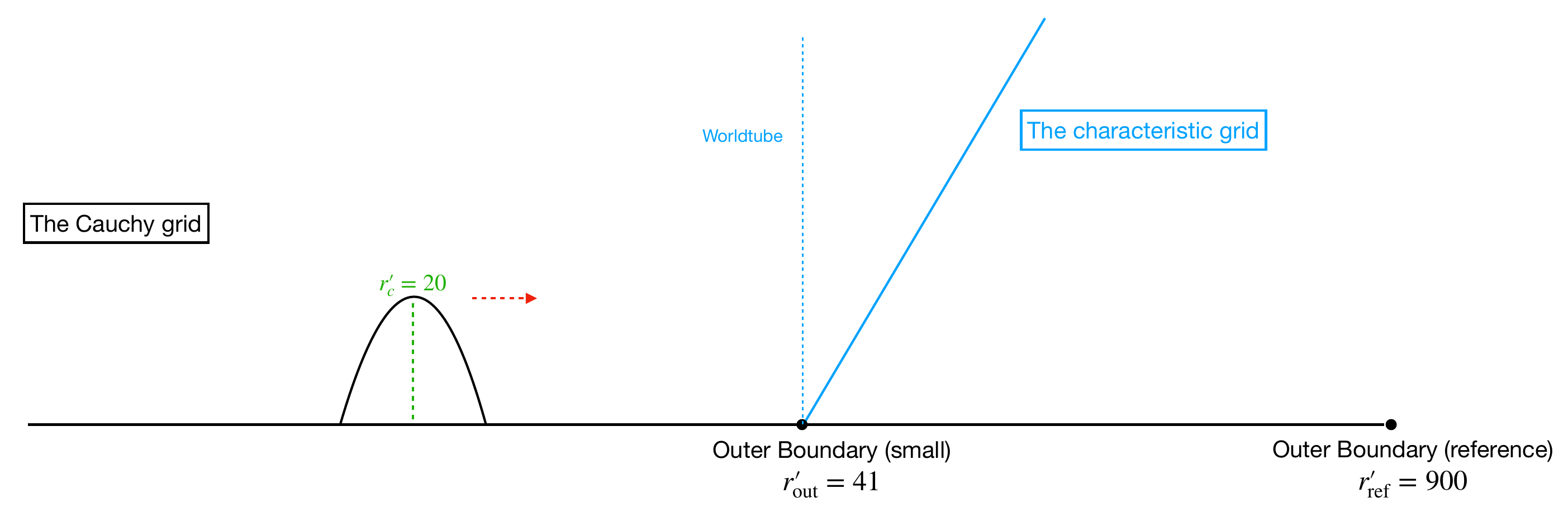}
  \caption{An illustration for the propagation of an outgoing Teukolsky wave within a flat background. The horizontal black line represents the spatial domain of the Cauchy grid. The Teukolsky wave is initially located at $r^\prime_c=20$ and has a width of $\tau=2$. We use CCE and CCM to evolve the system with a small Cauchy domain $(r^{\prime}_{\rm out}=41)$.  For comparison, we carry out a reference simulation without CCM, with a distant outer boundary $r^{\prime}_{\rm ref}$ at 900 to ensure it is causally disconnected from the system throughout the entire simulation. The worldtube for the characteristic system is situated at $r^{\prime}_{\rm out}=41$, indicated by the vertical blue dashed line.}
 \label{fig:Teukolsky_wave}
\end{figure*}

The inverse problem, namely $\hat{x}^{\hat{A}}$ as functions of $x^A$, has been worked out while we are constructing the worldtube data for the characteristic system \cite{Moxon:2020gha,Moxon:2021gbv} --- the partially flat Bondi-like angular coordinates  $\hat{x}^{\hat{A}}$ are evolved with respect to Bondi-like angular coordinates  $x^A$ using Eq.~\eqref{eq:du_xhat}. In principle, one can invert the dependence numerically to fulfill our purpose, but the process might be numerically expensive. A cheaper way is to evolve $x^A$ as functions of $\hat{x}^{\hat{A}}$ simultaneously. The counterpart of Eq.~\eqref{eq:du_xhat} for the evolution of $x^A$ can be read off directly from the Jacobian in Eq.~\eqref{eq:jacobian_bondi_inertial}:
\begin{align}
    \partial_{\hat{u}}x^A=U^{(0)A}. \label{eq:duhat_x}
\end{align}
In practice, we find it is more convenient to convert $x^A$ to Cartesian coordinates $x^i$ on a unit sphere,
\begin{align}
    x^i=\left(\sin\theta\cos\phi,\sin\theta\sin\phi,\cos\theta\right),
\end{align}
since the spin-weight of $x^i$ is 0 and we can make use of the spin-weighted derivatives \cite{Gomez:1996ge}
\begin{align}
    &\eth x^i=q^BD_Bx^i, & \bar{\eth}x^i=\bar{q}^BD_Bx^i,
\end{align}
where $D_A$ denotes the covariant derivative associated with the metric $q_{AB}=1/2(q_A\bar{q}_B+\bar{q}_Aq_B)$. Then Eq.~\eqref{eq:duhat_x} can be written as 
\begin{align}
   \partial_{\hat{u}}x^i&= \frac{1}{2}\hat{\mathcal{U}}^{(0)}\bar{\eth} x^i+\frac{1}{2}\bar{\hat{\mathcal{U}}}^{(0)}\eth x^i, \label{eq:duhat_x_cartesian}
\end{align}
where we have introduced an auxiliary variable $\mathcal{U}^{(0)\hat{A}}$ such that 
\begin{align}
&\mathcal{U}^{(0)\hat{A}}\partial_{\hat{A}}x^B=U^{(0)B},\\
&\mathcal{U}^{(0)}=\mathcal{U}^{(0)\hat{A}}q_{\hat{A}}.
\end{align}
The two equations above imply 
\begin{align}
    \mathcal{U}^{(0)}=\frac{1}{2\hat{\omega}^2}\left(\hat{\bar{b}}U^{(0)}-\hat{a}\bar{U}^{(0)}\right),\label{eq:U_mathcal_U}
\end{align}
and its inverse
\begin{align}
    U^{(0)}=\frac{1}{2\omega^2}\left(\bar{b}\mathcal{U}^{(0)}-a\bar{\mathcal{U}}^{(0)}\right).
\end{align}
For completeness, we also cast Eq.~\eqref{eq:du_xhat} into its Cartesian version
\begin{align}
&\partial_u\hat{x}^{\hat{i}}=-\frac{1}{2}U^{(0)}\bar{\eth} \hat{x}^{\hat{i}}-\frac{1}{2}\bar{U}^{(0)}\eth  \hat{x}^{\hat{i}}. \label{eq:du_xhat_cartesian}
\end{align}

In practice, Eq.~\eqref{eq:duhat_x_cartesian} is evolved numerically along with the evolution of Eq.~\eqref{eq:du_xhat_cartesian} and the evolution of the characteristic metric components. This determines the map $x^A(\hat{u},\hat{x}^{\hat{A}})$. We then adopt the spin-weighted Clenshaw algorithm \cite{Moxon:2021gbv} to perform the angular interpolation of $m_{\mu^\prime}$ and $\psi_0^\prime$ to the Cauchy grid, and we assemble these interpolated quantities into the incoming characteristics $w_{\mu^\prime \nu^\prime}^{-}$ using  Eq.~\eqref{eq:wab}.

\mycomment{We remark that the dependence $x^A(\hat{x}^{\hat{A}},\hat{u})$ constructed from Eq.~\eqref{eq:duhat_x} fully determines its inverse $\hat{x}^{\hat{A}}(x^A,u)$, which needs to be consistent with the relationship constructed from Eq.~\eqref{eq:du_xhat}. The sufficient and necessary conditions to capture the consistency are listed in Eq.~\eqref{eq:a_ahat_b_bhat}, through their Jacobian factors. They lead to two new constraints for the CCM system:
\begin{align}
    &C_a=a+\frac{\hat{a}}{\hat{\omega}^2}, && C_b=\frac{\bar{\hat{b}}}{b\hat{\omega}^2}-1. \label{eq:constraints_ca_cb}
\end{align}
In practice, we need to monitor their values and keep them as small as possible.
}

\section{Numerical Tests}
\label{sec:tests}
We now present numerical tests of our CCM algorithm using three physical systems. First, in Sec.~\ref{subsec:Teukolsky}, we examine the linear and nonlinear propagation of gravitational Teukolsky waves \cite{PhysRevD.26.745} on a flat background. Next, in Sec.~\ref{subsec:perturb_kerr}, we perturb a Kerr BH with a Teukolsky wave. Finally, in Sec.~\ref{subsec:pulse_on_CCE}, we initialize a GW pulse on the characteristic grid and inject it into the Cauchy domain. Throughout the simulations, we primarily focus on the Choice 2 algorithm outlined in Sec.~\ref{sec:tetrad_transformation_Scenario_II}, since this approach involves a single Lorentz transformation, simplifying its implementation and facilitating future code development. 

In the code, we use the third-order Adams-Bashforth time stepper for time integration, with the time step being fixed to $0.001$. The Cauchy domain, as detailed in \cite{SpECTREDomain}, is configured with a refinement level of 3. It is radially partitioned at radii of 6, 12, and 26. In each dimension, every domain element has 5 grid points. The numerical settings of CCE can be found in \cite{Moxon:2021gbv}. The angular resolution of the CCE domain is set to $l=24$, and there are 12 grid points in the radial direction.



\subsection{A Teukolsky wave propagating within a flat background}
\label{subsec:Teukolsky}
Following the tests in Refs.~\cite{Barkett:2019uae,Moxon:2021gbv}, we investigate the propagation of a Teukolsky wave \cite{PhysRevD.26.745} on a flat background. The initial data of the Cauchy system is
constructed utilizing the Extended Conformal Thin Sandwich (XCTS) formulation \cite{York:1998hy,Pfeiffer:2002iy}, which accounts for nonlinear effects that arise when the amplitude of the Teukolsky wave is large.  Subsequently, the system undergoes full nonlinear evolution. We carry out two tests: one with a small-amplitude Teukolsky wave that has an analytic perturbative solution, and one with a large-amplitude wave. In the perturbative regime, the metric is described by \cite{PhysRevD.26.745}:
\begin{align}
    ds^2&=-dt^{\prime\,2}+(1+Af_{r^{\prime}r^{\prime}})dr^{\prime\,2}+2Bf_{r^{\prime}\theta^{\prime}}r^{\prime}dr^{\prime}d\theta^{\prime} \notag \\
    &+2Bf_{r^{\prime}\phi^{\prime}}r^{\prime}\sin\theta^{\prime}dr^{\prime}d\phi^{\prime}+(1+Cf^{(1)}_{\theta^{\prime}\theta^{\prime}}+Af^{(2)}_{\theta^{\prime}\theta^{\prime}})r^{\prime\,2}d\theta^{\prime\,2} \notag \\
    &+2(A-2C)f_{\theta^{\prime}\phi^{\prime}}r^{\prime\,2}\sin\theta^{\prime}d\theta^{\prime}d\phi^{\prime} \notag \\
    &+(1+Cf^{(1)}_{\phi^{\prime}\phi^{\prime}}+Af^{(2)}_{\phi^{\prime}\phi^{\prime}})r^{\prime\,2}\sin^2\theta^{\prime}d\phi^{\prime\,2}, \label{eq:Teukolsky_metric}
\end{align}
with
\begin{subequations}
\label{eq:Teukolsky_ABC}
\begin{align}
    &A=3\left[\frac{F^{(2)}}{r^{\prime\,3}}+\frac{3F^{(1)}}{r^{\prime\,4}}+\frac{3F}{r^{\prime\,5}}\right],\\
    &B=-\left[\frac{F^{(3)}}{r^{\prime\,2}}+\frac{3F^{(2)}}{r^{\prime\,3}}+\frac{6F^{(1)}}{r^{\prime\,4}}+\frac{6F}{r^{\prime\,5}}\right], \\
    &C=\frac{1}{4}\left[\frac{F^{(4)}}{r^{\prime}}+\frac{2F^{(3)}}{r^{\prime\,2}}+\frac{9F^{(2)}}{r^{\prime\,3}}+\frac{21F^{(1)}}{r^{\prime\,4}}+\frac{21F}{r^{\prime\,5}}\right],
\end{align}
\end{subequations}
and
\begin{align}
    &f_{r^{\prime}r^{\prime}}=4\sqrt{\frac{\pi}{5}}Y_{20}(\theta^
    \prime,\phi^{\prime}),\quad  f_{r^{\prime}\theta^{\prime}}=2\sqrt{\frac{\pi}{5}}\partial_{\theta^{\prime}} Y_{20}(\theta^
    \prime,\phi^{\prime}), \notag \\
    &f_{r^{\prime}\phi^{\prime}}=0 , \quad f^{(2)}_{\theta^{\prime}\theta^{\prime}}=-1, \quad f_{\theta^{\prime}\phi^{\prime}}=0, \notag \\
    &f^{(1)}_{\theta^{\prime}\theta^{\prime}}=2\sqrt{\frac{\pi}{5}}\left(\partial_{\theta^{\prime}}^2-\cot\theta^\prime\partial_{\theta^{\prime}}-\frac{\partial^2_{\phi^{\prime}}}{\sin^2\theta^{\prime}}\right)Y_{20}(\theta^
    \prime,\phi^{\prime}), \notag \\
    &f^{(1)}_{\phi^{\prime}\phi^{\prime}}=-f^{(1)}_{\theta^{\prime}\theta^{\prime}},\quad f^{(2)}_{\phi^{\prime}\phi^{\prime}}=1-f_{r^{\prime}r^{\prime}}. \label{eq:Teukolsky_angular_basis}
\end{align}
The spherical harmonic $Y_{20}(\theta^\prime,\phi^{\prime})$ is given by
\begin{align}
    Y_{20}=\frac{1}{8}\sqrt{\frac{5}{\pi}}(1+3\cos2\theta^\prime).
\end{align}
We are free to specify the form of $F(u^\prime)$ in Eq.~\eqref{eq:Teukolsky_ABC}. Here we consider an outgoing Gaussian pulse:
\begin{align}
    F(u^\prime)=Xe^{-\frac{(u^\prime-r^\prime_c)^2}{\tau^2}}, \label{eq:Gaussian_pulse_F}
\end{align}
where $u^\prime=t^\prime-r^\prime$ is the retarded time, $r^\prime_c$ is the center of the pulse at $t^\prime=0$, $\tau$ is its width, and $X$ is its amplitude.
In Eq.~\eqref{eq:Teukolsky_ABC} the symbol $F^{(n)}$ denotes  the $n$th derivative of $F(u^\prime)$:
\begin{align}
    F^{(n)}\equiv\left[\frac{d^nF(u^\prime)}{du^{\prime\,n}}\right]_{u^\prime=t^\prime-r^\prime}. \label{eq:Teukolsky_wave_outgoing}
\end{align}

\begin{figure}[htb]
\includegraphics[width=\columnwidth,clip=true]{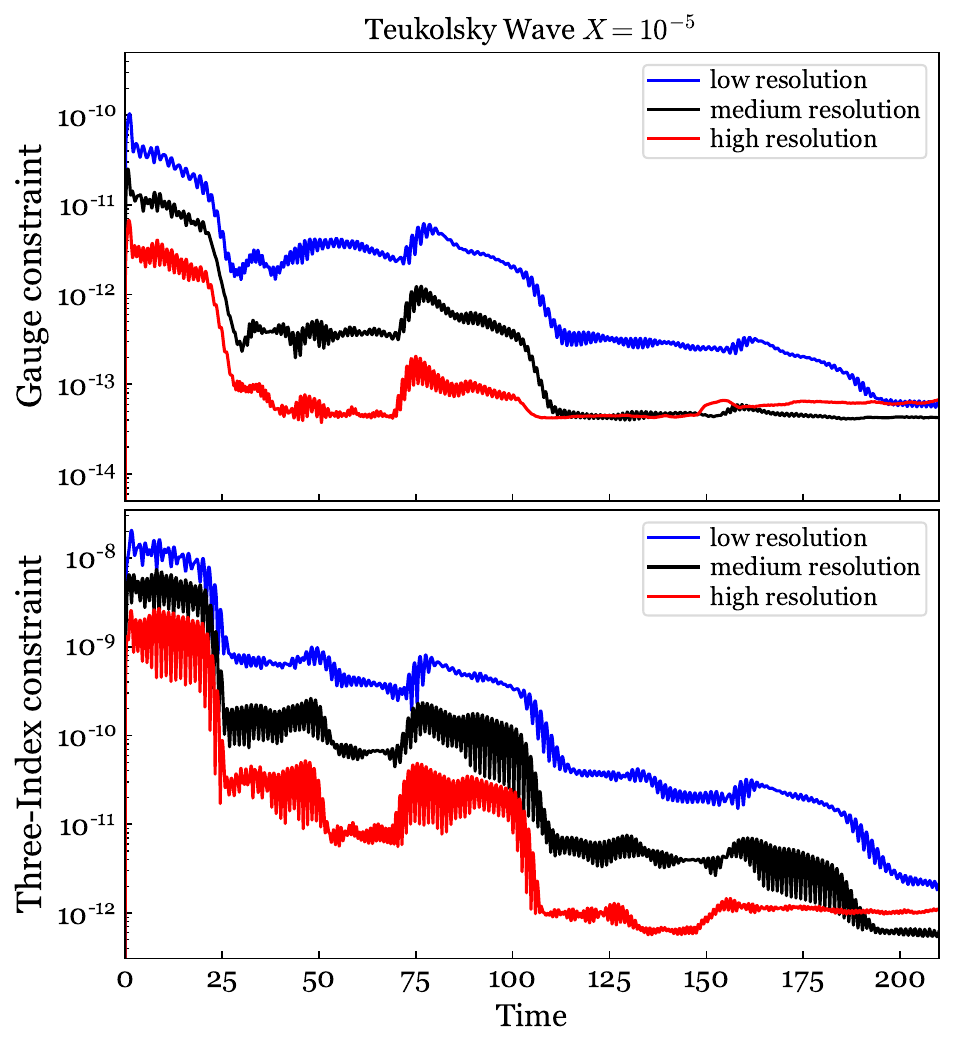}
  \caption{The evolution of the GH gauge constraint violation (top) and three-index constraint violation (bottom) for the CCM system, as defined in Eqs.~\eqref{eq:gauge_constraint} and \eqref{eq:three_constraint}.  The CCE system yields identical constraint violations as CCM. We simulate the propagation of an outgoing Teukolsky wave (depicted in Fig.~\ref{fig:Teukolsky_wave}) with three numerical resolutions. The amplitude of the wave $X$ is set to $10^{-5}$.
}
 \label{fig:low_amp_con}
\end{figure}

Our numerical setup is sketched in Fig.~\ref{fig:Teukolsky_wave}. Initially, the Gaussian pulse is centered at $r_c^\prime=20$, with a width of $\tau=2$. We begin by simulating the system with a small Cauchy domain, where the outer grid radius $r^{\prime}_{\rm out}=41$ is sufficiently small to highlight the impact of backscattering of GWs at the outer boundary, allowing the code to resolve the improvements provided by CCM. The evolution of the system is performed using both CCE and CCM, with the time-like worldtube consistently positioned at $r^{\prime}_{\rm out}=41$, coinciding with the outer boundary of the Cauchy domain. It is expected that the CCM system will provide more accurate boundary conditions at $r^{\prime}_{\rm out}$, better representing the true evolution of the system. Therefore, we need to establish a reference system that serves as an exact solution uncontaminated by numerical approximations such as inaccurate boundary conditions. This can be achieved differently under two separate scenarios.

Firstly, when the amplitude of the Teukolsky wave $X$ [Eq.~\eqref{eq:Gaussian_pulse_F}] is adequately small, we are in the perturbative regime, where the analytic solution if given by Eq.~\eqref{eq:Teukolsky_metric}. Therefore, we can compare the CCE and CCM simulations with the analytic results. This scenario is explored below in Sec.~\ref{subsec:Teukolsky_X_1e-5}, where $X$ is set to $10^{-5}$.

Next in Sec.~\ref{subsec:Teukolsky_X_2}, we address the second case where the amplitude $X$ is large, and nonlinear effects cannot be neglected. The reference system is chosen to be a CCE simulation with a larger Cauchy computational domain, and its outer boundary remains casually disconnected from the system throughout the simulation. It is worth noting that, despite the weakly hyperbolic nature of CCE \cite{Giannakopoulos:2020dih,Giannakopoulos:2021pnh,Giannakopoulos:2023zzm}, previous studies \cite{Moxon:2021gbv,Iozzo:2021vnq,Mitman:2020pbt,Mitman:2021xkq} have shown that the SpECTRE CCE system can yield high-quality waveforms at future null infinity, establishing it as a reasonable reference for comparison. To ensure the independence of the reference run, we employ another NR code, the Spectral Einstein Code (SpEC) \cite{SpECwebsite}, developed by the
Simulating eXtreme Spacetimes (SXS) collaboration \cite{SXSWebsite}. As illustrated in Fig.~\ref{fig:Teukolsky_wave}, we place the outer boundary of the reference system at $r^{\prime}_{\rm ref}=900$. The location of the worldtube for CCE wave extraction remains at $r^{\prime}_{\rm wt}=41$, consistent with the other two systems, facilitating fair comparisons.

\subsubsection{Perturbative regime: \texorpdfstring{$X=10^{-5}$}{x1e-5}}
\label{subsec:Teukolsky_X_1e-5}
We adopt CCE and CCM to evolve the system with three different numerical resolutions, spanning over 1000 code units. The duration of our simulation greatly exceeds the timescale of the physical process of interest, which involves the propagation of the Teukolsky wave from its initial location to null infinity within the first 50 code units. Throughout our investigation, we do not observe any numerical instabilities. As a standard numerical diagnostic, in Fig.~\ref{fig:low_amp_con}, we plot the pointwise Euclidean $L^2$ norm
of the GH gauge constraint $C_a$ [Eq.~(40) of \cite{Lindblom:2005qh}]:
\begin{align}
    {\rm{Gauge~ constraint}}=\left\lVert \sqrt{\sum_{a=0}^3C_a^2}\right\rVert, \label{eq:gauge_constraint}
\end{align}
and the three-index constraint [Eq.~(26) of \cite{Lindblom:2005qh}]:
\begin{align}
    {\rm{Three-Index~ constraint}}=\left\lVert \sqrt{\sum_{i=1}^3\sum_{a,b=0}^3C_{iab}^2}\right\rVert, \label{eq:three_constraint}
\end{align}
where $\lVert\cdot\rVert$ denotes the $L^2$ norm over grid points in the Cauchy domain. We find that the CCE and CCM systems yield nearly identical constraint evolutions. Moreover, as anticipated, the constraints decrease with increasing resolution.

\begin{figure}[htb]
\includegraphics[width=\columnwidth,clip=true]{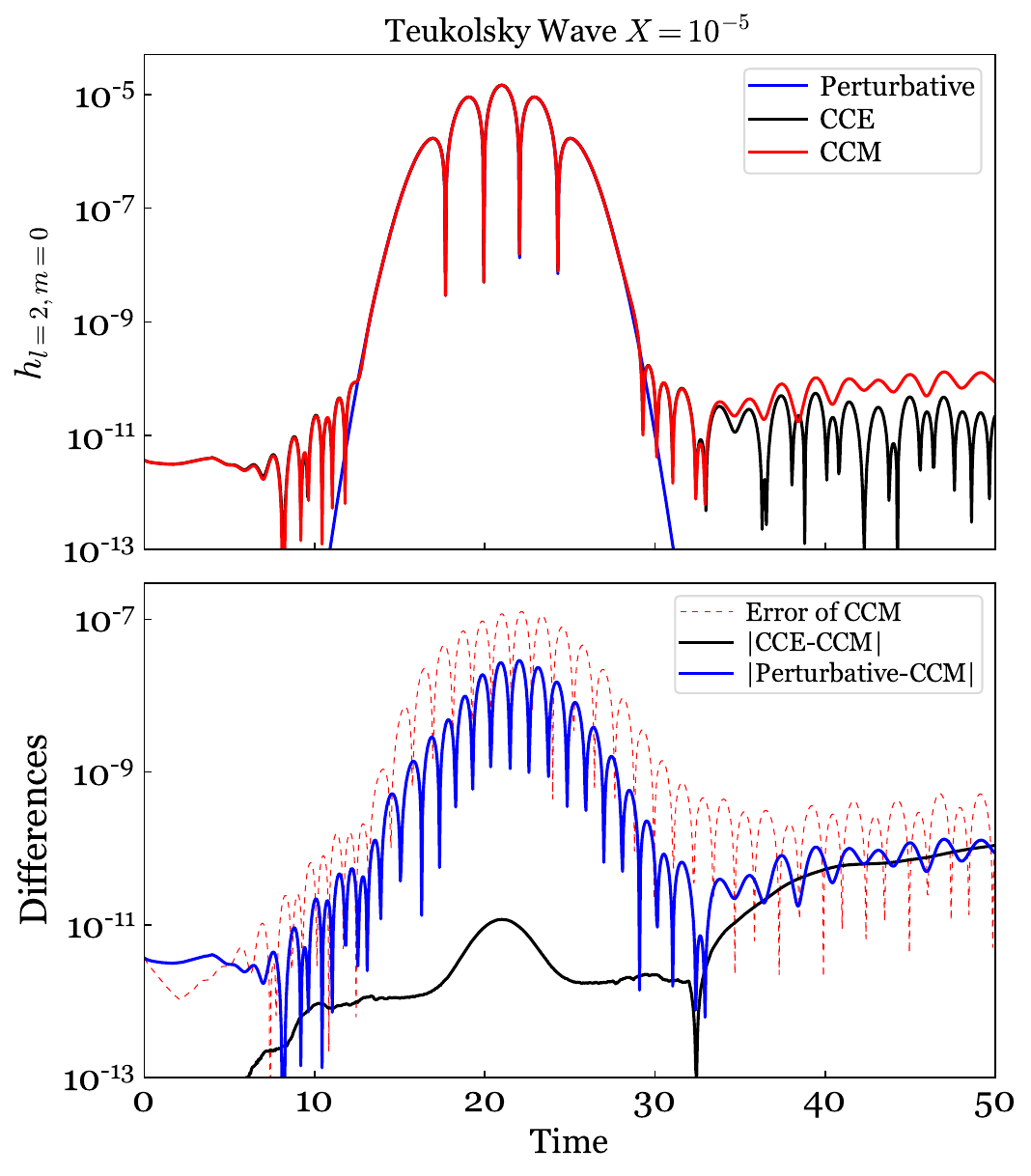}
  \caption{Top panel: The $(l=2,m=0)$ harmonic of the strain, measured at $\mathscr{I}^+$, of the Teukolsky wave with an amplitude of $X=10^{-5}$. The strain is extracted with CCE (in black) and CCM (in red), respectively. Both are compared to the perturbative expression given in Eq.~\eqref{eq:Teukolsky_h_20}, represented by the blue curve. Bottom panel: The differences between the waveforms. The numerical error of the CCM method (dashed red curve)
  is estimated by calculating the difference between two numerical resolutions.}
 \label{fig:low_amp_strain_only}
\end{figure}

Next, we analyze the strain $h$ measured at future null infinity. Considering our initial data in Eq.~\eqref{eq:Teukolsky_angular_basis}, which comprises a single $(l=2,m=0)$ harmonic, the only non-zero component of the strain is $h_{20}$. In the perturbative limit, 
its time evolution reads \cite{PhysRevD.26.745}
\begin{align}
    [rh_{20}]_{\mathscr{I}^+}=\sqrt{\frac{6\pi}{5}}F^{(4)}. \label{eq:Teukolsky_h_20}
\end{align}
In the top panel of Fig.~\ref{fig:low_amp_strain_only}, we compare the perturbative expression (in blue) with the results produced by CCE (in black) and CCM (in red), observing consistency among them. To provide a quantitative assessment, we estimate the error in the CCM waveform by comparing two numerical resolutions. The resulting difference is plotted as the red-dashed curve in the bottom panel of Fig.~\ref{fig:low_amp_strain_only}. Additionally, we compute the differences between CCM's and CCE's results (in black), as well as between CCM's result and the perturbative expression (in blue). It can be seen that the discrepancies across the three systems (CCE, CCM, and the perturbative limit) are smaller than the numerical error. This implies that the code cannot resolve the distinction between the CCE and CCM systems, which is reasonable because backscattered waves are effectively suppressed in the current perturbative limit, so that the new matching term in Eq.~\eqref{eq:bc_bjorhus} becomes negligible. On the other hand, the agreement between our numerical systems (CCE and CCM) and the analytic expression in Eq.~\eqref{eq:Teukolsky_h_20} serves as a cross-check for the accuracy of our numerical code.

Appendix \ref{app:1e-5} presents comprehensive comparisons for the Weyl scalars $\psi_{0...4}$ and the News $N$ for CCE and CCM versus the analytic solution, for this same case.
These findings align with the observations we made regarding $h_{20}$.

\begin{figure}[htb]
\includegraphics[width=\columnwidth,clip=true]{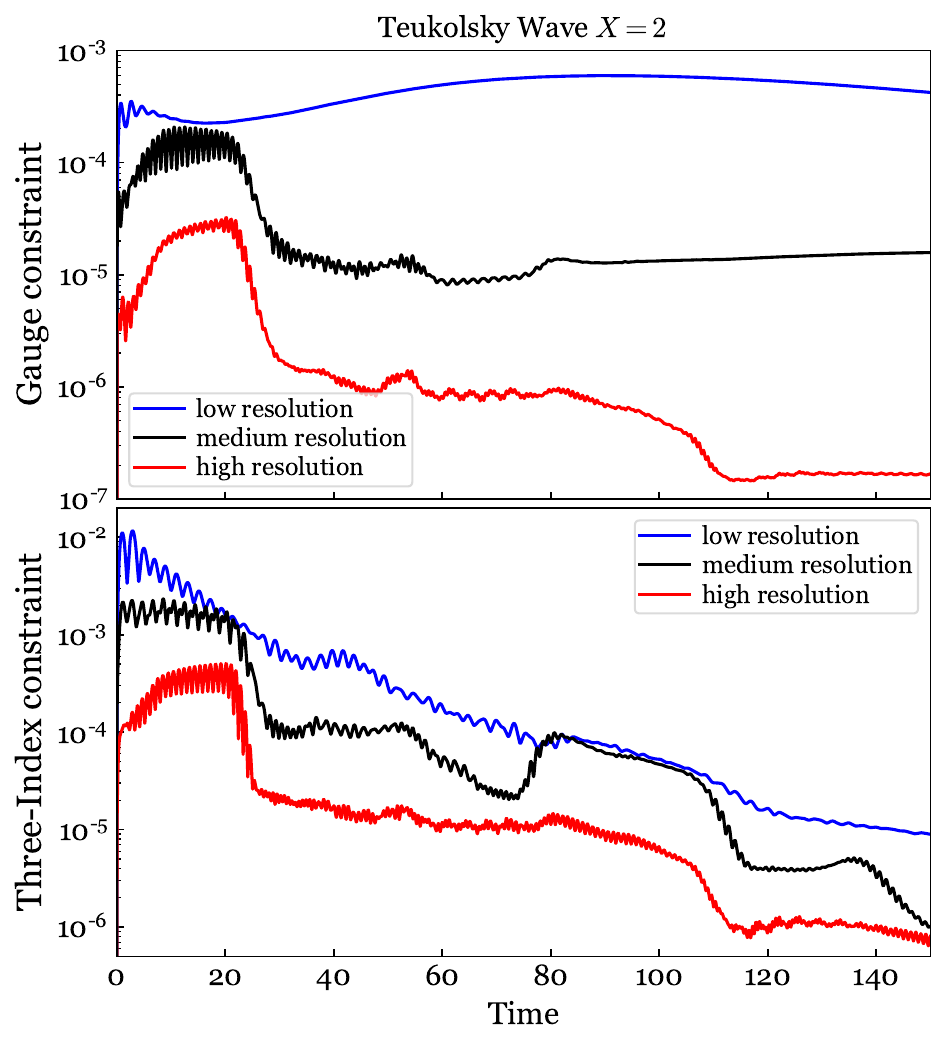}
  \caption{ (Similar to Fig.~\ref{fig:low_amp_con}) The evolution of the GH three-index constraint violation (top) and gauge constraint violation (bottom) for the CCM system, except that the amplitude of the Teukolsky wave $X$ is set to $2$. The CCE system yields identical constraint violations as CCM.}
 \label{fig:gauge_constraint_X2}
\end{figure}

\begin{figure*}[htb]
    \includegraphics[width=\columnwidth,clip=true]{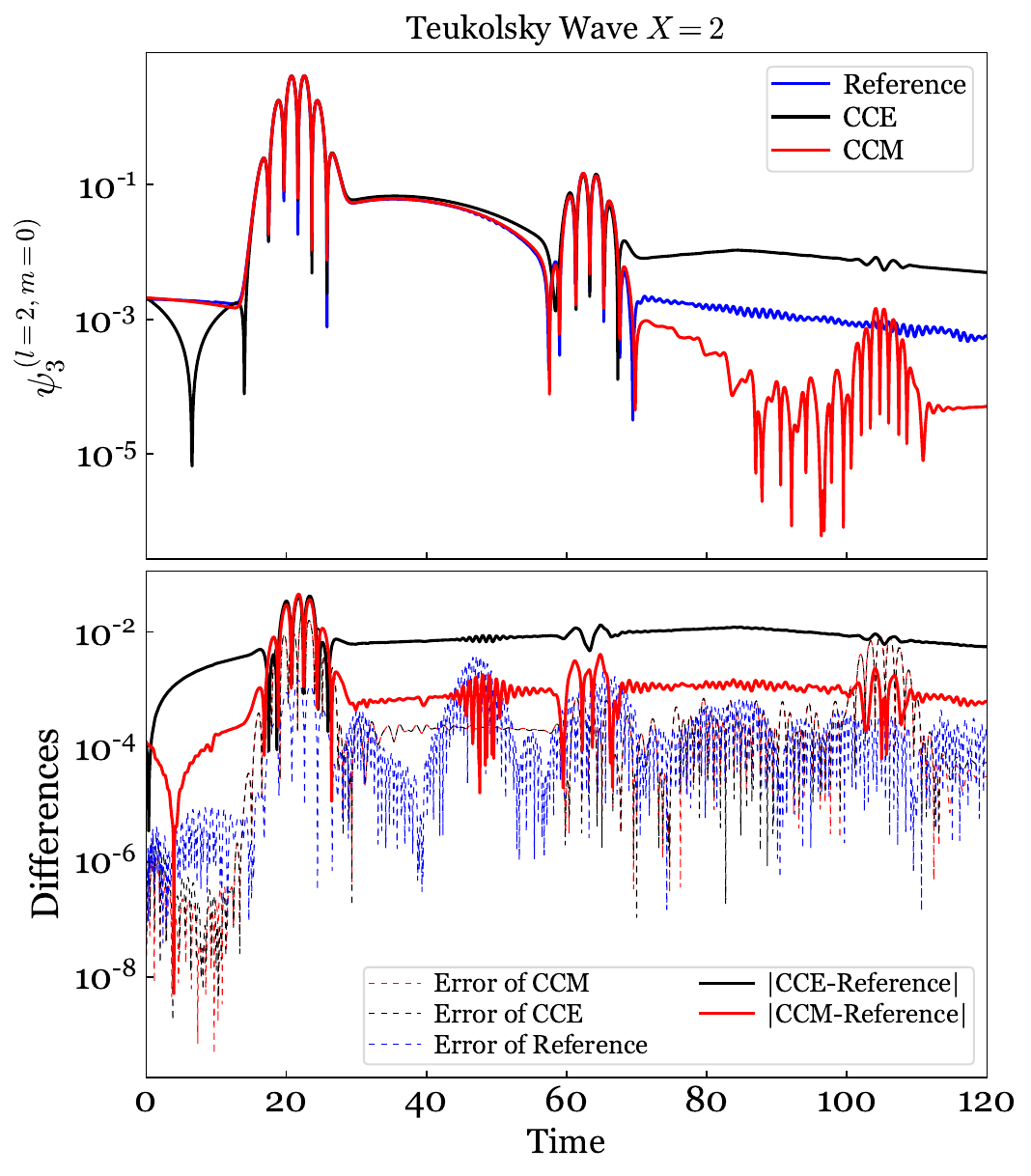} 
    \includegraphics[width=\columnwidth,clip=true]{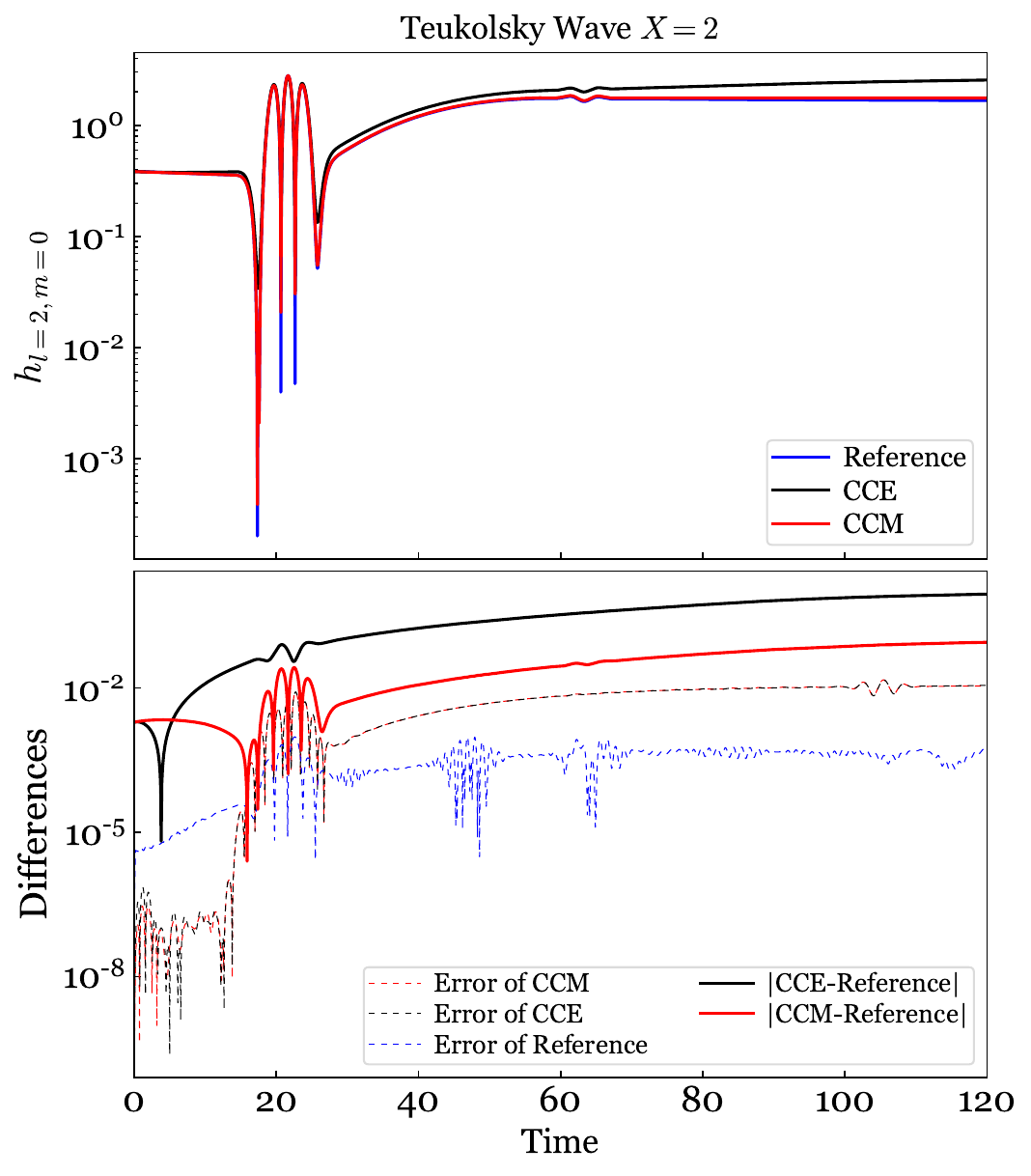}
  \caption{Top panels: The $(l=2,m=0)$ harmonic of the Weyl scalar $\psi_3$ (left) and strain (right), measured at $\mathscr{I}^+$, of the Teukolsky wave with an amplitude of $X=2$. They are extracted with CCE (in black) and CCM (in red), respectively. The third reference run (in blue) is performed independently with another NR code SpEC, where the outer boundary of the Cauchy domain remains causally disconnected from the system throughout the simulation.  Bottom panels: The differences between the waveforms. Solid curves show the deviation of the CCE's and CCM's results from the reference ones. For comparison, we also estimate the numerical error of the waveforms (dashed curves) by calculating the difference between two numerical resolutions.}
 \label{fig:large_amp_strain_only}
\end{figure*}

\subsubsection{Nonperturbative regime: \texorpdfstring{$X=2$}{X2}}
\label{subsec:Teukolsky_X_2}
Next, we switch our attention to the nonlinear case $X=2$. Again, the system can undergo stable evolution without encountering numerical instabilities. Furthermore, the application of CCM does not lead to any worsening or improvement in the constraint violations [see Eqs.~\eqref{eq:gauge_constraint} and \eqref{eq:three_constraint}] when compared to the CCE system, as shown in Fig.~\ref{fig:gauge_constraint_X2}. 

\begin{figure*}[htb]
\includegraphics[width=\columnwidth,clip=true]{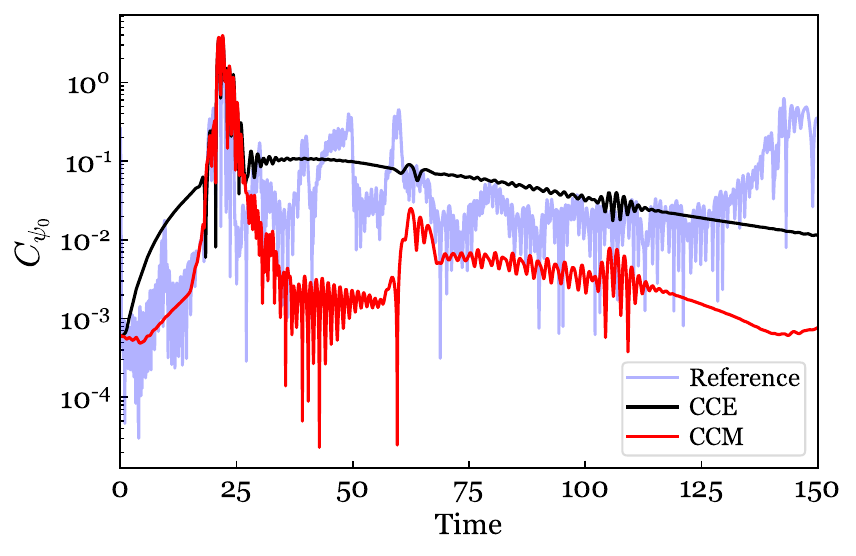}
\includegraphics[width=\columnwidth,clip=true]{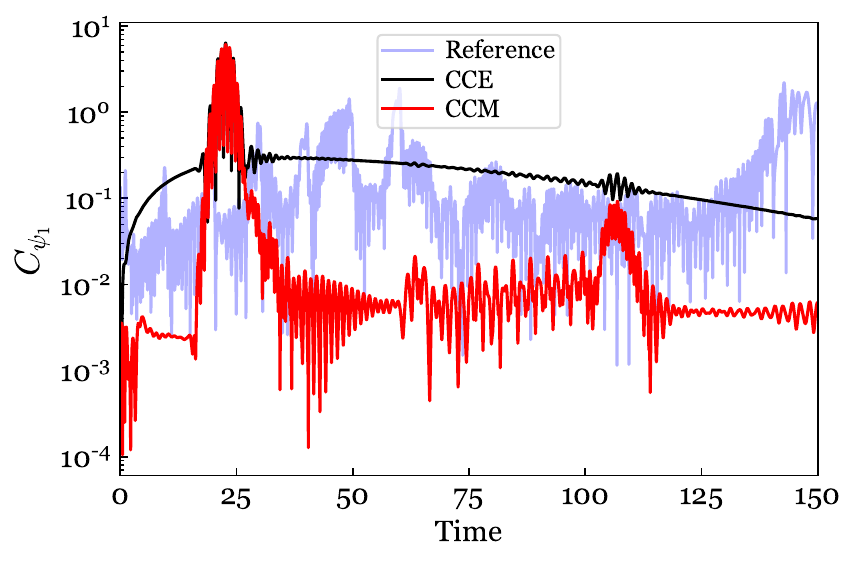} \\
\includegraphics[width=\columnwidth,clip=true]{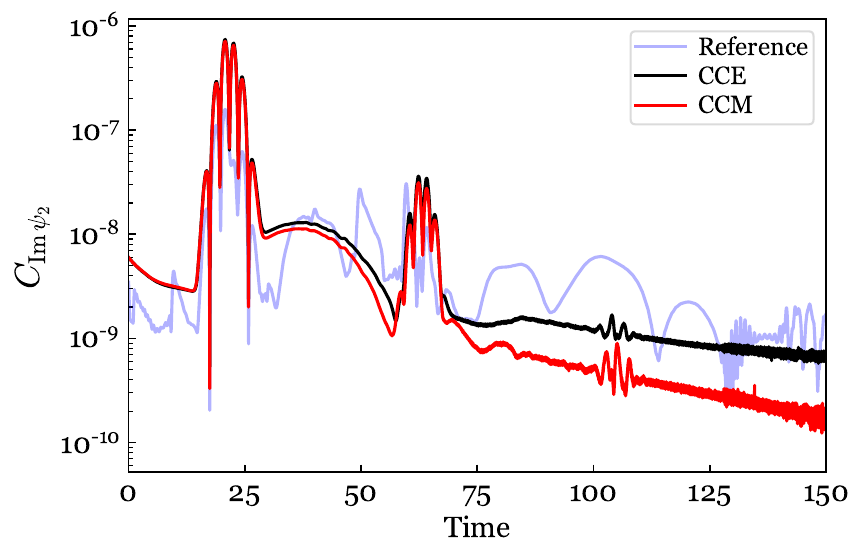}
\includegraphics[width=\columnwidth,clip=true]{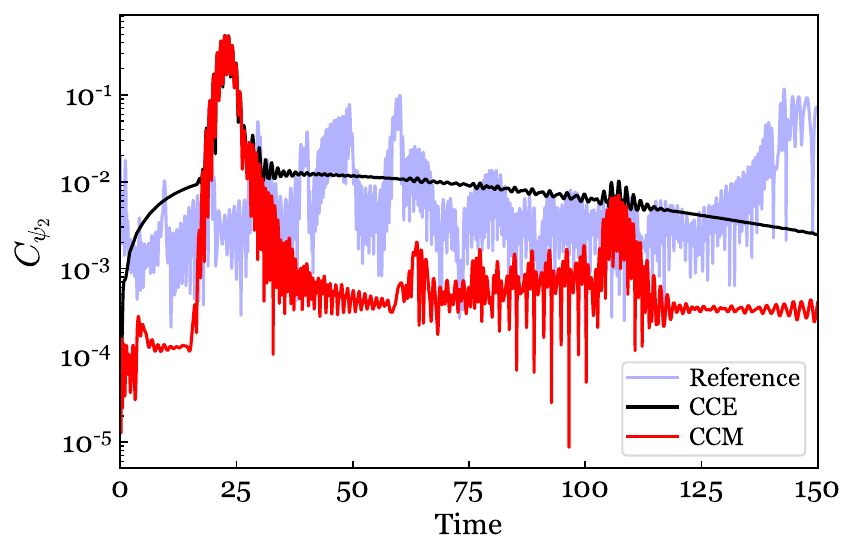} \\
\includegraphics[width=\columnwidth,clip=true]{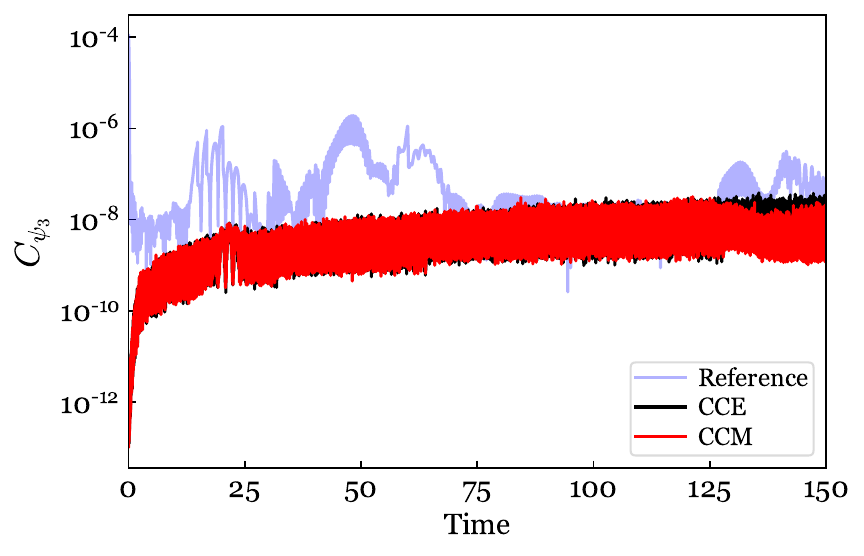}
\includegraphics[width=\columnwidth,clip=true]{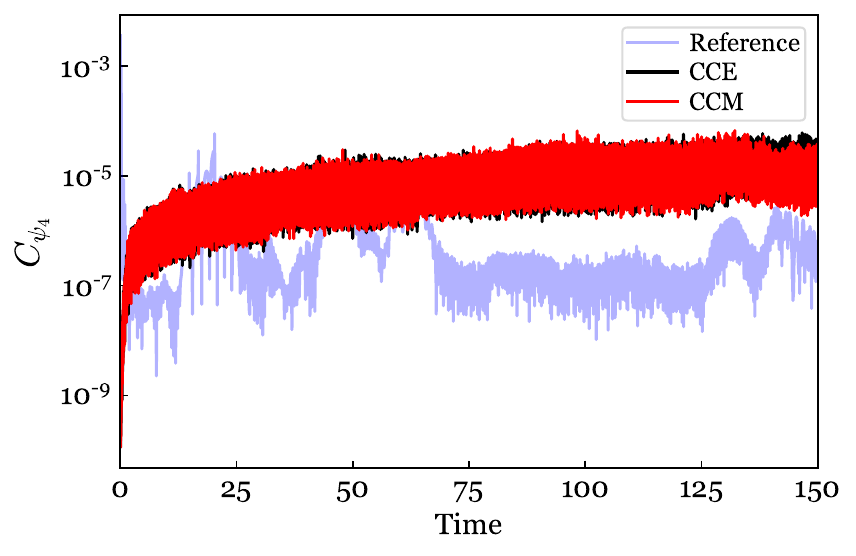}
  \caption{The norms of the violations of Bondi-gauge constraints [Eqs.~\eqref{eq:bondi_violation}] and the real-valued constraint for $\psi_2$ [Eq.~\eqref{eq:im_psi2}]. The propagation of a Teukolsky wave with an amplitude of 2 is considered. The system is evolved using CCE (in black) and CCM (in red), respectively. They are compared to the third reference run (in blue), which is independently performed with another NR code SpEC.}
 \label{fig:norms_of_bondi_violations}
\end{figure*}

At future null infinity, we first compute the evolution of the Weyl scalar $\psi_3^{(l=2,m=0)}$ and the strain $h_{20}$. In the top left panel of Fig.~\ref{fig:large_amp_strain_only}, we present the results obtained using the CCM method for $\psi_3^{(2,0)}$ (in red). This plot clearly illustrates the physical process: The primary outgoing Teukolsky pulse, initially located at $r_c^\prime$ (Fig.~\ref{fig:Teukolsky_wave}), reaches the Cauchy outer boundary $r_{\rm out}^\prime$ after a time interval of $r_{\rm out}^\prime-r_c^\prime=21$. Because of
the null slicing of the characteristic system, any outgoing null GW is \emph{instantaneously} transmitted to future null infinity as soon as it intersects the worldtube\footnote{Note that the instantaneity applies specifically to outgoing GWs when the characteristic system adopts outgoing null hypersurfaces. Conversely, ingoing GWs would be instantaneously transmitted if we had used a characteristic system that utilizes ingoing null hypersurfaces, see e.g., Ref.~\cite{Gomez:1997nqa}.}. Consequently, we observe that the main pulse also appears at $\mathscr{I}^+$ at a time of 21. On the other hand, since our constructed NR initial data is not a perfect error-free solution for a solely outgoing Teukolsky wave, there is a wave component that travels inward, commonly known as junk radiation. As the evolution progresses, this component falls towards the coordinate center, crosses it, and bounces back out. The junk wave eventually reaches the Cauchy outer boundary (which coincides with the worldtube) and $\mathscr{I}^+$ at a time of $r_{\rm out}^\prime+r_c^\prime=61$. This junk wave can be seen as a secondary pulse at time 61 in the top left panel of Fig.~\ref{fig:large_amp_strain_only}. Moreover, we find the appearance of a tertiary wave at a time of 100. This arises because a portion of the primary Teukolsky wave reflects off the numerical boundary between the Cauchy and characteristic systems. Subsequently, this reflected wave traverses the entire Cauchy region, with a propagation time of $2r_{\rm out}$, namely the domain's diameter. At $t=2r_{\rm out}+(r_{\rm out}^\prime-r_c^\prime)=103$, the reflected wave escapes to future null infinity. 

The presence of the tertiary reflected wave indicates that our current CCM algorithm cannot entirely eliminate the spurious reflection at the numerical boundary. We attribute this issue to the gauge part of the boundary conditions. As previously mentioned, the boundary conditions can be classified into three subsets \cite{Kidder:2004rw,Lindblom:2005qh}: constraint-preserving, physical, and gauge conditions. Although we have made progress in accounting for the physical degree of freedom by appropriately matching the backscattered GWs, the gauge aspect remains largely unexplored. Here we simply follow Ref.~\cite{Rinne:2007ui} and adopt Sommerfeld boundary conditions for the gauge subset, which could be a major source of the spurious numerical reflection. In principle, the gauge information is also encoded in the characteristic system, thereby offering a potential avenue to extend our current matching algorithm to encompass the gauge degrees of freedom. We leave the relevant discussions for future work.

Nevertheless, upon comparing the results obtained using CCE and CCM to the reference ones (in blue) in Fig.~\ref{fig:large_amp_strain_only}, we find that the new physical boundary conditions already lead to noticeable improvements. Specifically, the differences between the waveforms shown in the bottom panels of Fig.~\ref{fig:large_amp_strain_only} clearly demonstrate that the CCM method systematically reduces the deviation from the reference system by approximately one order of magnitude. This conclusion holds true for other Weyl scalars and the News function as well. For more detailed information, please refer to Appendix \ref{app:X=2}.

In addition to the gauge constraint violation defined in Eq.~\eqref{eq:gauge_constraint}, an analysis of the Bondi gauge provides an additional self-consistency test to assess the quality of the waveforms \cite{O-M-Moreschi_1986,Iozzo:2020jcu}: With the Bondi gauge, one can write the Bianchi identities in the Newman-Penrose formalism, which yields a set of constraint equations for the Weyl scalars and the strain \cite{O-M-Moreschi_1986,Iozzo:2020jcu}
\begin{subequations}
\label{eq:bondi_violation}
\begin{align}
    &C_{\psi_4}\equiv\psi_4+\ddot{h}=0, \quad  C_{\psi_3}\equiv\psi_3-\frac{1}{\sqrt{2}}\eth \dot{h}=0,\label{eq:bondi_violation_psi3}\\
    &C_{\psi_s}\equiv\dot{\psi}_{s}+\frac{1}{\sqrt{2}}\eth \psi_{s+1}-\frac{3-s}{4}\bar{h}\psi_{s+2}=0, 
\end{align}
\end{subequations}
with $s=0, 1, 2$.
Moreover, the requirement for the Bondi mass aspect to be real-valued introduces an additional constraint \cite{Flanagan:2015pxa}
\begin{align}
    C_{\rm{Im} \psi_2 }\equiv{\rm{Im}} \psi_2+{\rm{Im}} \left(\frac{1}{2}\eth^2h+\frac{1}{4}\bar{h}\dot{h}\right)=0. \label{eq:im_psi2}
\end{align}
We compute the norms of these constraints with the $\texttt{PYTHON}$ package $\texttt{scri}$ \cite{scri,mike_boyle_2020_4041972,Boyle:2013nka,Boyle:2014ioa,Boyle:2015nqa}, and plot the results in Fig.~\ref{fig:norms_of_bondi_violations}. We see the CCM method systematically reduces the constraint violations compared to CCE, with the exception of the constraints given in Eqs.~\eqref{eq:bondi_violation_psi3}, for which CCM and CCE are similar. 

\subsection{Perturbing a Kerr BH with a Teukolsky wave}
\label{subsec:perturb_kerr}
Our second type of test involves perturbing a Kerr BH with a Teukolsky wave. In this case, we set the dimensionless spin of the BH, denoted as $\chi$, to 0.5. To maintain conciseness, we assume a BH mass of 1 
throughout the discussions.  

We still build the initial data nonlinearly by solving the XCTS formulation. In particular, we use spherical Kerr-Schild coordinates \cite{Chen:2021rtb} to compute the conformal metric. With these coordinates, the outer horizon of the Kerr BH remains spherical, with a coordinate radius of $r_+=1+\sqrt{1-\chi^2}=1.87$. We excise the Cauchy domain at 1.8, slightly inside the horizon. In addition, we set the worldtube at $r^\prime_{\rm out}=150$, aligned with the outer boundary.  For the Teukolsky wave, we replace the free function $F(u^\prime)$ in Eqs.~\eqref{eq:Teukolsky_ABC} and \eqref{eq:Gaussian_pulse_F} with the following expression:
\begin{align}
    F(v^\prime)=Xe^{-\frac{(v^\prime-r^\prime_c)^2}{\tau^2}}, 
\end{align}
where $v^\prime=t^\prime+r^\prime$ is the advanced time. This replacement ensures the Teukolsky wave falls into the BH rather than escaping. Initially, the center of the pulse is located at $r_c^\prime=40$. We choose $X=0.01$ and $\tau=2$.

\begin{figure}[htb]
        \includegraphics[width=\columnwidth,clip=true]{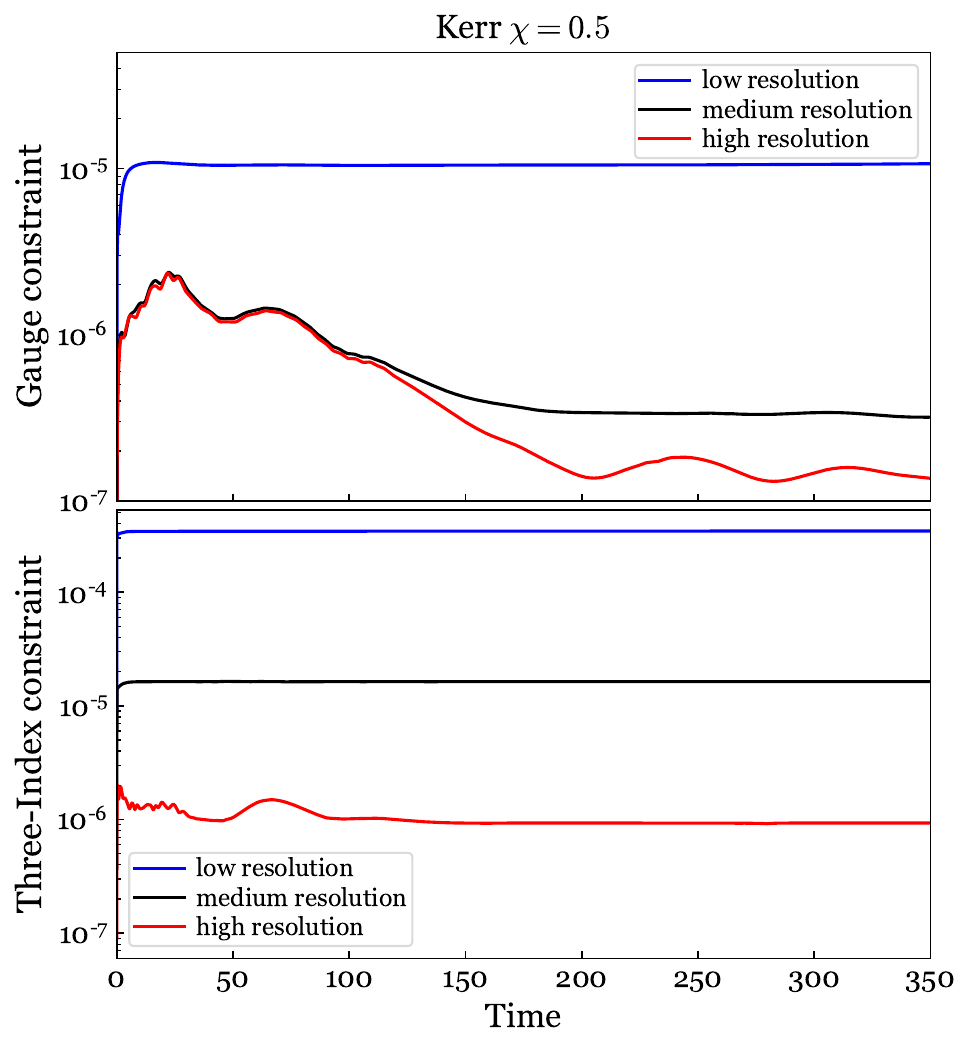}
  \caption{The evolution of the GH gauge constraint violation (top) and three-index constraint violation (bottom) for the CCM system, as defined in Eqs.~\eqref{eq:gauge_constraint} and \eqref{eq:three_constraint}.  The CCE system yields identical constraint violations as CCM.
  We perturb a Kerr BH, whose dimensionless spin is 0.5, with an ingoing Teukolsky wave. Three numerical resolutions are adopted.}
 \label{fig:Kerr_con}
\end{figure}

\begin{figure}[htb]
\includegraphics[width=\columnwidth,clip=true]{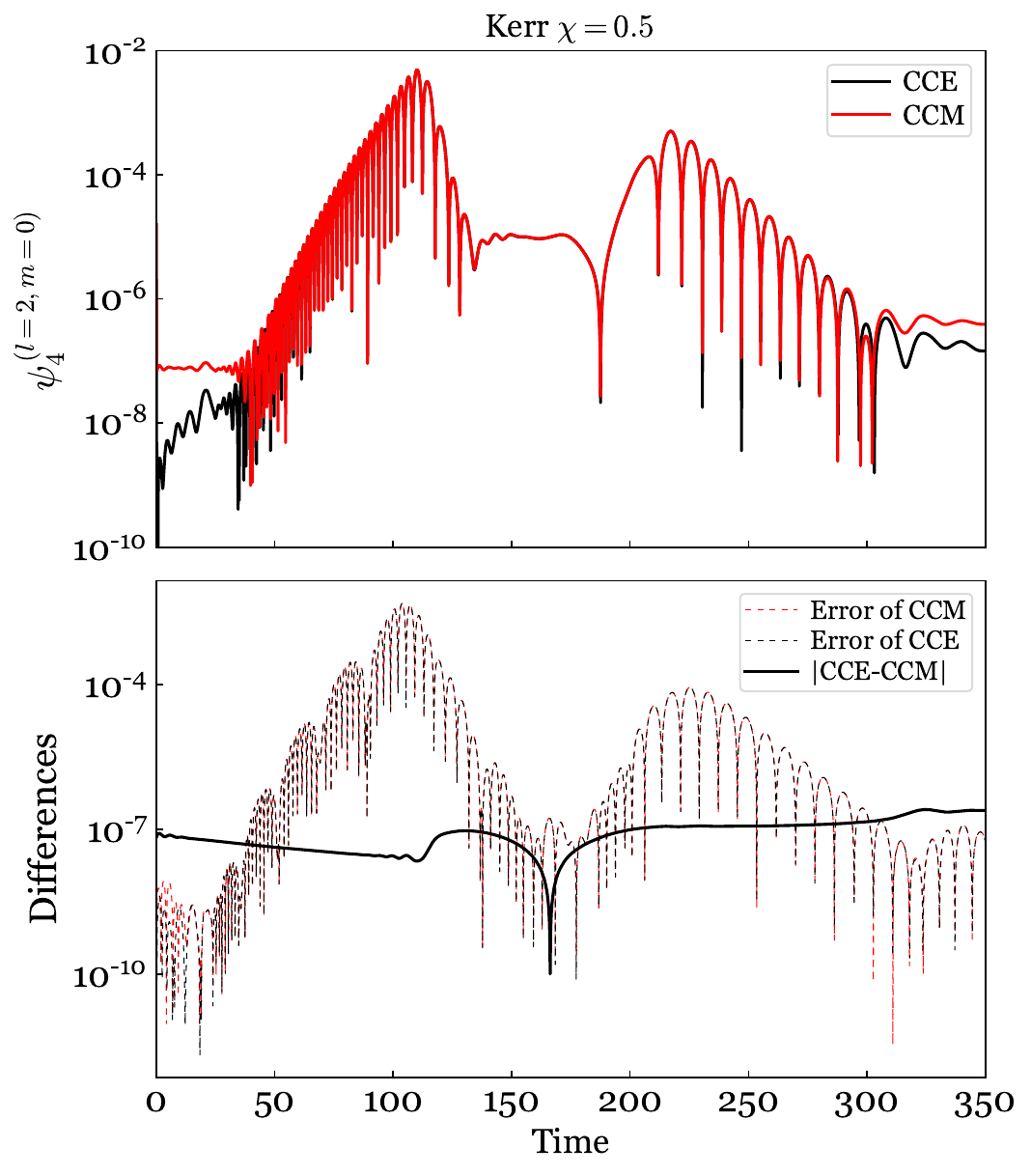}
  \caption{Top panel: The $(l=2,m=0)$ harmonic of the Weyl scalar $\psi_4$, measured at $\mathscr{I}^+$, emitted by a Kerr BH with a dimensionless spin of 0.5. The BH is perturbed by an ingoing Teukolsky wave. The waveform is extracted with CCE (in black) and CCM (in red), respectively. Bottom panel: The differences between the waveforms. The numerical errors of CCM and CCE are computed by calculating the difference between two numerical resolutions.}
 \label{fig:kerr_waveform}
\end{figure}

We evolve the system with three numerical resolutions. Figure \ref{fig:Kerr_con} displays the evolution of the gauge and three-index constraints. Similarly to the previous tests, we find that the CCM method produces identical constraint evolutions compared to CCE. We also check that CCM does not introduce extra numerical instabilities --- the system can be evolved stably for more than $1000 M$. Next, in Fig.~\ref{fig:kerr_waveform}, we present the News measured at $\mathscr{I}^+$. The first peak at a time of $\sim 110$ corresponds to the junk radiation: Since our numerically constructed initial data is not a perfectly ingoing GW pulse, a fraction of the wave travels outward once the evolution starts, and it reaches the worldtube at a time of $r^\prime_{\rm out}-r_c^\prime=110$. On the other hand, the ingoing Teukolsky wave collides with the Kerr BH and excites its quasinormal ringing. This ringdown signal disperses to null infinity at a time of $r_c^\prime+r^\prime_{\rm out}=190$. Utilizing the $\texttt{PYTHON}$ package $\texttt{qnm}$ \cite{Stein:2019mop}, we verified that the frequency and damping rate of the quasinormal ringing align with the prediction of BH perturbation theory. 

As shown in the bottom panel of Fig.~\ref{fig:kerr_waveform}, the difference between CCE and CCM (solid curve) is smaller than the numerical errors (dashed curves), which suggests that the influence of the matching term in the boundary condition [Eq.~\eqref{eq:bc_bjorhus}] is insignificant for the present system. A similar conclusion can also be drawn from other waveform quantities. Please refer to Appendix \ref{app:kerr_bh} for more comparisons. This is because the decay of the ingoing component of the Weyl scalar $\psi_0$ is rapid with distance, specifically, following a fifth power law in the context of BH perturbation theory (see Table I of Ref.~\cite{1974ApJ...193..443T}) and the Teukolsky wave [see Eq.~\eqref{eq:Teukolsky_bulk_psi0}]. Consequently, the rapid decay significantly reduces the strength of the backscattered wave at the outer boundary $r^\prime_{\rm out}=150$. To accentuate the difference, it would be necessary to bring the outer boundary closer. However, this would also require a more precise gauge boundary condition. Detailed discussions on these aspects are reserved for future studies.

\subsection{Initializing a GW pulse on the characteristic grid}
\label{subsec:pulse_on_CCE}
\begin{figure}[htb]
        \includegraphics[width=\columnwidth,clip=true]{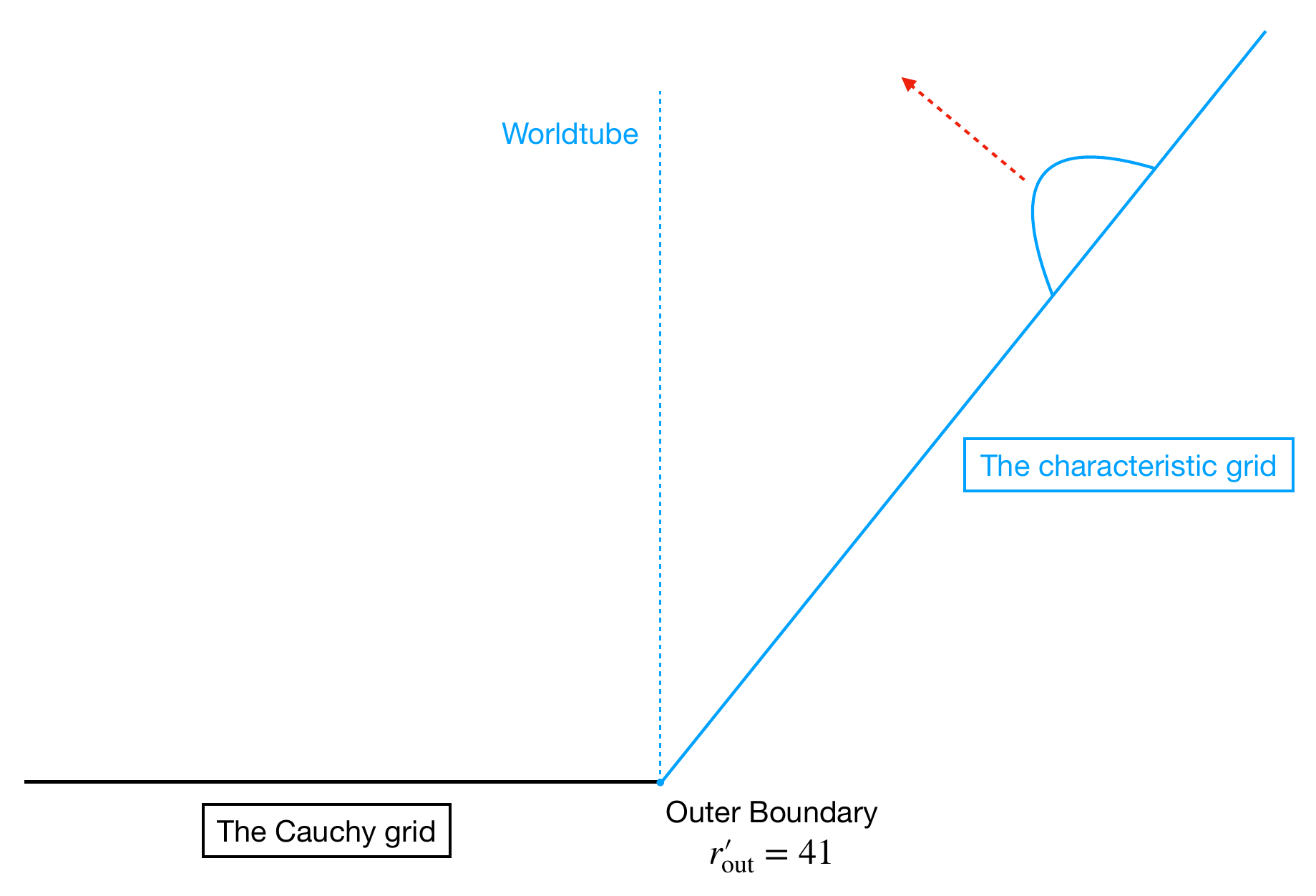}
  \caption{An illustration depicting the propagation of a GW originating from an initial location on the characteristic grid. The inner Cauchy region is initialized with Minkowski spacetime. The worldtube is positioned at $41$, coinciding with the outer boundary of the Cauchy grid.}
 \label{fig:Teukolsky_wave_on_cce}
\end{figure}

Our final test is to initialize a GW pulse on the characteristic grid. To do this, we adopt the following initial data for the Bondi variable $\hat{J}$ on the initial null slice:
\begin{align}
    \hat{J}(\hat{y},\hat{\theta},\hat{\phi})=
    \begin{cases}
    0, & \hat{y}\leq \hat{y}_{\min}, \\
    \tensor[_{+2}]{Y}{_{20}}(\hat{\theta},\hat{\phi})\mathcal{J}(\hat{y}), &\hat{y}_{\min} \leq \hat{y} \leq \hat{y}_{\max}, \\
    0, & \hat{y}\geq \hat{y}_{\max},
    \end{cases}
\end{align}
where $\hat{y}=1-2\hat{R}/\hat{r}$, and $\hat{R}$ is the partially flat Bondi-like radius of the worldtube. The spin weight of the pulse is set to 2 in order to match with that of $\psi_0$. The radial profile $\mathcal{J}(y)$ reads
\begin{align}
    \mathcal{J}(\hat{y})=4Z\frac{(\hat{y}_{\max}-\hat{y})(\hat{y}-\hat{y}_{\min})}{(\hat{y}_{\max}-\hat{y}_{\min})^2}e^{-\frac{(\hat{y}-\hat{y}_c)^2}{\tau^2}}.
\end{align}
Meanwhile, the inner Cauchy domain is initialized to a flat (Minkowski) spacetime. Figure \ref{fig:Teukolsky_wave_on_cce} depicts our numerical setup. The center of the pulse $\hat{y}_c$ is initially at 0. In addition, we choose $y_{\min}=-0.8,y_{\max}=0.8,\tau=0.15,Z=10^{-3}$. Here, the amplitude of the pulse $Z$ is small enough to ensure it does not collapse into a BH. 
Finally, the outer boundary of the Cauchy grid is 41, coinciding with the worldtube.

\begin{figure}[htb]
        \includegraphics[width=\columnwidth,clip=true]{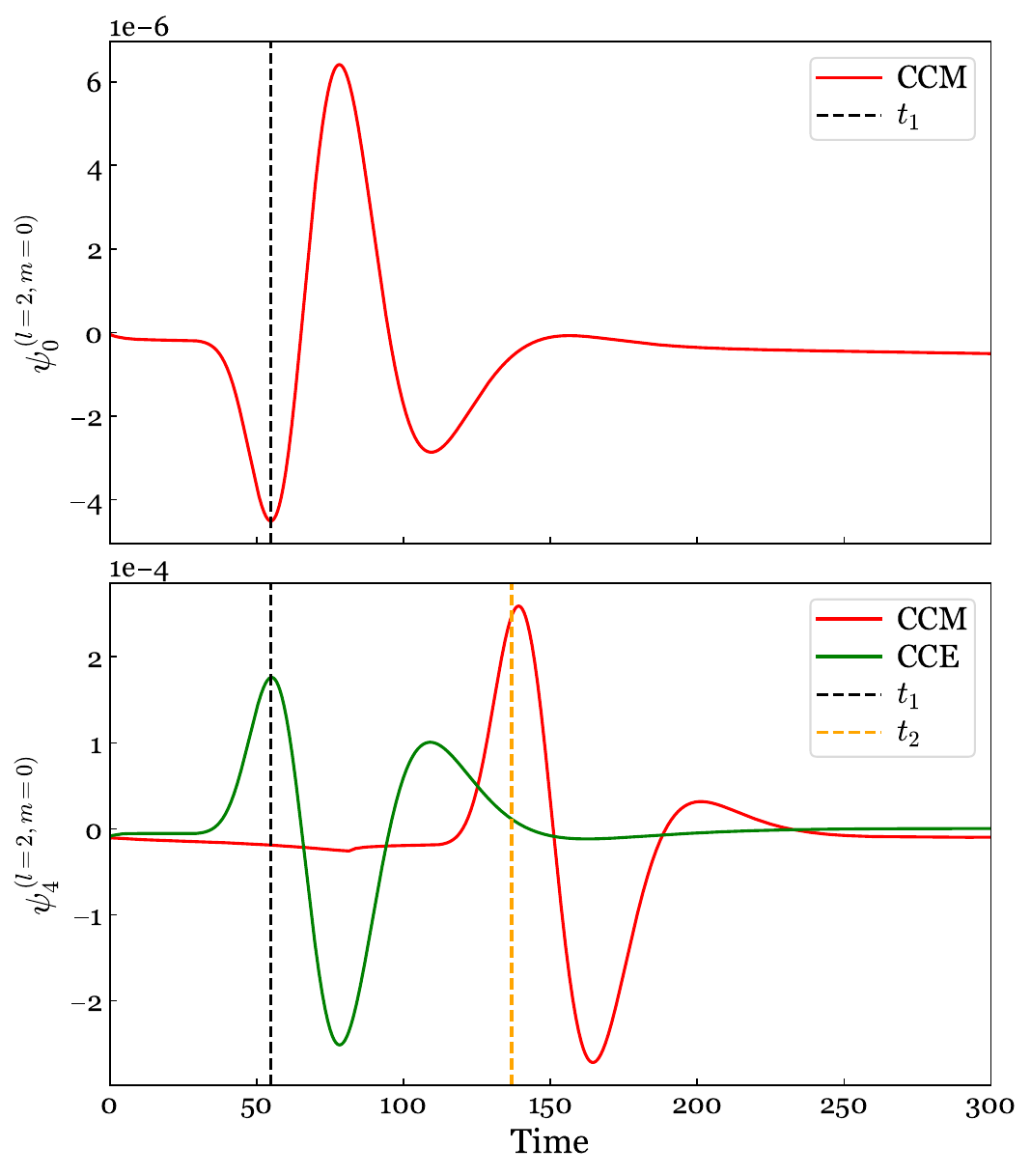}
  \caption{The $(l=2,m=0)$ harmonic of the Weyl scalar $\psi_0$ measured at the worldtube (top), and $\psi_4$ extracted at null infinity (bottom), from the system shown in Fig.~\ref{fig:Teukolsky_wave_on_cce}. The first trough of $\psi^{(l=2,m=0)}_{0}$, marked by the vertical blue-dashed line $(t_1)$, represents the time when the pulse reaches the outer boundary of the Cauchy domain. The vertical yellow-dashed line $(t_2=t_1+2R)$ indicates the time when the pulse again reaches the outer boundary of the Cauchy domain after passing through the origin and propagating back outward. } 
 \label{fig:pluse_on_cce_waveform}
\end{figure}

Since the pulse is imposed on an outgoing null surface, it naturally propagates inwards once we allow the system to evolve. By using the CCM algorithm, the interface between the Cauchy and characteristic grid becomes transparent to this incoming pulse. As a result, the GW is transmitted to the Cauchy domain through the matching term in the boundary condition [Eq.~\eqref{eq:bc_bjorhus}]. Subsequently, the pulse descends toward the center and bounces back. Eventually, it leaves the inner Cauchy region and disperses to null infinity after the crossing time of the Cauchy domain (namely its diameter). On the contrary, if we deactivate the matching term and evolve the system with the standard CCE algorithm, the worldtube will become a perfectly reflective mirror.  As a result, the inner Cauchy system remains unaffected by the incoming pulse, and Minkowski spacetime persists.

In the top panel of Fig.~\ref{fig:pluse_on_cce_waveform}, we plot the evolution of $\psi^{(l=2,m=0)}_{0}$ on the outer boundary of the Cauchy domain, which characterizes the incoming GW received by the inner Cauchy system. We identify the moment when $\psi^{(l=2,m=0)}_{0}$ reaches its first trough, labeled as $t_1$, as the point at which the pulse enters the Cauchy domain. The bottom panel of Fig.~\ref{fig:pluse_on_cce_waveform} exhibits the evolution of $\psi_{4}^{(l=2,m=0)}$ at future null infinity. In the absence of the matching (CCE), the ingoing pulse is entirely reflected by the worldtube --- the first peak of the reflected wave shows up at $t_1$. In contrast, for CCM, the reflected wave is significantly suppressed at that moment. After the crossing time of the inner Cauchy domain, which is its diameter $(2R)$, the pulse exits the Cauchy grid at $t_2=t_1+2R$ and disperses to null infinity. This result verifies that our CCM algorithm successfully directs the characteristic pulse into the Cauchy system.

To close this section, we present the evolution of the GH three-index and gauge constraints [Eq.~\eqref{eq:three_constraint} and \eqref{eq:Teukolsky_h_20}] in Fig.~\ref{fig:pluse_on_cce_waveform_con}. We see the constraints exhibit an initial rise soon after the simulation starts, followed by a plateau that persists throughout the duration of the simulation time under consideration. In addition, the constraints converge with the numerical resolutions.

\begin{figure}[htb]
        \includegraphics[width=\columnwidth,clip=true]{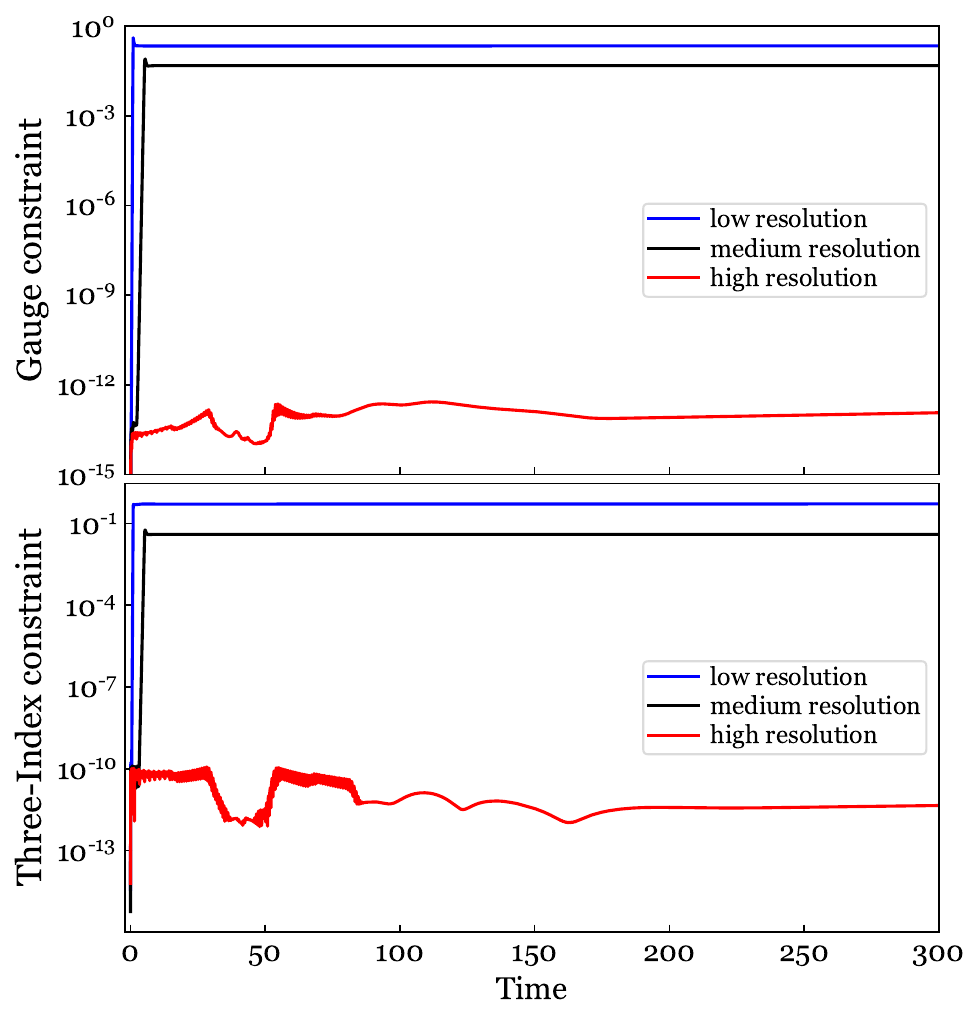}
  \caption{The evolution of the GH gauge constraint violation (top) and three-index constraint violation (bottom), as defined in Eqs.~\eqref{eq:gauge_constraint} and \eqref{eq:three_constraint}. We initialize a GW pulse on the characteristic grid (depicted in Fig.~\ref{fig:Teukolsky_wave_on_cce}), which subsequently propagates to the Cauchy grid. The system is evolved with three numerical resolutions.}
 \label{fig:pluse_on_cce_waveform_con}
\end{figure}

\section{Conclusions}
\label{sec:conclusion}
In this paper, we implemented a fully relativistic 3D CCM algorithm for physical degrees of freedom in the numerical relativity code SpECTRE. The method is generic and applicable to any physical system. Core steps towards matching involve (a) evaluating the Weyl scalar $\psi_0$ and the Newman-Penrose tetrad vector $m_{\mu}$ at the outer boundary of the Cauchy domain; (b) performing tetrad transformations to the Cauchy tetrad; (c) interpolating the quantities to the Cauchy grid; (d) computing the physical subset of the Cauchy Bjørhus boundary conditions.

To evaluate the performance and correctness of the CCM algorithm, various physical systems were designed and tested\footnote{Their simulation data are available in a Github repository \cite{CCMData}.}. These tests include the propagation of Teukolsky waves within a flat background, the perturbation of a Kerr BH with a Teukolsky wave, and the injection of a GW pulse from the characteristic grid. Notably, no numerical instabilities were found in the CCM simulations. 

Comparing the CCM algorithm to CCE, it was observed that the CCM algorithm did not alter the evolution of GH constraint violations (e.g., gauge and three-index constraints). However, it demonstrated the ability to suppress violations of Bondi-gauge constraints. When testing the propagation of a Teukolsky wave with a large amplitude $(X=2)$, where nonlinear effects are significant, the CCM waveforms exhibited better agreement with reference results, indicating more accurate boundary conditions for the Cauchy evolution. Additionally, the test involving the initialization of a GW pulse on the characteristic grid demonstrated the successful directing of characteristic information into the Cauchy system.

Our preliminary tests show that CCM allows for placing the outer boundary of a Cauchy system at smaller radii without significant loss of waveform precision, thus improving computational efficiency. However, the gauge aspect of the boundary conditions poses a major limitation in moving the outer boundary further inward. Currently we adopt the Sommerfeld condition that minimizes the reflection of a gauge wave at the outer boundary \cite{Rinne:2007ui}. However, it is less physically motivated and may lead to spurious reflections under some scenarios. Since the gauge information is in principle encoded in the characteristic system, future work could generalize the matching algorithm to include the gauge subset, which will enable the construction of more accurate boundary conditions for Cauchy evolution.

On the other hand, it has been shown that CCM is only weakly hyperbolic \cite{Giannakopoulos:2020dih,Giannakopoulos:2021pnh,Giannakopoulos:2023zzm}, suggesting potential numerical instabilities. In our simulations that involve only smooth data, we did not observe any instabilities in the tests conducted. It is possible that instabilities may manifest over longer timescales or during more violent gravitational processes, such as BH mergers, or processes that might produce high-frequency bursts, such as neutron-star mergers. Therefore, robust stability tests with high frequency data \cite{Calabrese:2005ft,soton393564,Cao:2011fu} would be a valuable avenue for future research.

Nevertheless, if CCM proves capable of handling BBH collisions, it will have the potential to enhance the accuracy and computational efficiency of future NR simulations. This advancement would be particularly significant for third-generation GW detectors \cite{Punturo:2010zz,Hild:2010id,Evans:2016mbw,Reitze:2019iox}.

\begin{acknowledgments}
We thank Keefe Mitman and Justin Ripley for useful discussions.  This
work was supported in part by the Sherman Fairchild Foundation and by
NSF Grants PHY-2011961, PHY-2011968, PHY-2309231, and OAC-2209655 at
Caltech.  The computations presented here were conducted using the
Resnick High Performance Computing Center, a facility supported by
Resnick Sustainability Institute at the California Institute of
Technology. Marceline S.~Bonilla and Geoffrey Lovelace acknowledge
support from NSF awards PHY-1654359 and PHY-2208014, the Dan Black
Family Trust, and Nicholas and Lee Begovich.
\end{acknowledgments}

\appendix
\counterwithin*{equation}{section}
\renewcommand\theequation{\thesection\arabic{equation}}

\section{Teukolsky wave in the perturbative limit: $X=10^{-5}$}
\label{app:1e-5}
In the perturbative limit, it is straightforward to obtain analytic expressions for the Weyl scalars, News, and strain for a Teukolsky wave:
\begin{subequations}
\label{eq:Teukolsky_weyl_strain_news}
\begin{align}
    &\psi_0=-\sqrt{\frac{2\pi}{15}}\tensor[_{+2}]{Y}{_{20}}\left[(6\ddot{C}-3\ddot{A})+\frac{1}{2}r(3\dddot{B}+\dddot{A})\right], \\
    &\psi_1=\frac{1}{2}\sqrt{\frac{2\pi}{15}}\tensor[_{+1}]{Y}{_{20}}\left[r\dddot{A}+3\ddot{B}\right] ,\\
    &\psi_2=-\sqrt{\frac{\pi}{5}}Y_{20}\ddot{A}, \\
    &\psi_3=\frac{1}{2}\sqrt{\frac{2\pi}{15}}\tensor[_{-1}]{Y}{_{20}}\left[r\dddot{A}-3\ddot{B}\right] ,\\
    &\psi_4=\sqrt{\frac{2\pi}{15}}\tensor[_{-2}]{Y}{_{20}}\left[(3\ddot{A}-6\ddot{C})+\frac{1}{2}r(3\dddot{B}+\dddot{A})\right], \\
    &N=-\sqrt{\frac{2\pi}{15}}\tensor[_{-2}]{Y}{_{20}}\left[(3\dot{A}-6\dot{C})+\frac{1}{2}r(3\ddot{B}+\ddot{A})\right],\\
    &h=-\sqrt{\frac{2\pi}{15}}\tensor[_{-2}]{Y}{_{20}}\left[(3A-6C)+\frac{1}{2}r(3\dot{B}+\dot{A})\right].
\end{align}
\end{subequations}
Here we have used the Appendix of \cite{PhysRevD.26.745}. The (spin-weighted) spherical harmonics $\tensor[_{s}]{Y}{_{lm}}$ are given by
\begin{align}
    &\tensor[_{-2}]{Y}{_{20}}=\tensor[_{+2}]{Y}{_{20}}=\frac{1}{4}\sqrt{\frac{15}{2\pi}}\sin^2\theta,\notag \\
    &\tensor[_{-1}]{Y}{_{20}}=\tensor[_{+1}]{Y}{_{20}}=-\frac{1}{4}\sqrt{\frac{15}{2\pi}}\sin2\theta,\notag\\
    &Y_{20}=\frac{1}{8}\sqrt{\frac{5}{\pi}}(1+3\cos2\theta).\notag
\end{align}
By substituting Eqs.~\eqref{eq:Teukolsky_ABC}, Eqs.~\eqref{eq:Teukolsky_weyl_strain_news} can be simplified at future null infinity $\mathscr{I}^+$:
\begin{subequations}
\label{eq:Teukolsky_scri_weyl_strain_news}
\begin{align}
    &rh|_{\mathscr{I}^+}=\sqrt{\frac{6\pi}{5}}F^{(4)}\times\tensor[_{-2}]{Y}{_{20}}, \\
    &rN|_{\mathscr{I}^+}=\sqrt{\frac{6\pi}{5}}F^{(5)}\times\tensor[_{-2}]{Y}{_{20}}, \\
    &r\psi_4|_{\mathscr{I}^+}=-\sqrt{\frac{6\pi}{5}}F^{(6)}\times\tensor[_{-2}]{Y}{_{20}}, \\
    &r^2\psi_3|_{\mathscr{I}^+}=\sqrt{\frac{6\pi}{5}}F^{(5)}\times\tensor[_{-1}]{Y}{_{20}}, \\
    &r^3\psi_2|_{\mathscr{I}^+}=-\sqrt{\frac{9\pi}{5}}F^{(4)}\times Y_{20}, \\
    &r^4\psi_1|_{\mathscr{I}^+}=\sqrt{\frac{27\pi}{10}}F^{(3)}\times \tensor[_{+1}]{Y}{_{20}},\\
    &r^5\psi_0=-\sqrt{\frac{27\pi}{10}}F^{(2)}\times\tensor[_{+2}]{Y}{_{20}}. \label{eq:Teukolsky_bulk_psi0}
\end{align}
\end{subequations}
Note that Eq.~\eqref{eq:Teukolsky_bulk_psi0} is in fact valid throughout the entire spacetime. It shows that $\psi_0$ follows  a fifth power decay law with distance. Consequently,
the backscattered wave
decays quickly as one moves away from the origin.  

We focused on comparing the strain in Sec.~\ref{subsec:Teukolsky_X_1e-5}. Below in Figs.~\ref{fig:low_amp_ccm_results1} and \ref{fig:low_amp_ccm_results2}, we complete the comparisons by providing the results of the Weyl scalars $\psi_{0...4}$ and the News $N$.

\begin{figure*}[htb]
    \includegraphics[width=\columnwidth,clip=true]{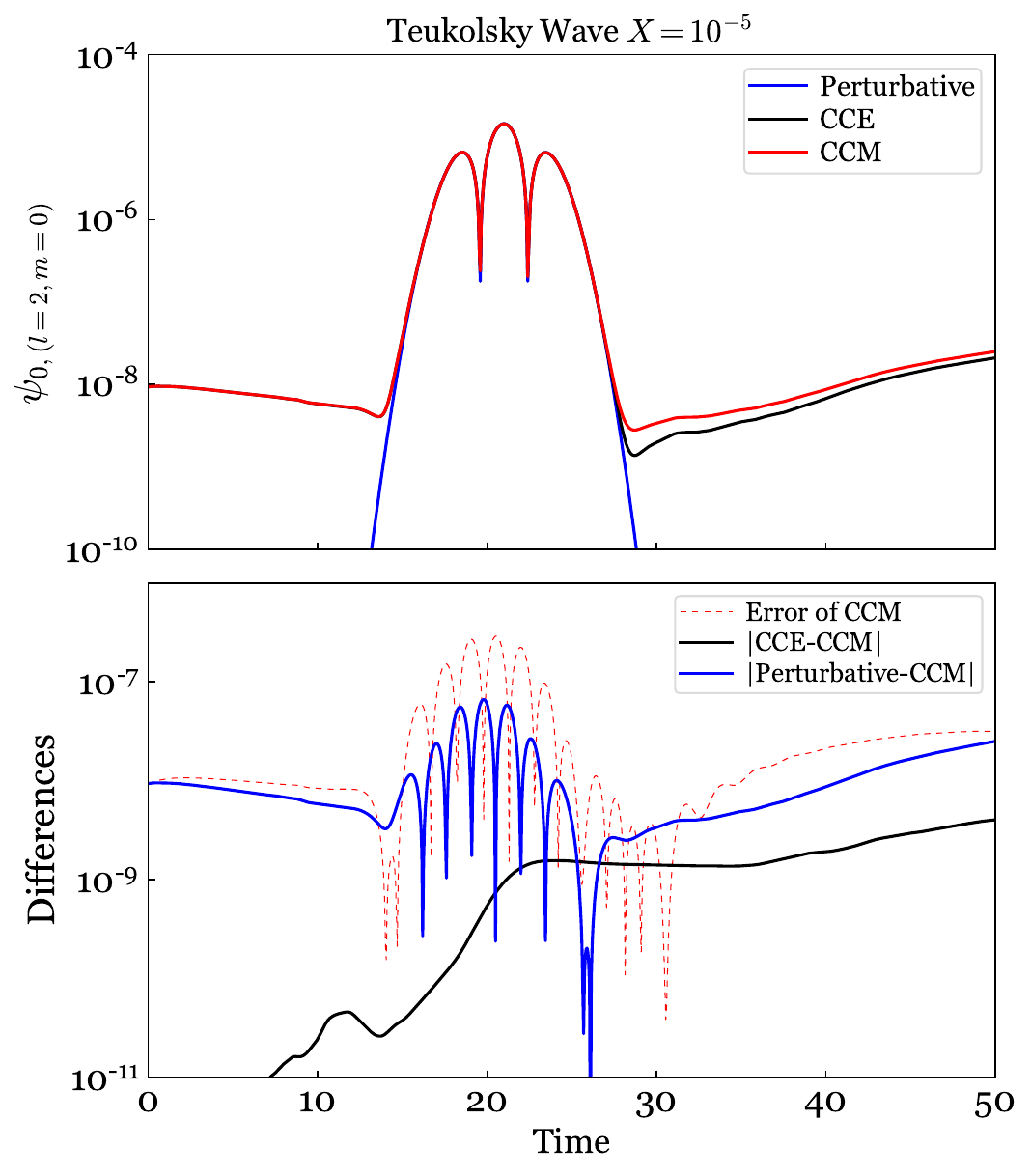}
    \includegraphics[width=\columnwidth,clip=true]{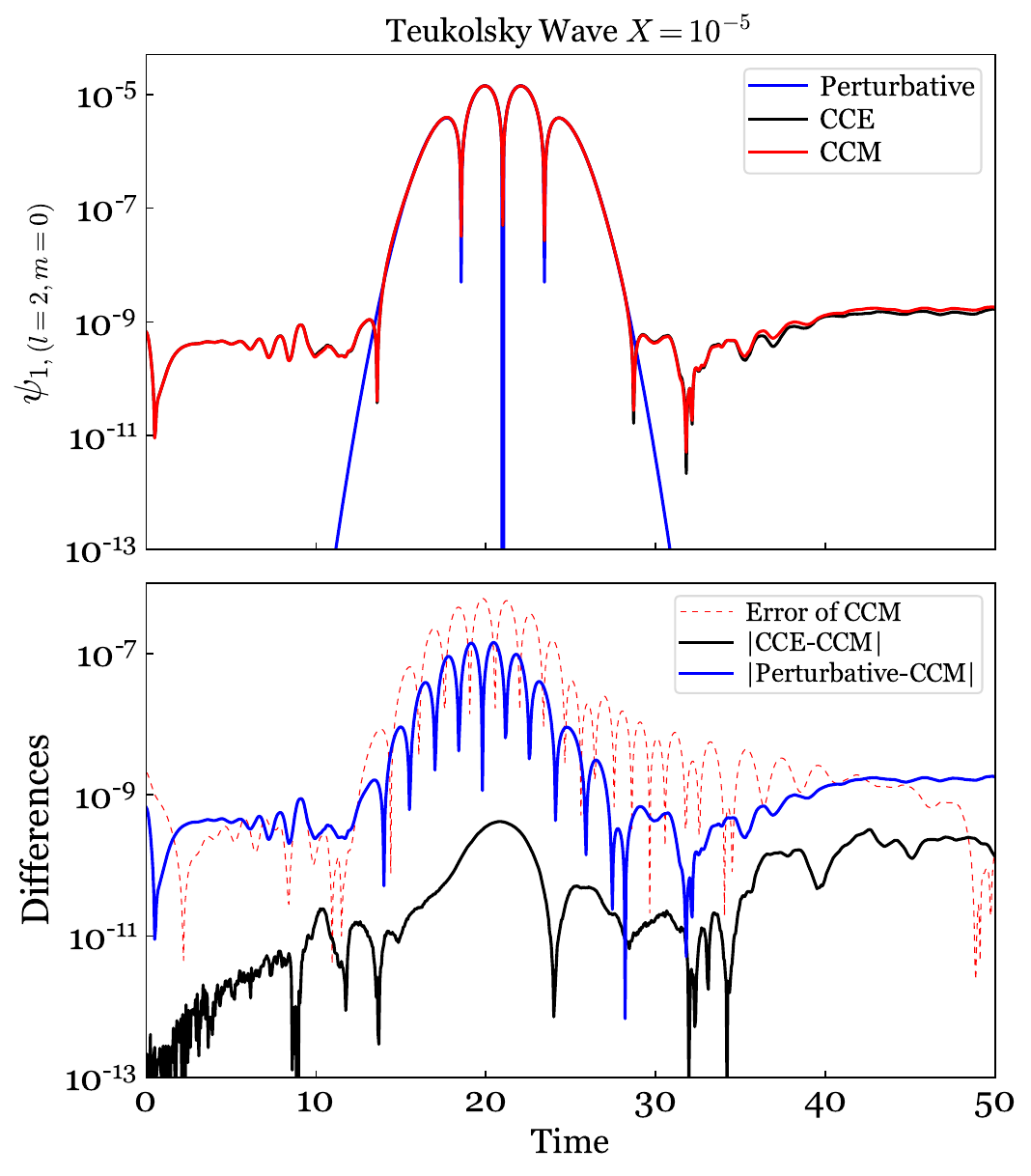} \\
    \includegraphics[width=\columnwidth,clip=true]{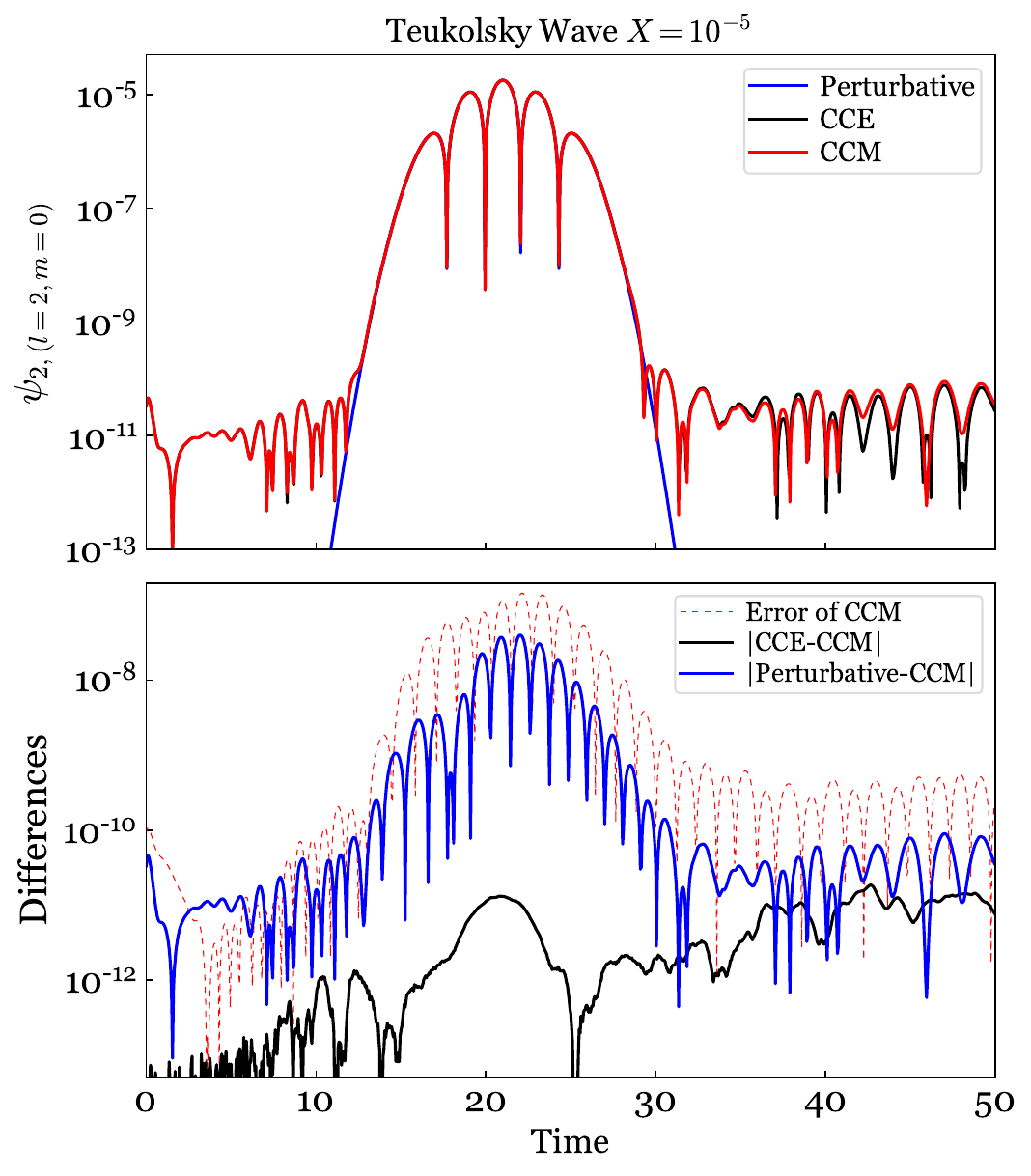}
    \includegraphics[width=\columnwidth,clip=true]{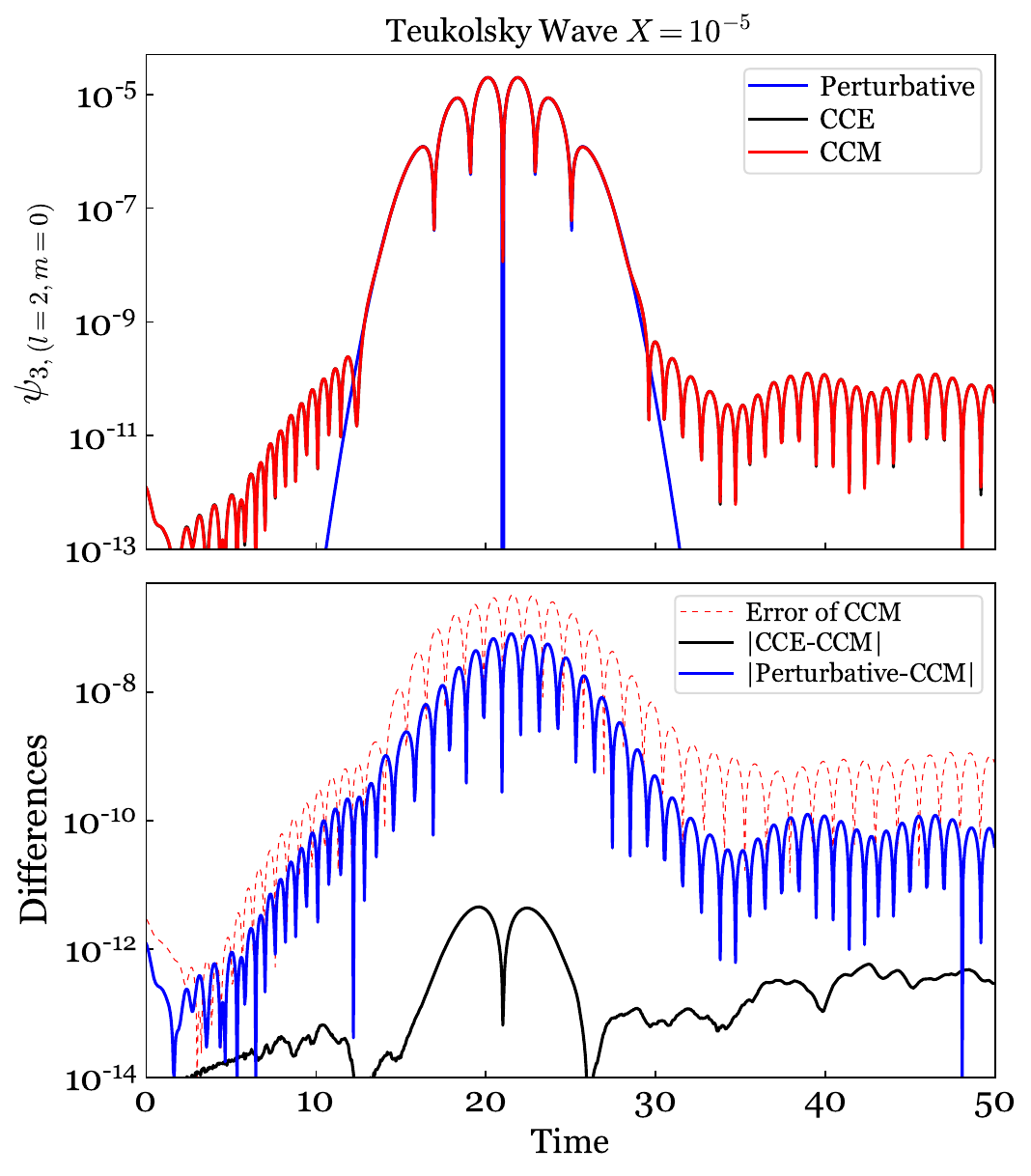}
  \caption{Continuation of Fig.~\ref{fig:low_amp_strain_only}. The Weyl scalars of the Teuskolsky wave with an amplitude of $X=10^{-5}$. }
 \label{fig:low_amp_ccm_results1}
\end{figure*}

\begin{figure*}[htb]
    \includegraphics[width=\columnwidth,clip=true]{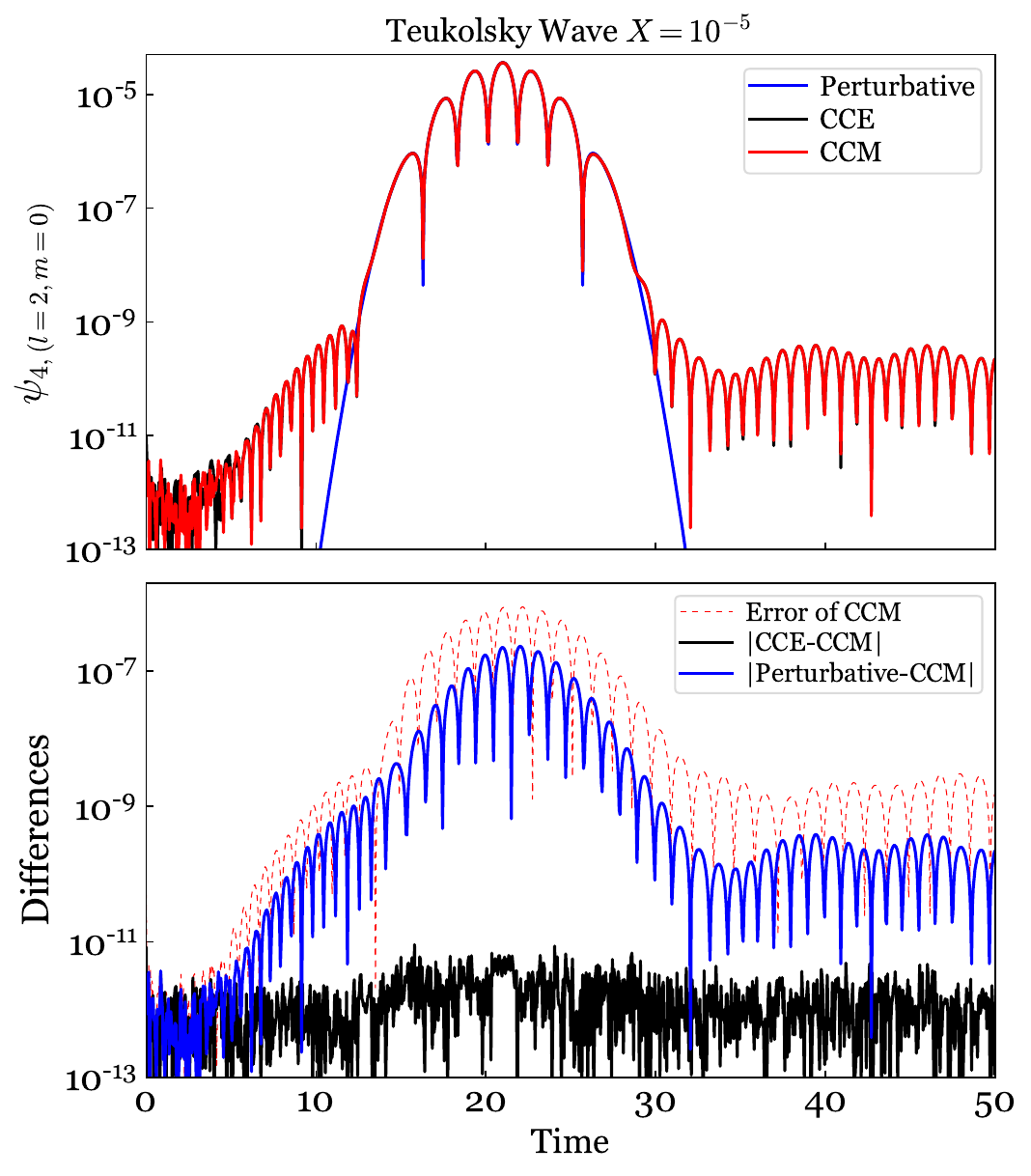} 
    \includegraphics[width=\columnwidth,clip=true]{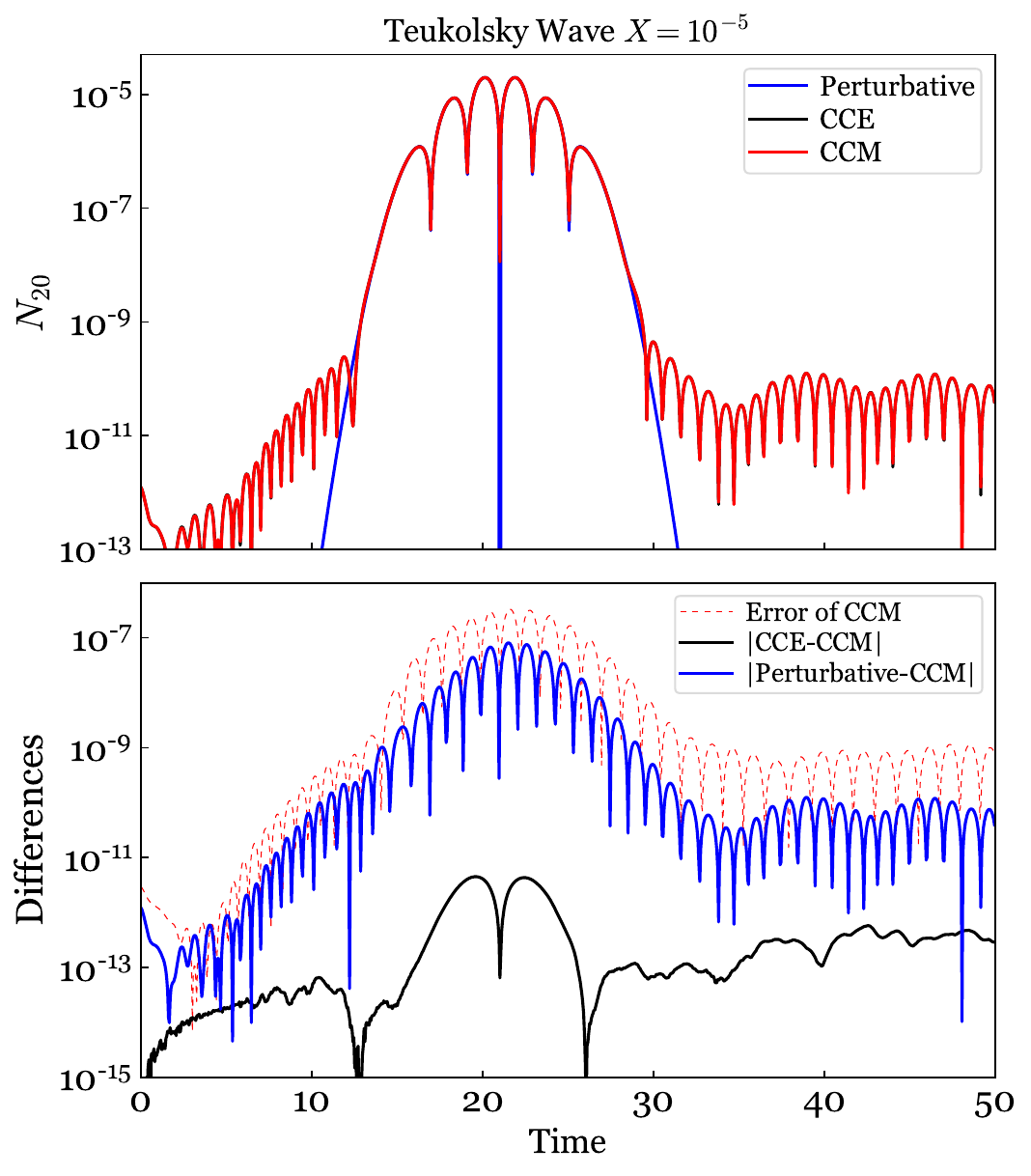}
  \caption{Continuation of Fig.~\ref{fig:low_amp_ccm_results1}.  The Weyl scalars of the Teuskolsky wave with an amplitude of $X=10^{-5}$.}
 \label{fig:low_amp_ccm_results2}
\end{figure*}

\begin{figure*}[htb]
    \includegraphics[width=\columnwidth,clip=true]{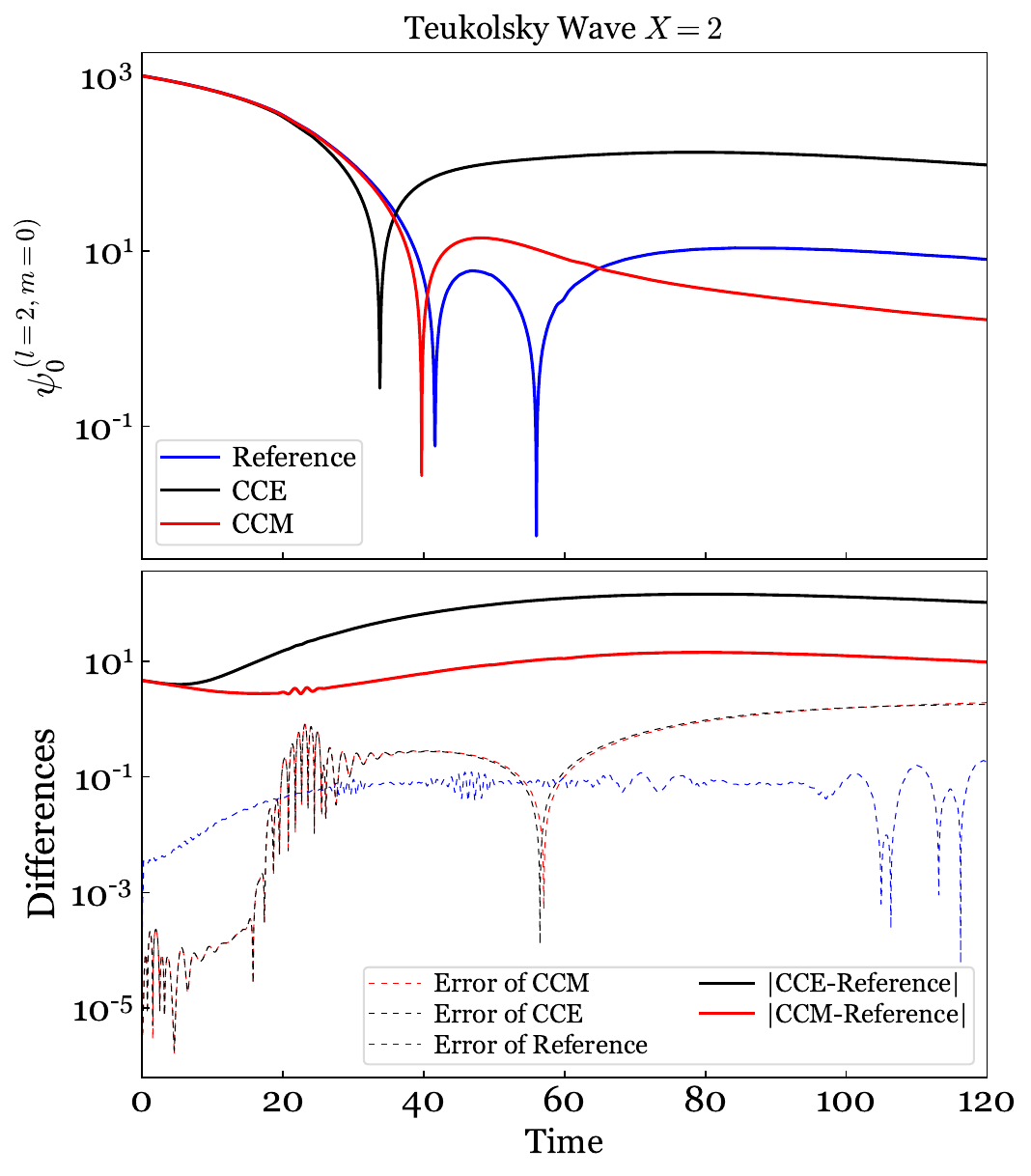}
    \includegraphics[width=\columnwidth,clip=true]{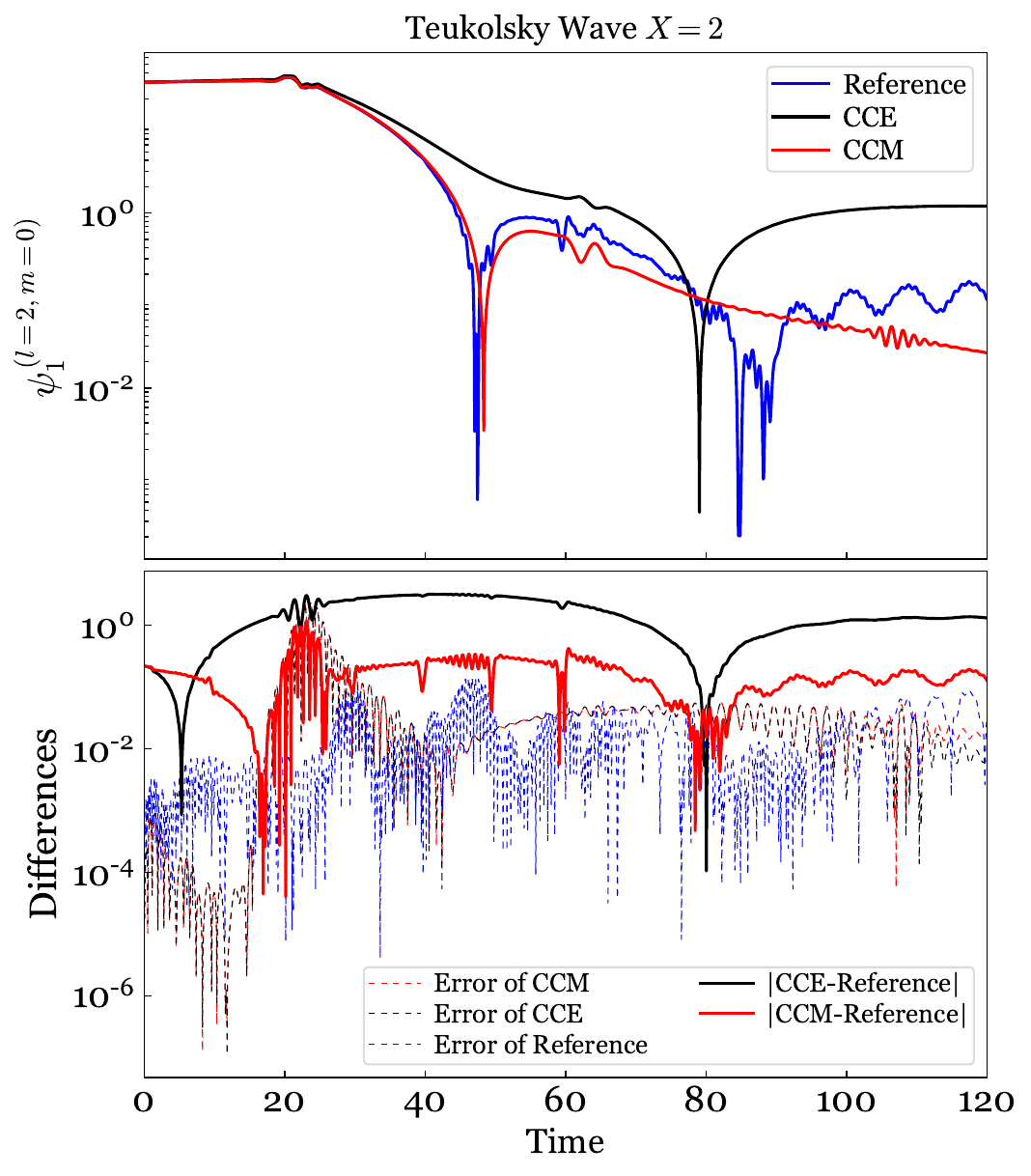} \\
    \includegraphics[width=\columnwidth,clip=true]{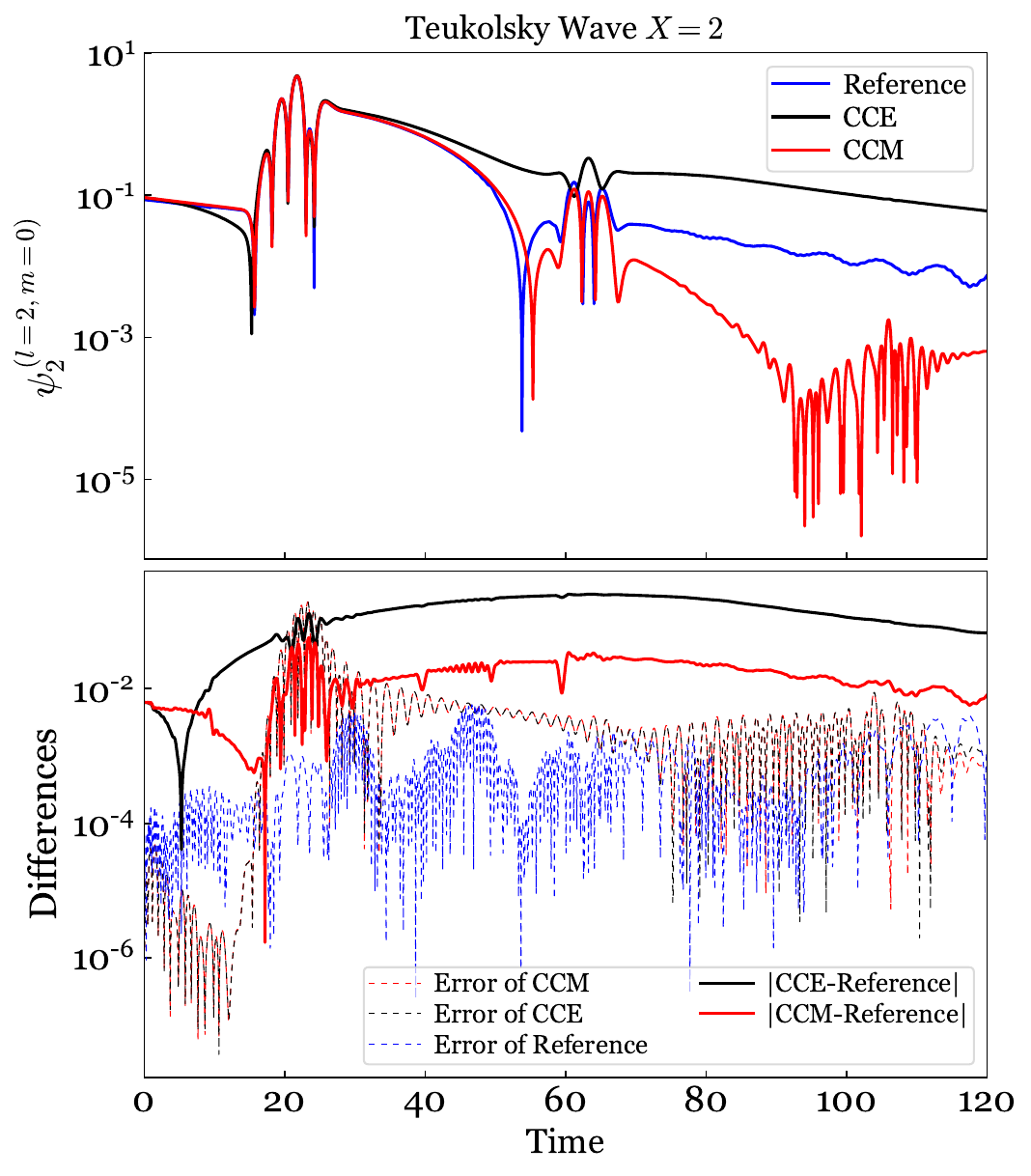}
    \includegraphics[width=\columnwidth,clip=true]{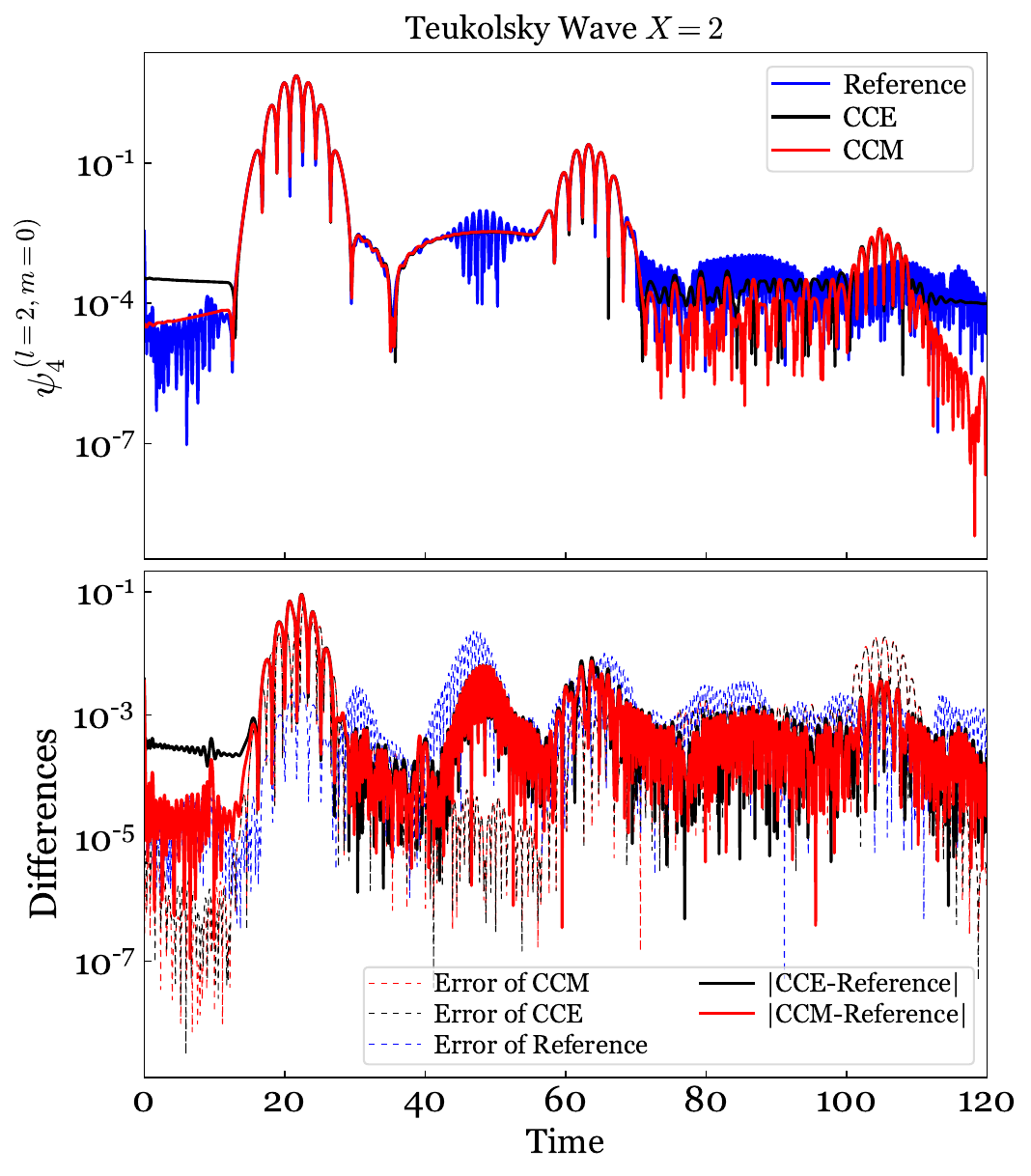} 
  \caption{Continuation of Fig.~\ref{fig:large_amp_strain_only}.  The Weyl scalars of the Teuskolsky wave with an amplitude of $X=2$.}
 \label{fig:ccm_results1}
\end{figure*}

\section{Teukolsky wave in the nonlinear regime: $X=2$}
\label{app:X=2}
Figures \ref{fig:ccm_results1} and \ref{fig:ccm_results2} display the Weyl scalars and News of the Teukolsky wave with an amplitude of $X=2$, complementing the discussions in Fig.~\ref{fig:large_amp_strain_only}.

\begin{figure}[htb]
    \includegraphics[width=\columnwidth,clip=true]{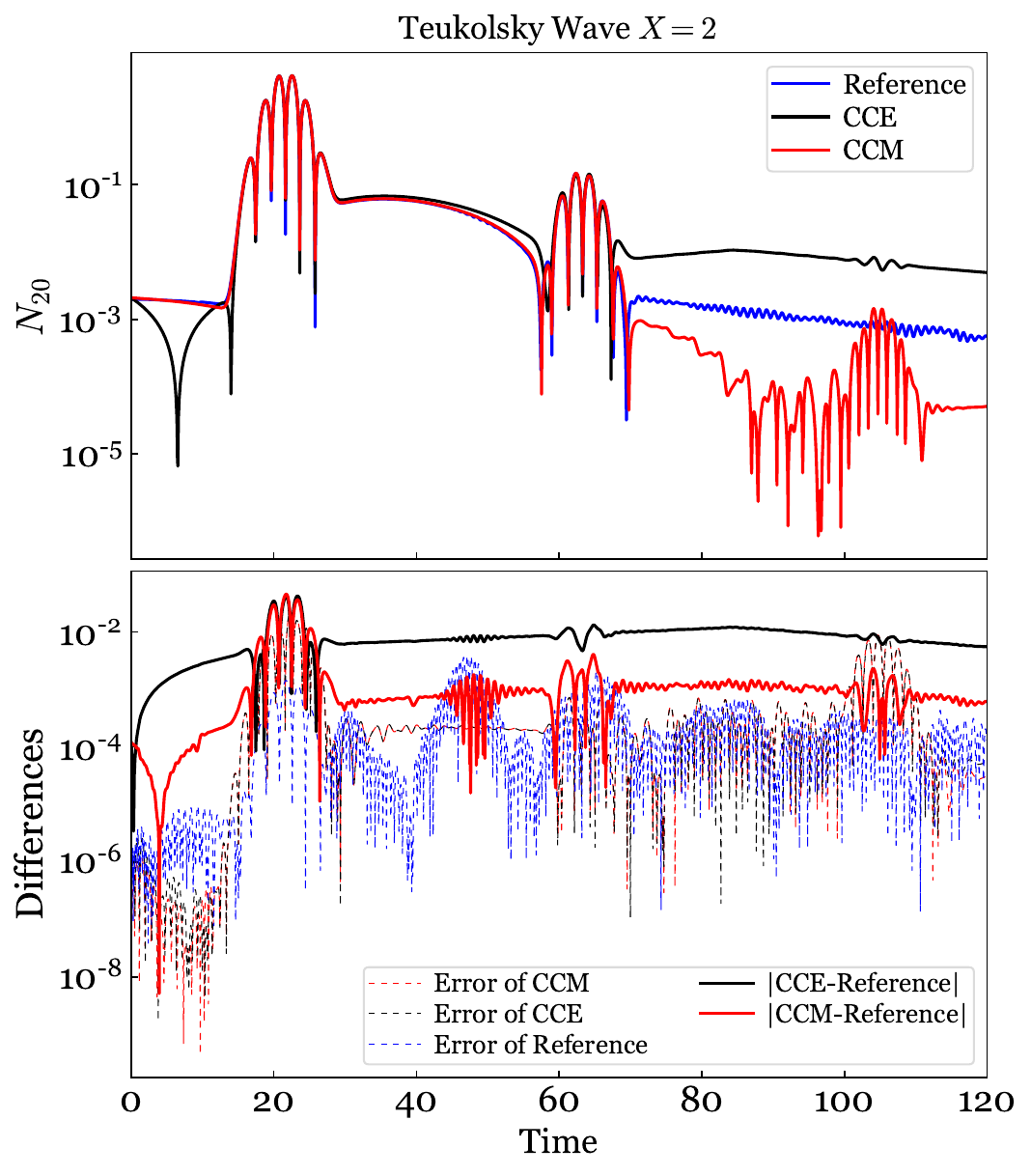}
  \caption{Continuation of Fig.~\ref{fig:ccm_results1}. The News of the Teuskolsky wave with an amplitude of $X=2$.}
 \label{fig:ccm_results2}
\end{figure}

\section{Perturbing a Kerr BH with a Teukolsky wave}
\label{app:kerr_bh}
Figures \ref{fig:app_kerr1} and \ref{fig:app_kerr2} display the Weyl scalars, News, and strain of the $\chi=0.5$ Kerr BH perturbed by the  Teukolsky wave, complementing the discussion in Fig.~\ref{fig:kerr_waveform}.

\begin{figure*}[htb]
    \includegraphics[width=\columnwidth,clip=true]{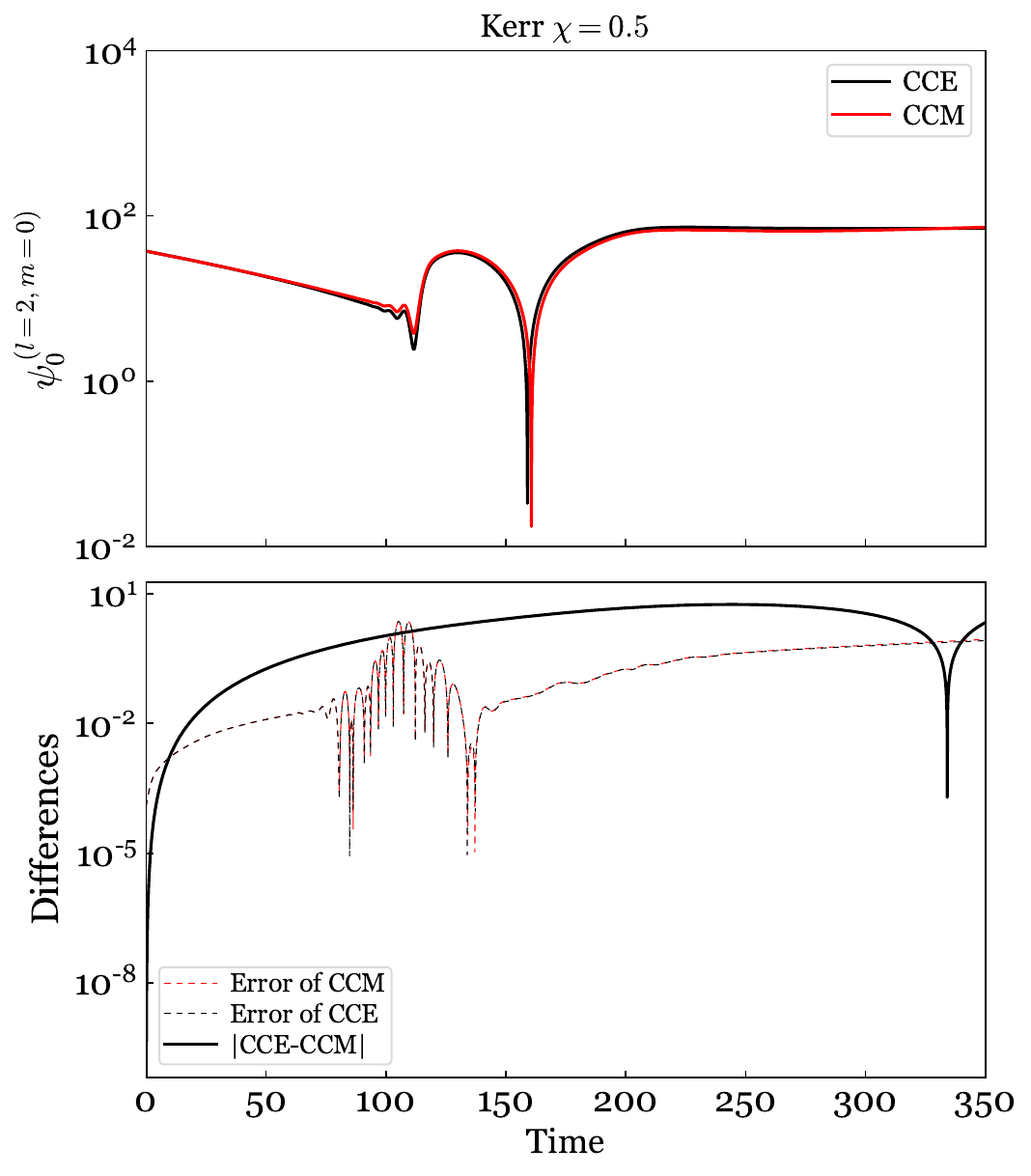}
    \includegraphics[width=\columnwidth,clip=true]{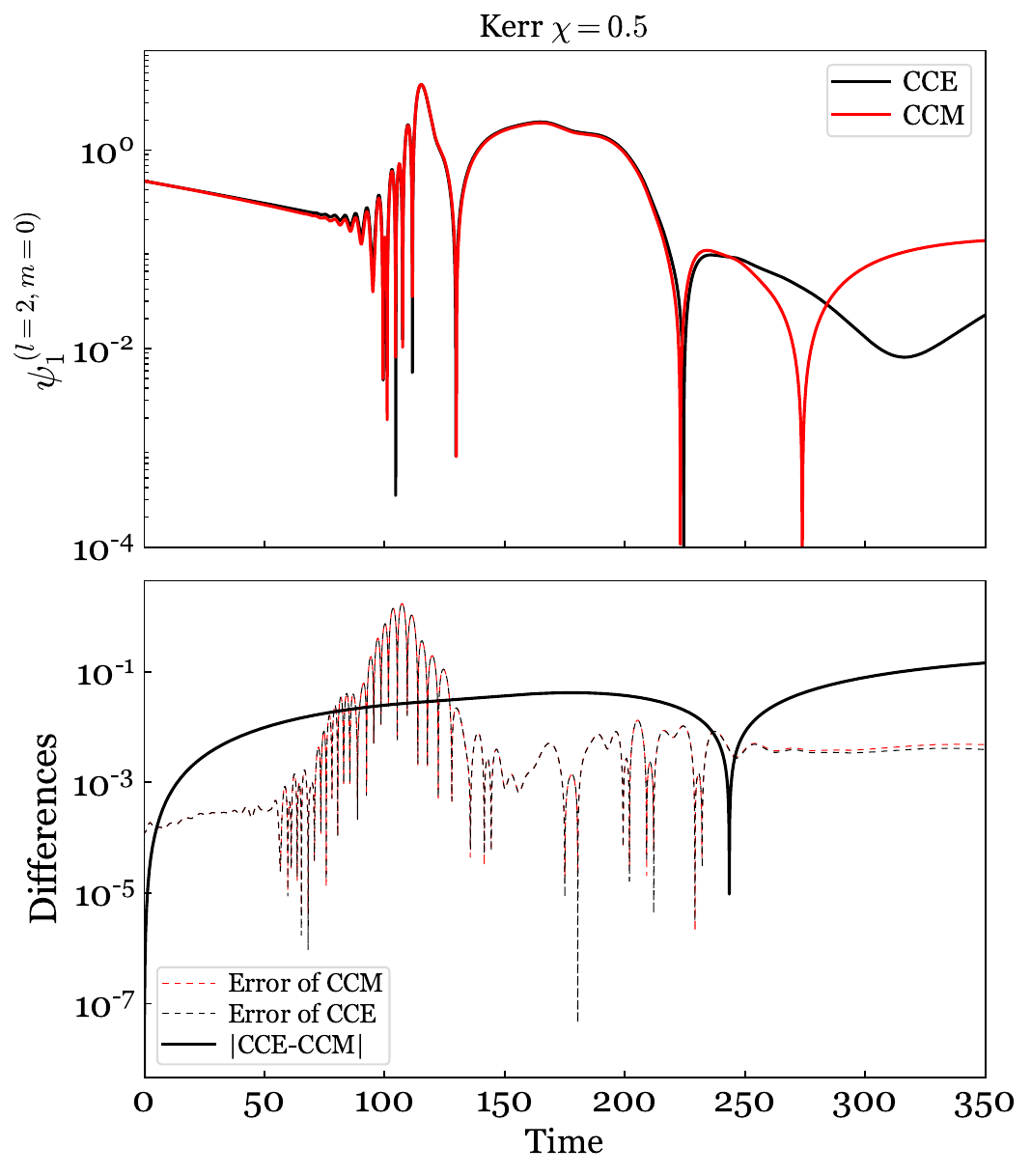} \\
    \includegraphics[width=\columnwidth,clip=true]{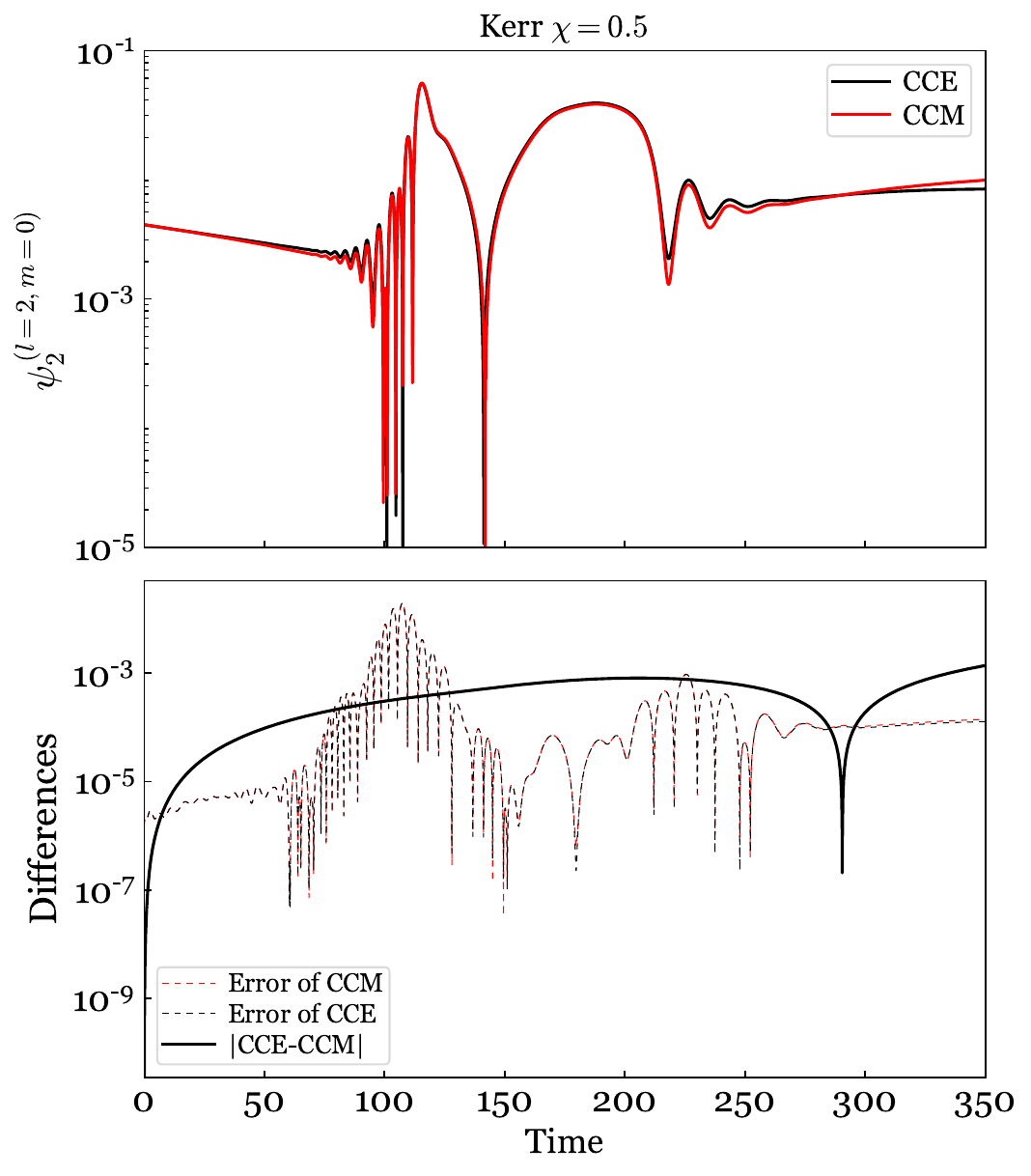}
    \includegraphics[width=\columnwidth,clip=true]{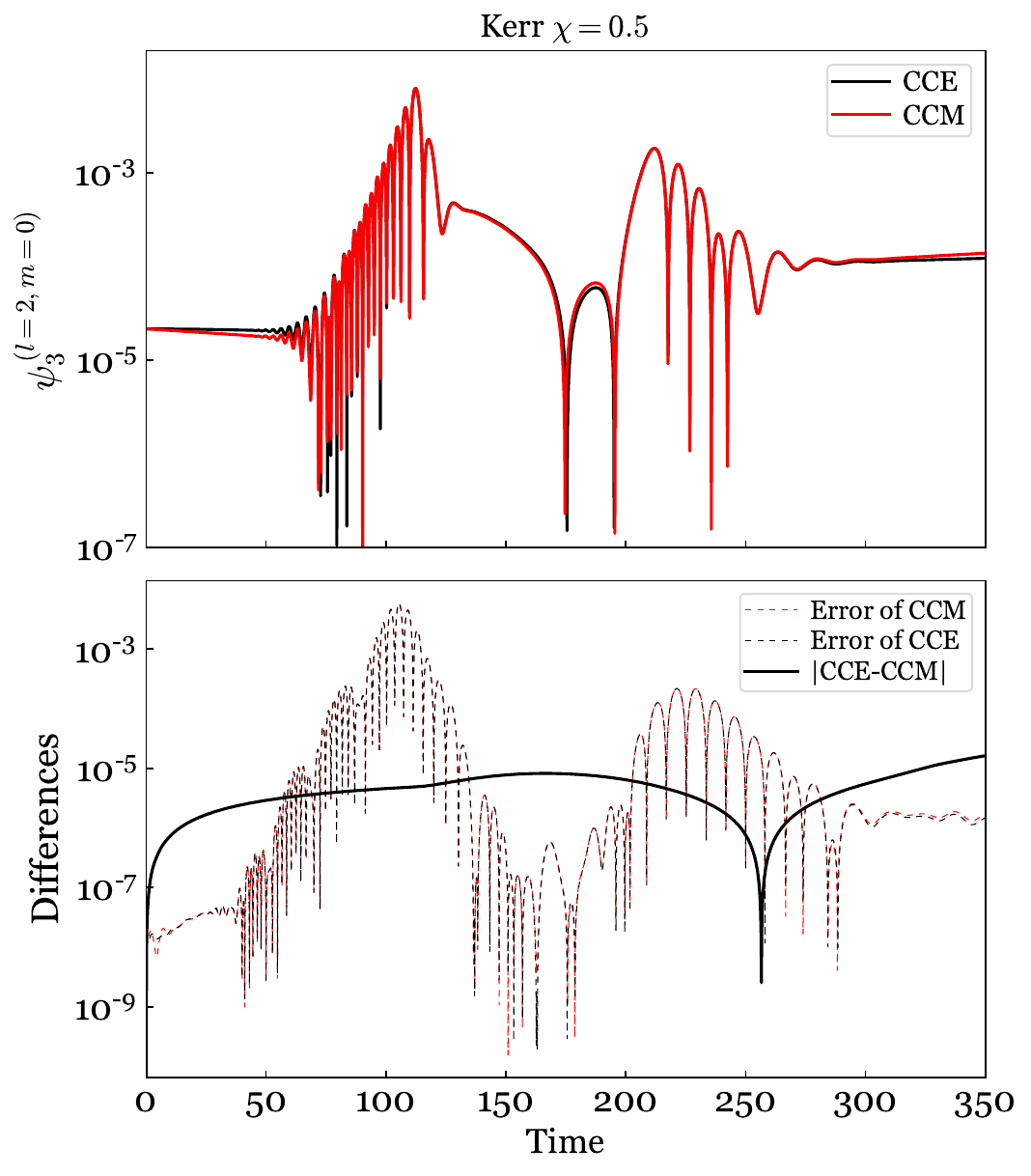}
  \caption{Continuation of Fig.~\ref{fig:kerr_waveform}. The Weyl scalars of the $\chi=0.5$ Kerr BH perturbed by the Teukolsky wave.}
 \label{fig:app_kerr1}
\end{figure*}

\begin{figure*}[htb]
    \includegraphics[width=\columnwidth,clip=true]{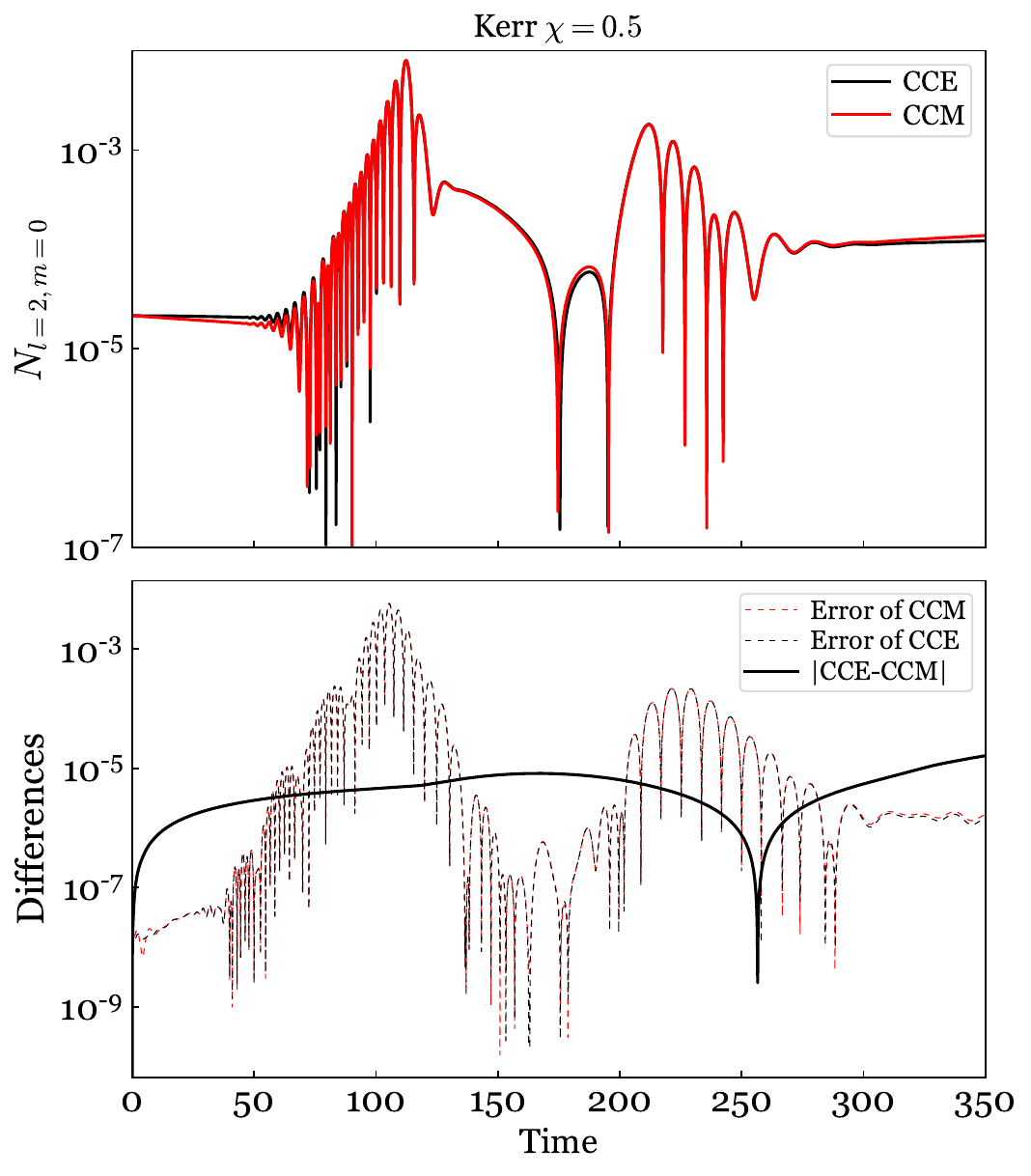}
    \includegraphics[width=\columnwidth,clip=true]{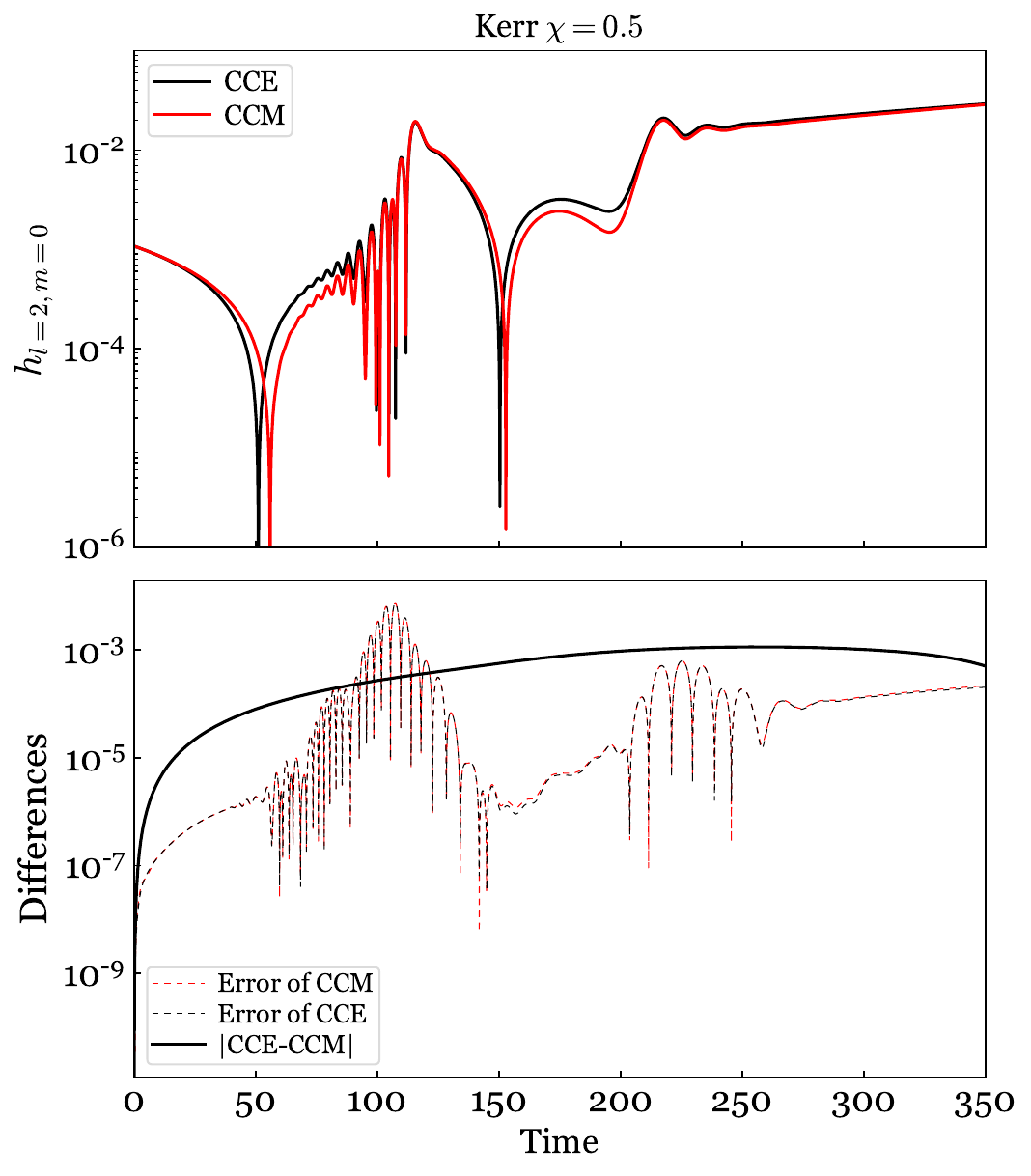} 
  \caption{Continuation of Fig.~\ref{fig:app_kerr1}. The News and strain of the $\chi=0.5$ Kerr BH perturbed by the Teukolsky wave.}
 \label{fig:app_kerr2}
\end{figure*}


\def\bibsection{\section*{References}}
\bibliography{References}

\end{document}